\documentclass[11pt]{article}

\usepackage{amsmath}
\usepackage{graphicx}
\usepackage{indentfirst}
\usepackage{amssymb}
\usepackage{cite}
\usepackage{color}
\usepackage{subfigure}
\usepackage{varwidth}
\usepackage{subfigure}
\usepackage[colorlinks=true, linkcolor=red, citecolor=blue, urlcolor=magenta]{hyperref}
\usepackage{xcolor}%参考文献
\usepackage{float} 
\usepackage{caption}
\begin{document}
\title{Optical Characteristics of the Kerr-Bertotti-Robinson Black Hole}

\date{}
\maketitle

\begin{center}
\author{Xiao-Xiong Zeng}$^{a,}$\footnote{E-mail: xxzengphysics@163.com},
\author{Chen-Yu Yang}$^{b,}$\footnote{E-mail: chenyuyang\_2024@163.com},
\author{Hao Yu}$^{c,}$\footnote{E-mail: yuhaocd@cqu.edu.cn (Corresponding author)}
\\

\vskip 0.25in
$^{a}$\it{College of Physics and Electronic Engineering, Chongqing Normal University, Chongqing 401331, China}\\
$^{b}$\it{Department of Mechanics, Chongqing Jiaotong University, Chongqing 400000, China}\\
$^{c}$\it{Physics Department, Chongqing University, Chongqing 401331, China}\\
\end{center}

\abstract{The Kerr-Bertotti-Robinson (Kerr-BR) black hole, a theoretical model of a rotating black hole immersed in a uniform magnetic field, has been proposed recently by Podolsky and Ovcharenko. This study investigates the optical characteristics of the Kerr-BR black hole based on the exact solution. We analyze the black hole shadow and the optical image under two illumination models: a celestial light source and a geometrically thin accretion disk. We reveal distinct roles for the fundamental parameters in the model. Specifically, it is found that under both illumination models, the affect of the rotation parameter on the optical image of the Kerr-BR black hole is significant different from that of the magnetic field. As the magnetic field increases, the radii of both the shadows and the Einstein rings enlarge. We also attempt to use the data from M87* and Sgr A* to constrain the magnetic field. These results enhance our comprehension on the optical characteristics of the Kerr-BR black hole and establish a theoretical foundation for interpreting future observations on the optical image of the black hole immersed in a uniform magnetic field. Finally, we point out that with advances in the resolution of black hole images, it is possible to detect potentially existing BR-like magnetic fields around black holes.}

\thispagestyle{empty}
\newpage
\setcounter{page}{1}

%%%%%%%%%%%%%%%%%%%%%%%%%%%%%%%%%%%%%%%%%%%%%%%%%%%%%%%%%%%%%%%%%%%%%%%%
%%%%%%%%%%%%%%%%%%%%%%%%%%%%%%%%%%%%%%%%%%%%%%%%%%%%%%%%%%%%%%%%%%%%%%%%
%%%%%%%%%%%%%%%%%%%%%%%%%%%%%%%%%%%%%%%%%%%%%%%%%%%%%%%%%%%%%%%%%%%%%%%%

\section{Introduction}
The groundbreaking images of the supermassive black holes M87* and Sgr A* released by the Event Horizon Telescope (EHT) have ushered in a new era of observational black hole astrophysics~\cite{EventHorizonTelescope:2019dse,EventHorizonTelescope:2022wkp}. These images, revealing characteristic dark shadows surrounded by bright, asymmetric emission rings, provide unprecedented opportunities to test general relativity in the strong-field regime and probe complex plasma environments surrounding compact objects~\cite{Gralla:2019xty}. The observed black hole shadow encodes important information about the black hole's spacetime geometry, spin orientation, and surrounding matter distribution~\cite{Falcke:1999pj,EventHorizonTelescope:2020qrl,Zeng:2022pvb,Zeng:2021mok,Zeng:2021dlj,Zeng:2025nmu}. Consequently, establishing accurate theoretical models of black hole shadows under realistic astrophysical conditions has become a frontier in gravitational physics. 

The Kerr metric~\cite{Kerr:1963ud}, which describes an isolated, rotating black hole in vacuum, has long served as the standard theoretical framework for interpreting shadow observations. Extensive studies have established that the Kerr shadow's characteristic deformation from circularity is governed by the rotation parameter, while its overall size scales with mass~\cite{Chandrasekhar:1985kt,Synge:1966okc,Hioki:2009na,Atamurotov:2013sca,Gralla:2017ufe,Wang:2017hjl,Tsukamoto:2017fxq,Chen:2020qyp,Wei:2019pjf,Chang:2021ngy,Afrin:2021imp,Kuang:2022xjp,Zheng:2024ftk}. The observed brightness asymmetry, particularly the prominent ``photon ring'' substructure and ``crescent-like'' features, arises from relativistic effects including Doppler boosting, gravitational lensing, and frame-dragging within the accretion flow~\cite{Luminet:1979nyg, Cui:1997zs,Bisnovatyi-Kogan:2017kii,Kraniotis:2019ked,Wong:2020ziu,Kuang:2022ojj}. However, real astrophysical black holes are not isolated and reside in magnetized environments. Strong, ordered magnetic fields are ubiquitous in active galactic nuclei and X-ray binaries, playing crucial roles in launching jets, governing accretion dynamics, and enabling energy extraction mechanisms such as the Blandford-Znajek process~\cite{Blandford:1977ds,Begelman:1984mw,Gammie:2003rj}. Ignoring these fields potentially limits the accuracy of spacetime parameter estimation from shadow observations. Additionally, progress has been made in studying black hole shadows in magnetic fields, such as investigations of the Schwarzschild-Melvin solution that reveal the distorting effects of magnetic fields on shadow shapes~\cite{Junior:2021dyw}. Wang et al. conducted a systematic study on the influence of the magnetic field on the shadow and stable photon orbits when a Kerr black hole is immersed in a Melvin magnetic field~\cite{Wang:2021ara}. Further, the influence of the magnetic field on the black hole shadow could be investigated in complex scenarios, such as adding new effects~\cite{Zhong:2021mty}, studying more complex black holes~\cite{Junior:2021svb}, considering other geometric backgrounds~\cite{Taylor:2025ixw}, and so forth. There are also some studies on magnetized black hole shadows focusing on perturbative approaches or numerical spacetimes~\cite{Yumoto:2012kz}, or specific illumination models~\cite{Hu:2020usx}.

Recently, Podolsky and Ovcharenko provided an elegant exact solution within Einstein theory to describe a Kerr black hole immersed in a uniform, axis-aligned magnetic field~\cite{Bertotti:1959pf,Robinson:1959ev,Kunduri:2013gce}, called as the Kerr-Bertotti-Robinson (Kerr-BR) spacetime~\cite{Podolsky:2025tle}. This new spacetime is still of algebraic type D~\cite{Griffiths:2005qp,Podolsky:2006px,Podolsky:2022xxd,Wu:2024tuh,Ovcharenko:2025fxg}. This solution interpolates between key limiting cases: the pure Kerr metric, the Schwarzschild metric in a magnetic field, and the static BR universe. In fact, research on a black hole being in a uniform magnetic field can be traced back to 1974 proposed by Wald~\cite{Wald:1974np}. But in Wald's work, the background spacetime is not the BR spacetime. It is worth noting that although black holes are all in magnetic fields, the spacetime structures differ essentially due to the different configurations of the magnetic field. For example, for the magnetic field configuration in the BR spacetime, the symmetry satisfies AdS$_2\times$S$^2$~\cite{Bertotti:1959pf,Robinson:1959ev,Kunduri:2013gce,Garfinkle:2011mp}, whereas for the Melvin magnetic field~\cite{Melvin:1963qx,Melvin:1965zza,Ernst:1976bsr} (a cylindrical symmetry magnetic field), the spacetime only has the cylindrical symmetry. As we known, black holes in different magnetic fields exhibit entirely different optical properties~\cite{Wang:2021ara, Junior:2021dyw, Zhong:2021mty}. Although the BR spacetime is a highly idealized model, studying BR-like spacetimes is of great significance for understanding the physics near black hole horizons, because the near-horizon geometry of a non-rotating, extremally charged black hole can decompose into the direct product AdS$_2$ × S$^2$~\cite{Maldacena:1998uz,Clement:2000ms,Clement:2001gia,deCesare:2024csp}. Moreover, due to its unique geometric configuration in BR spacetime, it has not only garnered significant attention in (two-dimensional) AdS/CFT correspondence~\cite{Maldacena:1998uz, Strominger:1998yg, Spradlin:1999bn} and string theory~\cite{Cadoni:1994uf, Navarro-Salas:1999zer, Caldarelli:2000xk}, but also holds substantial importance for our understanding of quantum field theory~\cite{Ottewill:2012mq}. Therefore, detecting BR(-like) magnetic fields in astronomy holds particular significance. With the emergence of an exact solution of the Kerr-BR spacetime, the energy extraction, geodesic and shadow of the Kerr-BR black hole have drawn attention~\cite{Zeng:2025olq,Wang:2025vsx}. In this work, we investigate the optical image of the Kerr-BR black hole, aiming to analyze the influence of the magnetic field on the optical image and provide insights for future detection of potential BR-like magnetic fields around black holes.

In addition,  systematic analysis of how the combined effects of the rotation parameter and the external magnetic field manifest in the shadow's size, shape, and brightness under different illumination in the Kerr-BR black hole is lacking. This gap also motivates our present work. We conduct a comprehensive study of the optical characteristics of the Kerr-BR black hole. Our primary objective is distinguishing the roles of the rotation parameter and the magnetic field in the optical image of the black hole. We will consider two astrophysically relevant illumination scenarios: celestial light source illumination and thin accretion disk illumination, where both prograde and retrograde configurations are condidered.

The structure of the paper is as follows. In Section II, we briefly review the exact solution of the Kerr-BR black hole. Section III is about the shadow of the Kerr-BR black hole and we analyze the influence of the rotational parameter and the magnetic field on the shadow contour. In Sections IV and V, we study the optical images of the Kerr-BR black hole under celestial light source illumination and thin accretion disk illumination, respectively. In Section VII, we constrain the magnetic field by the data from M87* and Sgr A*. The last section presents a summary of this work.

\section{An Exact Solution of the Kerr-Bertotti-Robinson Black Hole}
Within the framework of Einstein's gravity, a Kerr black hole immersed in a uniform magnetic (or electric) field aligned with its rotation axis, referred to as a Kerr-BR black hole, admits an exact solution~\cite{Podolsky:2025tle}:
\begin{equation}
	\mathrm{d}s^2=\frac{1}{\Omega^2}\Bigg[-\frac{Q}{\rho^2}\left(\mathrm{d}t-a\sin^2\theta\mathrm{d}\varphi\right)^2+\frac{\rho^2}{Q}\mathrm{d}r^2+\frac{\rho^2}{P}\mathrm{d}\theta^2+\frac{P}{\rho^2}\sin^2\theta\left(a\,\mathrm{d}t-(r^2+a^2)\mathrm{d}\varphi\right)^2\Bigg].\label{eq:ma}
\end{equation}
According to Ref.~\cite{Podolsky:2025tle}, the functions in this metric are given by 
\begin{align}
	&\rho^{2}=r^{2}+a^{2}\cos^{2}\theta,\\
	&P=1+B^{2}\left(M^{2}\frac{I_{2}}{I_{1}^{2}}-a^{2}\right)\cos^{2}\theta,\\
	&Q=\left(1+B^{2}r^{2}\right)\Delta,\\
	&\Omega^{2}=\left(1+B^{2}r^{2}\right)-B^{2}\Delta\cos^{2}\theta,\\
	&\Delta=\left(1-B^{2}M^{2}\frac{I_{2}}{I_{1}^{2}}\right)r^{2}-2M\frac{I_{2}}{I_{1}}r+a^{2},
\end{align}
with
\begin{equation}
	I_1=1-\frac{1}{2}B^2a^2,\quad I_2=1-B^2a^2.
\end{equation}
There are three independent parameters, which can be physically interpreted as the mass $M$ of the black hole, the rotation parameter $a$, and the external uniform magnetic field $B$, respectively. When $M=0$, it degenerates into the BR spacetime. When $a=0$, the metric describes a Schwarzschild black hole in BR spacetime. When $B=0$, it is simplified to a Kerr metric in Boyer-Lindquist  coordinates. To determine the location of the horizon, setting $Q=0$ yields two roots
\begin{equation}
	r_{\pm}=\frac{M\,I_2\pm\sqrt{M^2I_2-a^2I_1^2}}{I_1^2-B^2M^2I_2}I_1,
\end{equation}
which correspond to the outer and inner horizons of the black hole, respectively. When $B=0$, we have $I_1=1=I_2$, and they reduce to the case of the Kerr black hole. Furthermore, if $a=0$, then there is only one Schwarzschild horizon.

In the extremal limit, where the rotation parameter $a$ becomes large and the condition $a^2 I_1^2 = M^2 I_2$ is satisfied, the magnetic field $B_{\text{extr}}$ is given by
\begin{equation}
	B^2_{\text{extr}}=\frac{2}{a^4}\left(M-\sqrt{M^2-a^2}\right)\sqrt{M^2-a^2}.
\end{equation}
And the two horizons coincide at
\begin{equation}
	r_{\text{extr}}=\frac{M}{I_1}=\frac{a^2}{M-\sqrt{M^2-a^2}}.
\end{equation}
If $a=M$, then we obtain the extreme Kerr case, i.e., $r_{\text{extr}}=M$. 

\section{Outline of the Kerr-BR Black Hole Shadow}
With the exact solution of the Kerr-BR black hole, we can investigate its shadow characteristics. In this section, we focus primarily on how the rotation parameter $a$ and the magnetic field $B$ influence the shadow's outline. For convenience, we can set $M=1$. The null geodesics of photons around a black hole are governed by the following Hamilton-Jacobi equation~\cite{Chandrasekhar:1985kt}
\begin{equation}
	\frac{\partial\mathcal{J}}{\partial\sigma}=-\frac12g^{\alpha\beta}\frac{\partial\mathcal{J}}{\partial x^\alpha}\frac{\partial\mathcal{J}}{\partial x^\beta},\label{eq:Jac1}
\end{equation}
where $\sigma$ is the affine parameter. The action  $\mathcal{J}$ of the photons can be expressed as
\begin{equation}
	\mathcal{J}=-\hat{E}\,t+\hat{L}\,\varphi+B_r(r)+B_\theta(\theta),\label{eq:Jac2}
\end{equation}
where $\hat{E}=-p_{t}$ represents the energy of a photon and $\hat{L}=p_{\varphi}$ represents the angular momentum of a photon. From the metric~(\ref{eq:ma}), it can be seen that these two quantities are both conserved. Therefore, substituting Eq.~(\ref{eq:Jac2}) into Eq.~(\ref{eq:Jac1}), we obtain the equations of motion for the photons around the Kerr-BR black hole~\cite{Wang:2025vsx}
\begin{align}
	&\frac{\rho^{2}}{\Omega^{2}}\frac{dt}{d\sigma}=\frac{a}{P}(\hat{L}-a\,\hat{E}\sin^2\theta)+\frac{r^2+a^2}{Q}\left[\hat{E}\left(r^2+a^2\right)-a\,\hat{L}\right],\\
	&\frac{\rho^{2}}{\Omega^{2}}\frac{dr}{d\sigma}=\pm_r\sqrt{\mathcal{R}(r)},\\
	&\frac{\rho^{2}}{\Omega^{2}}\frac{d\theta}{d\sigma}=\pm_\theta\sqrt{\Theta(\theta)},\\
	&\frac{\rho^{2}}{\Omega^{2}}\frac{d\varphi}{d\sigma}=\frac{\hat{L}}{P\,\sin^2\theta}-\frac{a\,\hat{E}}{P}+\frac{a}{Q}\left[\hat{E}\left(r^2+a^2\right)-a\,\hat{L}\right],
\end{align}
with
\begin{align}
	&\mathcal{R}(r) = \left[\hat{E}\left(r^2+a^2\right)-a\,\hat{L}\right]^2-\Delta\left[\mathcal{C}+\left(\hat{L}-a\,\hat{E}\right)^2\right]-B^2r^2\Delta \left[\mathcal{C}+(\hat L-a\,\hat E)^2\right],\\
	&\Theta(\theta) = \mathcal{C}-\left(\hat{L}^2\csc^2\theta-a^2\hat{E}^2\right)\cos^2\theta +B^2\left(\frac{I_2}{I_1^2}-a^2\right)\left[\mathcal{C}+(\hat L -a\,\hat E)^2\right]\cos^2\theta.
\end{align}
Here, $\mathcal{C}$ a constant related to the geometric structure of this spacetime. Using the equations, the trajectories of photons around the Kerr-BR black hole can be determined. Among all possible trajectories, there exists a special class corresponding to circular orbits with fixed radius $r$, known as photon rings, which are directly related to the size and boundary of the black hole shadow. For the trajectory of a photon ring, we have  $\dot{r}=0=\ddot{r}$, where `` $\dot{}$ '' indicates the derivative of the affine parameter $\sigma$ This condition  is equivalent to $\mathcal{R}(r)=0=\partial_r\mathcal{R}(r)$. Moreover, with the two conserved quantities $\hat{E}$ and $\hat{L}$, one can define the following impact parameters to describe the motion of photons
\begin{equation}
	\alpha=\frac{\hat{L}}{\hat{E}},\quad\beta=\frac{\mathcal{C}}{\hat{E}^2}.
\end{equation}
These two impact parameters can be directly solved by the condition $\mathcal{R}(r)=0=\partial_r\mathcal{R}(r)$. 

We now consider a zero-angular-momentum observer (ZAMO) positioned at spatial infinity, with coordinates fixed as $(t_{obs}=0,r_{obs},\theta_{obs},\varphi_{obs}=0)$ in the Boyer-Lindquist coordinates ($t,r,\theta,\varphi$). For such an observer, one can introduce the orthonormal tetrad as~\cite{Hu:2020usx}
\begin{align}
	\eta_0&=\eta_{(t)}=\left(\sqrt{\frac{g_{\varphi\varphi}}{g_{t\varphi}^2-g_{tt}g_{\varphi\varphi}}},\ 0,\ 0,\ -\frac{g_{t\varphi}}{g_{\varphi\varphi}}\sqrt{\frac{g_{\varphi\varphi}}{g_{t\varphi}^2-g_{tt}g_{\varphi\varphi}}}\right),\\
	\eta_1&=-\eta_{(r)}=\left(0,\ -\frac{1}{\sqrt{g_{rr}}},\ 0,\ 0\right),\\
	\eta_2&=\eta_{(\theta)}=\left(0,\ 0,\ \frac{1}{\sqrt{g_{\theta\theta}}},\ 0\right),\\
	\eta_3&=-\eta_{(\varphi)}=\left(0,\ 0,\ 0,\ -\frac{1}{\sqrt{g_{\varphi\varphi}}}\right),
\end{align}
where $\eta_0$ can be regarded as the observer's four-velocity and $\eta_1$ points toward the black hole center.

To describe the trajectory of a photon from the ZAMO's point of view, it is necessary to introduce celestial coordinates $(\mu, \nu)$. Consider the observer positioned at the origin $O$ of the celestial coordinate system, where the vector $\overrightarrow{OQ}$ corresponds to the tangent vector of the null geodesic passing through  $O$. The celestial coordinate $\mu$ represents the angle between $\overrightarrow{OQ}$ and $\eta_1$, while $\nu$ represents the angle between $\overrightarrow{OQ}$ and $\eta_2$. For a null geodesic $S(\sigma) = (t(\sigma), r(\sigma), \theta(\sigma), \varphi(\sigma))$, its tangent vector can be expressed as a linear combination of $\eta_0, \eta_1, \eta_2, \eta_3$, with the explicit expression given as~\cite{Hu:2020usx}
\begin{equation}
	\dot{S}=|\overrightarrow{OQ}|(-\eta_0+\cos\mu\,\eta_1+\sin\nu\cos\mu\,\eta_2+\sin\mu\sin\nu\,\eta_3).
\end{equation}
Here, the negative sign indicates that the tangent vector points toward the past. In the ZAMO's frame, the photon's four-momentum can be expressed as $p_{(\alpha)}=p_\beta \eta^{\beta}{}_{(\alpha)}$. In the celestial coordinates, it can be represented as~\cite{Hu:2020usx}
\begin{equation}
	\cos\mu=\frac{p^{(1)}}{p^{(0)}},\quad\tan\nu=\frac{p^{(3)}}{p^{(2)}}.
\end{equation}
In the ZAMO's frame, we can also establish a standard Cartesian coordinate system $(x,y)$, whose relationship with the celestial coordinates is given by
\begin{equation}
	x=-2\tan\frac\mu2\sin\nu,\quad y=-2\tan\frac\mu2\cos\nu.\label{proco}
\end{equation}

\begin{figure}[H]
	\centering 
	\subfigure[$B=0.001$]{\includegraphics[scale=0.4]{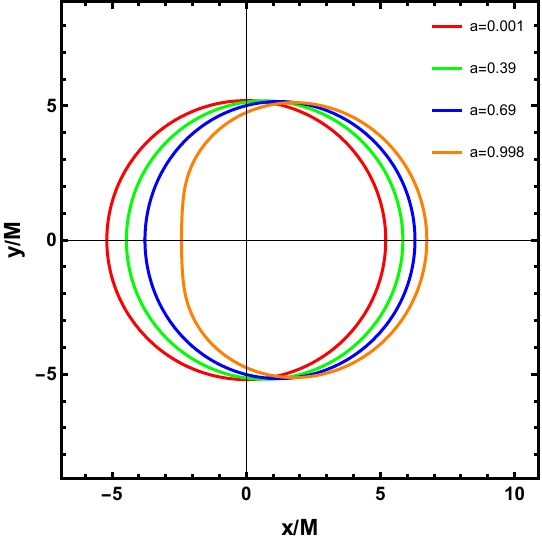}}
	\subfigure[$B=0.1$]{\includegraphics[scale=0.4]{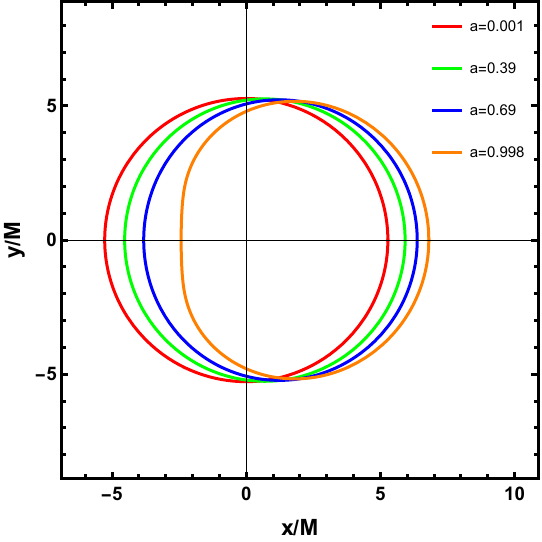}}
	\subfigure[$a=0.39$]{\includegraphics[scale=0.4]{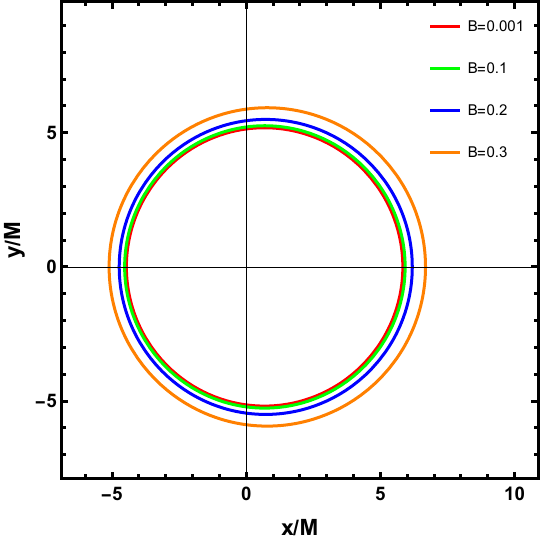}}
	\subfigure[$a=0.998$]{\includegraphics[scale=0.4]{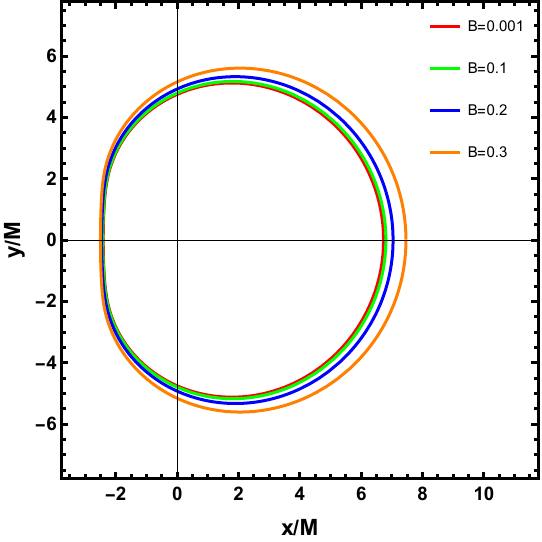}}

	\caption{Outline of the Kerr-BR black hole shadow with different values of the rotation parameter $a$ and magnetic field $B$.}
	\label{fig1}
\end{figure}

With these preparations, we can plot the outline of the Kerr-BR black hole in the Cartesian coordinate system. In Fig.~\ref{fig1}, we present the influence of the rotation parameter $a$ and magnetic field $B$ on the black hole shadow. The results indicate that as $a$ increases, the shadow evolves from a circular shape to a ``D-shaped'' outline (see the left two panels), while increasing $B$ causes almost no change in the outline's shape but enlarges its overall size (see the right two panels). Clearly, the effects of the rotation parameter $a$ on the black hole shadow is consistent with that observed in typical rotating black holes~\cite{Cunha:2018acu}. However, the magnetic field $B$ in the Kerr-BR black hole can produce a influence on the shadow size, which is a key distinguishing feature of the Kerr-BR black hole compared to purely rotating black holes. To further explore the impact of the rotation parameter $a$ and magnetic field $B$ on the optical characteristics of the Kerr-BR black hole, we next examine its shadow under different illumination conditions, primarily focusing on celestial light source illumination and thin accretion disk illumination.

\section{Celestial Light Source Illumination}

In this section, we employ the backward ray-tracing method to plot the optical image of the Kerr-BR black hole under a celestial light source. In such a light source model, the black hole is located at the center of the celestial sphere, with its rotation axis pointing toward the north pole. The black hole's size is much smaller than the diameter of the celestial sphere. For ease of distinction, the celestial sphere is equally divided into four quadrants, marked with red, green, blue, and purple, respectively.

We use the stereographic projection technique, namely  the fisheye camera model,  to visualize the optical image of the Kerr-BR black hole. In the fisheye camera model, the optical image is dependent on four parameters: the rotation parameter $a$, magnetic field $B$, observation inclination angle $\theta_o$, and field angle $\alpha_{\mathrm{fov}}$. The observation inclination angle $\theta_o$ is the angle between the black hole’s central axis and the ZAMO's line of sight. The field angle $\alpha_{\mathrm{fov}}$ is the angle between the projection screen and the point of the observer in space~\cite{Lee:2022rtg,He:2024qka,Guo:2024mij}. The field angle $\alpha_{\mathrm{fov}}$ determines the camera's viewing range. We denote the Cartesian coordinate system of the imaging plane as $(x, y)$. For convenience, we set the field angle in both the $x$ and $y$ directions to  $\alpha_{\mathrm{fov}}/2$, thereby defining a square screen with a side length of
\begin{equation}
	L = 2\left|\overrightarrow{OQ}\right|\tan\frac{\alpha_{\mathrm{fov}}}{2}.
\end{equation}
We divide the imaging plane into $n \times n$ pixels, with each pixel having a side length of
\begin{equation}
	l = \frac{L}{n} = \frac{2}{n}\left|\overrightarrow{OQ}\right|\tan\frac{\alpha_{\mathrm{fov}}}{2}.
\end{equation}
We use coordinates $(i,j)$ to represent the center position of a pixel, with the bottom-left pixel defined as $(1,1)$ and the top-right pixel defined as $(n,n)$, where $i$ and $j$ range from $1$ to $n$. The relationship between the Cartesian coordinate $(x,y)$ and the pixel coordinates $(i,j)$ is given by
\begin{equation}
	x = l\left(i - \frac{n+1}{2}\right),\quad 
	y = l\left(j - \frac{n+1}{2}\right). \label{proco2}
\end{equation}
Then, by comparing Eqs.~(\ref{proco}) and~(\ref{proco2}), the relationship between the pixel coordinates $(i,j)$ and the celestial sphere coordinates $(\mu,\nu)$ is given by 
\begin{align}
	\tan\frac{\mu}{2} &= \frac{1}{n}\tan\left(\frac{\alpha_{\mathrm{fov}}}{2}\right) 
	\sqrt{\left(i - \frac{n+1}{2}\right)^2+\left(j - \frac{n+1}{2}\right)^2},\nonumber \\
	\tan\nu &= \frac{2j-(n+1)}{2i-(n+1)}.
\end{align}

\begin{figure}[H]
	\centering 
	\subfigure[$a=0.001,B=0.002$]{\includegraphics[scale=0.5]{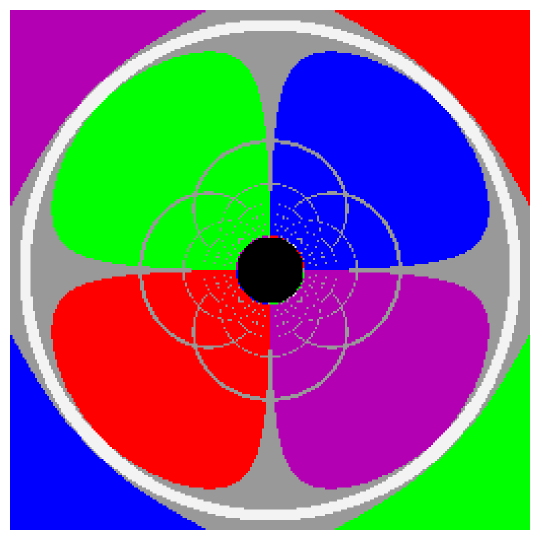}}
	\subfigure[$a=0.5,B=0.002$]{\includegraphics[scale=0.5]{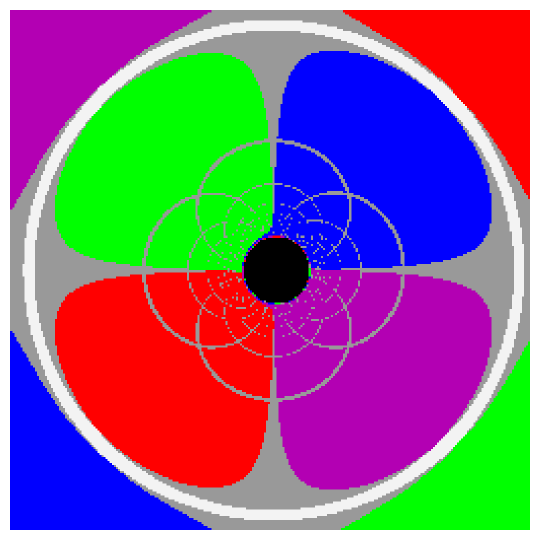}}
	\subfigure[$a=0.998,B=0.002$]{\includegraphics[scale=0.5]{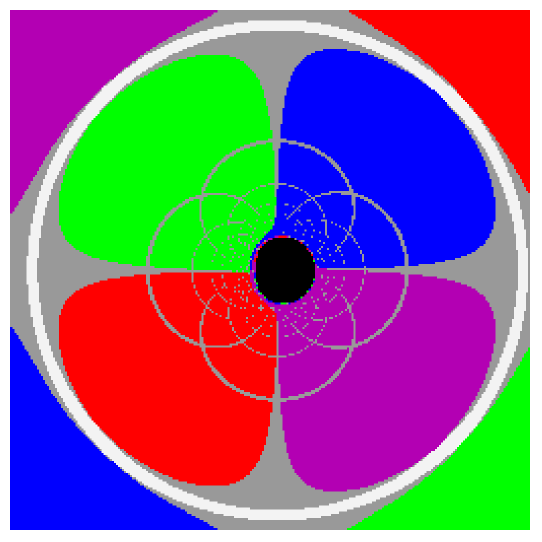}}
	
	\subfigure[$B=0.001,a=0.998$]{\includegraphics[scale=0.5]{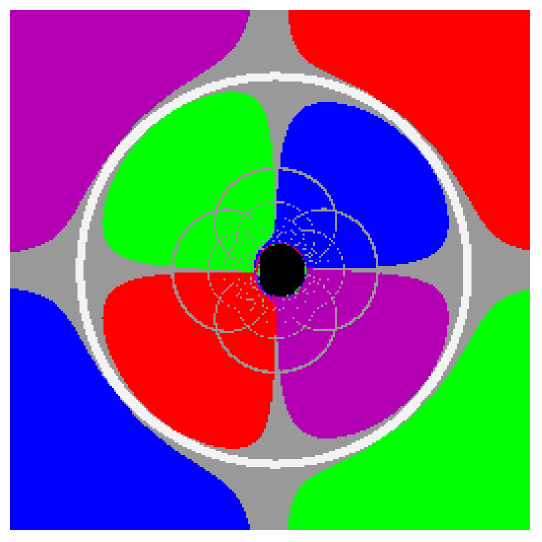}}
	\subfigure[$B=0.0015,a=0.998$]{\includegraphics[scale=0.5]{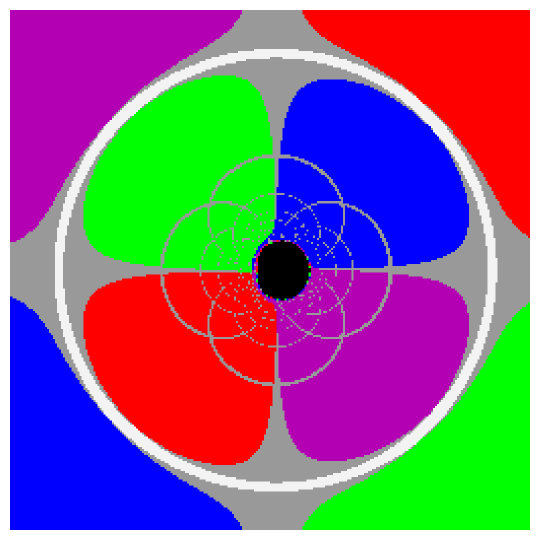}}
	\subfigure[$B=0.002,a=0.998$]{\includegraphics[scale=0.5]{Back8.pdf}}
	
	\caption{Optical images of the Kerr-BR black hole under a celestial light source with different values of the rotation parameter $a$ and magnetic field $B$. For all pictures, we set $\theta_o=80^\circ$ and $\alpha_{\mathrm{fov}}=13^\circ$}
	\label{fig2}
\end{figure}

In Fig.~\ref{fig2}, we plot the optical images of the Kerr-BR black hole with different values of the rotation parameter $a$ and magnetic field $B$. The figures reveal the effects of the rotation parameter $a$ and the magnetic field $B$ on the shadow of the Kerr-BR black hole. The central dark region corresponds to the black hole shadow, while the bright white ring structure manifests as the gravitationally lensed Einstein ring. From the above panel (with $B$ fixed), it can be observed that when $a \to 0$, the black hole shadow appears as a perfect circle, and the intersection line between the blue and green regions within the Einstein ring is perpendicular to the horizontal axis, indicating that the black hole is approximately static, with no significant space-dragging effect. As $a$ increases, the shadow contour gradually evolves from a circular shape to a ``D'' shape and a slight dragging effect becomes apparent, which can also be reflected through the rightward shift of the Einstein ring's position in the figures. On the other hand, as $B$ increases (see the below panel with $a$ fixed), one can observe that the size of the shadow contour increases, and the radius of the Einstein ring also significantly expands. Therefore, the influence of the magnetic field $B$ on the optical image of the Kerr-BR black hole is primarily manifested in the size of the shadow contour and Einstein ring under a celestial light source.

\section{Thin Accretion Disk Illumination}

In realistic scenarios, the light source of a black hole is usually the accretion disk around it. Thus, studying the optical images of black holes under the accretion disk illumination is particularly significant. To simplify the analysis, we can assume the accretion disk around the Kerr-BR black hole is optically and geometrically thin, located in the equatorial plane. The observer is situated at a sufficiently large distance. The accretion disk consists of freely moving, electrically neutral plasma following timelike geodesics along the equatorial plane. Currently, our understanding of the properties of accretion disks is incomplete, but in terms of gravity, there are some properties that have reached a consensus among researchers. For example, matter accreted by the black hole will fall into the event horizon if it is too close to the black hole, while matter sufficiently far from the black hole may orbit it along stable trajectories. To describe this phenomenon more intuitively, we introduce the concept of the innermost stable circular orbit (ISCO). The ISCO is located at the inner edge of the accretion disk, representing the last region where matter can safely orbit without falling into the event horizon. The accretion disk is divided into two regions by the ISCO. Inside the ISCO, matter falls into the event horizon along plunging orbits, while outside the ISCO, matter moves along Keplerian orbits.

In this work, we consider the accretion disk model presented in Ref.~\cite{Hou:2022eev}. In this model, the thin accretion disk extends radially inward to the Kerr-BR black hole's event horizon $r_h$ and outward to a sufficiently large distance, denoted as $r_f$. The observer's radial position $r_{obs}$ satisfies $r_h \ll r_{obs} < r_f$. To obtain the optical image of the Kerr-BR black hole under such thin accretion disk illumination, the position of the ISCO must be determined. Typically, the position of the ISCO is determined by the following conditions:
\begin{equation}
	\mathcal{V}_{eff}\bigg|_{r=r_{I}} = 0, \quad
	\partial_r \mathcal{V}_{eff}\bigg|_{r=r_{I}} = 0, \quad
	\partial_r^2 \mathcal{V}_{eff}\bigg|_{r=r_{I}} = 0.
\end{equation}
Here, $r_{I}$ is the position of the ISCO, and $\mathcal{V}_{eff}$ is a function of the radial distance $r$, representing the effective potential for a massive neutral particle. Generally, the effective potential $\mathcal{V}_{eff}$ can be expressed as
\begin{equation}
	\mathcal{V}_{eff}(r,\tilde{\mathcal{E}},\tilde{\mathcal{L}})=(1+g^{tt}\tilde{\mathcal{E}}^2+g^{\varphi\varphi}\tilde{\mathcal{L}}^2-2g^{t\varphi}\tilde{\mathcal{E}}\tilde{\mathcal{L}})\big|_{\theta=\pi/2}
\end{equation}
with
\begin{align}
	\tilde{\mathcal{E}}&=-\frac{g_{tt}+g_{t\varphi}\bar{\omega}}{\sqrt{-g_{tt}-2g_{t\varphi}\bar{\omega}-g_{\varphi\varphi}\bar{\omega}^2}},\\
	\tilde{\mathcal{L}}&=\frac{g_{t\varphi}+g_{\varphi\varphi}\bar{\omega}}{\sqrt{-g_{tt}-2g_{t\varphi}\bar{\omega}-g_{\varphi\varphi}\bar{\omega}^2}}.
\end{align}
They represent the specific energy and specific angular momentum of the massive neutral particle, respectively. The variable 
$\bar{\omega}$ has the dimensions of angular momentum, which is given by 
\begin{equation}
	\bar{\omega} = \frac{d\varphi}{dt} \equiv \frac{ \partial_r g_{t\varphi} + \sqrt{ \partial_r^2 g_{t\varphi} - (\partial_r g_{tt}) (\partial_r g_{\varphi\varphi}) } }{ \partial_r g_{\varphi\varphi} }.
	\label{av}
\end{equation}
When $r = r_{I}$, there are two conserved quantities, denoted as $\tilde{\mathcal{E}}{I}$ and $\tilde{\mathcal{L}}{I}$. When $r > r_{I}$, the matter in the accretion disk moves along Keplerian orbits, with its four-velocity given by 
\begin{equation}
	U_{out}^{a}=\sqrt{\frac{1}{-g_{tt}-2g_{t\varphi}\bar{\omega}-g_{\varphi\varphi}\bar{\omega}^2}}\left(1,0,0,\bar{\omega}\right)\bigg|_{\theta=\pi/2}.
\end{equation}
When $r_{h} < r < r_{I}$, matter freely falls from $r_{I}$ toward the black hole's event horizon. To simplify calculations, we assume that $\tilde{\mathcal{E}}$ and $\tilde{\mathcal{L}}$ remain constant and are equal to the values at the ISCO. In this case, the four-velocity of the freely falling matter can be given by
\begin{align}
	&U_{in}^{t}=\left(-g^{tt}\tilde{\mathcal{E}}_{I}+g^{t\varphi}\tilde{\mathcal{L}}_{I}\right)\bigg|_{\theta=\pi/2},\nonumber \\
	&U_{in}^{r}=-\sqrt{-\frac{g_{tt}U_{in}^{t}U_{in}^{t}+2g_{t\varphi}U_{in}^{t}U_{in}^{\varphi}+g_{\varphi\varphi}U_{in}^{\varphi}U_{in}^{\varphi}+1}{g_{rr}}}\bigg|_{\theta=\pi/2},\nonumber \\
	&U_{in}^{\theta}=0,\nonumber \\
	&U_{in}^{\varphi}=\left(-g^{t\varphi}\tilde{\mathcal{E}}_{I}+g^{\varphi\varphi}\tilde{\mathcal{L}}_{I}\right)\bigg|_{\theta=\pi/2}.
\end{align}
Note that the negative sign in front of $U_{in}^{r}$ indicates that the particle's motion is directed toward the black hole's event horizon.

Since light rays may intersect multiple times with the accretion disk and each intersection can change the light intensity on the observation screen, the complete optical image of the black hole should be the sum of all intersections. We denote the position of the $n$-th intersection of a light ray with the accretion disk as $r_n$ ($n=1, 2, 3, \dots$). When $n=1$, the resulting optical image is called the direct image. When $n=2$, it is called the lensed image. When $n>2$, it is referred to as the higher-order image. In this work, we focus on the cases where $n=1$ and $n=2$. As we know, when a light ray intersects with the accretion disk, the changes in the light intensity are mainly due to the photon emission and absorption. To simplify the model, we neglect reflection effects. Under these assumptions, the changes in the light intensity on the observation screen can be calculated by the following equation~\cite{Zhang:2023bzv}
\begin{equation}
	\frac{d}{d\sigma}\left(\frac{\mathcal{S}_\nu}{\nu^3}\right)=\frac{\mathcal{E}_\nu-\mathcal{A}_\nu \mathcal{S}_\nu}{\nu^2},\label{ligstr}
\end{equation}
where $\sigma$ remains the affine parameter of the null geodesic. The variables $\mathcal{E}{_\nu}$, $\mathcal{A}{_\nu}$, and $\mathcal{S}{_\nu}$ represent the emission coefficient, absorption coefficient and specific intensity at frequency $\nu$, respectively. When photons undergo neither absorption nor emission, $\mathcal{S}{_\nu}/\nu^3$ is conserved along the null geodesic. Moreover, since the thin accretion disk is located at the equatorial plane, outside the equatorial plane we have $\mathcal{E}_\nu = \mathcal{A}_\nu = 0$. In this case, the total light intensity at each position on the observation screen can be expressed as
\begin{equation}
	\mathcal{I}_{o}=\sum\limits_{n=1}^{N_{\max}} f_n(\chi_n)^3 \mathcal{E}_n,\label{eq:io}
\end{equation}
where $n=1,\ldots,N_{\max}$ represents the number of times a light ray intersects the equatorial plane. The parameter $f_n$ is a correction factor, which needs to be determined based on the specific accretion disk model. Since $f_n$ has a limited influence on the optical image of the black hole, we set it to 1 following Refs.~\cite{Hou:2022eev,Yang:2024nin}. The parameter $\chi_n \equiv \nu_0 / \nu_n$ is the redshift factor, where $\nu_0$ is the frequency measured on the observation screen, and $\nu_n$ is the frequency measured in the local rest frame co-moving with the accretion disk. We model the emission coefficient $\mathcal{E}_{\nu}$ as a second-order polynomial in logarithmic space, with the specific form given by~\cite{Zhong:2021mty}
\begin{equation}
	\mathcal{E}_{\nu}(r)=\exp\left(-\frac12k^2-2k\right),\quad k  =\log\frac r{r_h}.
\end{equation}
Since the accretion disk is composed of the massive neutral particle and moves along timelike geodesics with the specific energy $\tilde{\mathcal{E}}$ and specific angular momentum $\tilde{\mathcal{L}}$, for $r > r_{I}$, its redshift factor can be expressed as
\begin{equation}
	\chi_n^{out}=\frac{\alpha\left(1-\gamma\frac{p_\varphi}{p_t}\right)}{\beta\left(1+\bar{\omega}\frac{p_\varphi}{p_t}\right)}\Bigg|_{r=r_n}\quad (r>r_{I})
\end{equation}
with
\begin{equation}
	\gamma=\frac{g_{t\varphi}}{g_{\varphi\varphi}},\quad
	\alpha=\sqrt{\frac{-g_{\varphi\varphi}}{g_{tt}g_{\varphi\varphi}-g_{t\varphi}^{2}}},\quad
	\beta=\sqrt{\frac{-1}{g_{tt}+2g_{t\varphi}\bar{\omega}+g_{\varphi\varphi}\bar{\omega}^{2}}}.
\end{equation}
Note that the variable $\bar{\omega}$ is given by Eq.~(\ref{av}). The ratio of the energy measured on the observation screen to the energy propagating along the null geodesic is
\begin{equation}
	z=\alpha\left(1-\gamma\frac{p_{\varphi}}{p_{t}}\right).
\end{equation}
For an asymptotically flat spacetime, when the observer is at infinity, we have $z=1$.  When $r<r_{I}$, matter falls into the event horizon along a plunging orbit with two conserved quantities $\tilde{\mathcal{E}}{I}$ and $\tilde{\mathcal{L}}{I}$, with a radial velocity of $U_{in}^{r}$. At this point, the redshift factor can be expressed as
\begin{equation}
	\chi_{n}^{in}=\frac1{U_{in}^rp_r/p_t-\tilde{\mathcal{E}}_{I}(g^{tt}-g^{t\varphi}p_\varphi/p_t)+\tilde{\mathcal{L}}_{I}(g^{\varphi\varphi}p_\varphi/p_t+g^{t\varphi})}\Bigg|_{r=r_n}\quad (r<r_{I}).
\end{equation}

\begin{figure}[H]
	\centering 
	\subfigure[$a=0.001,\theta_o=0^\circ$]{\includegraphics[scale=0.35]{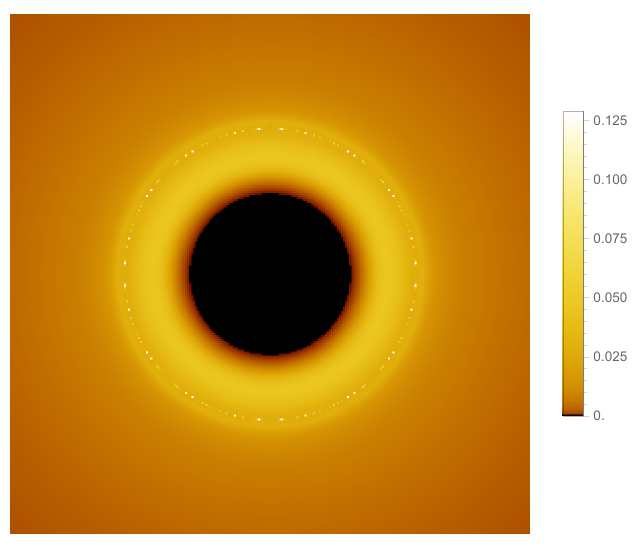}}
	\subfigure[$a=0.001,\theta_o=17^\circ$]{\includegraphics[scale=0.35]{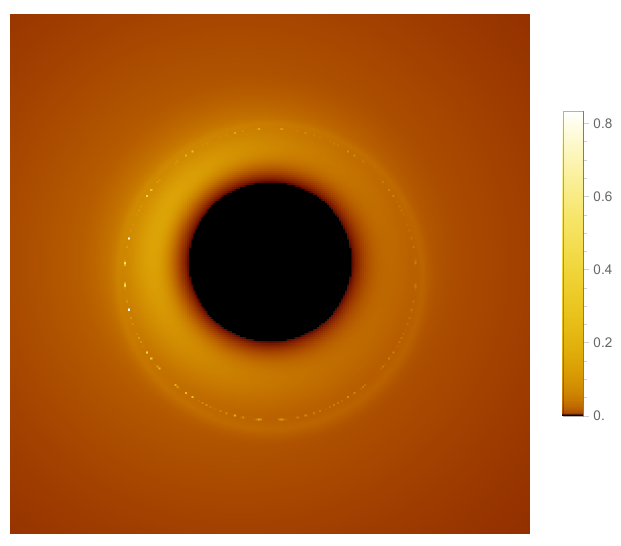}}
	\subfigure[$a=0.001,\theta_o=45^\circ$]{\includegraphics[scale=0.35]{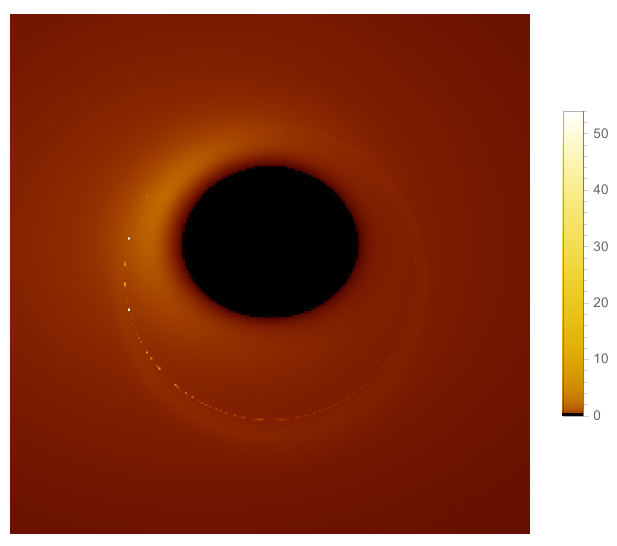}}
	\subfigure[$a=0.001,\theta_o=80^\circ$]{\includegraphics[scale=0.35]{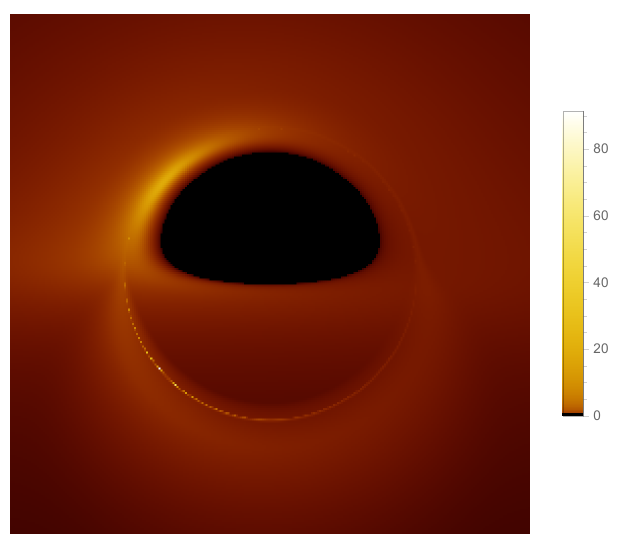}}
	
	\subfigure[$a=0.5,\theta_o=0^\circ$]{\includegraphics[scale=0.35]{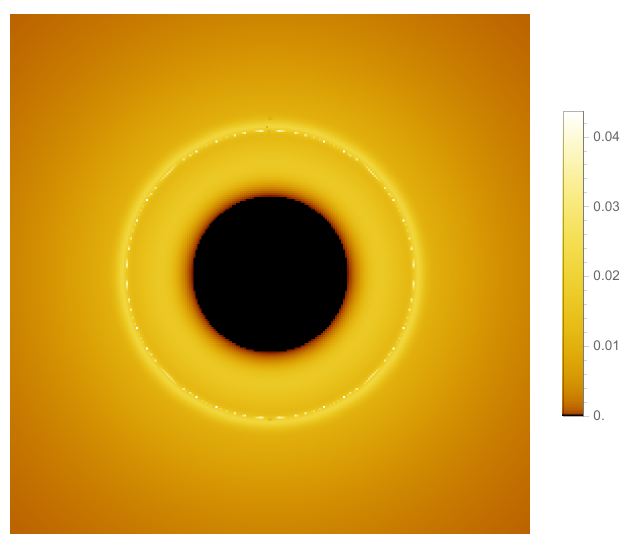}}
	\subfigure[$a=0.5,\theta_o=17^\circ$]{\includegraphics[scale=0.35]{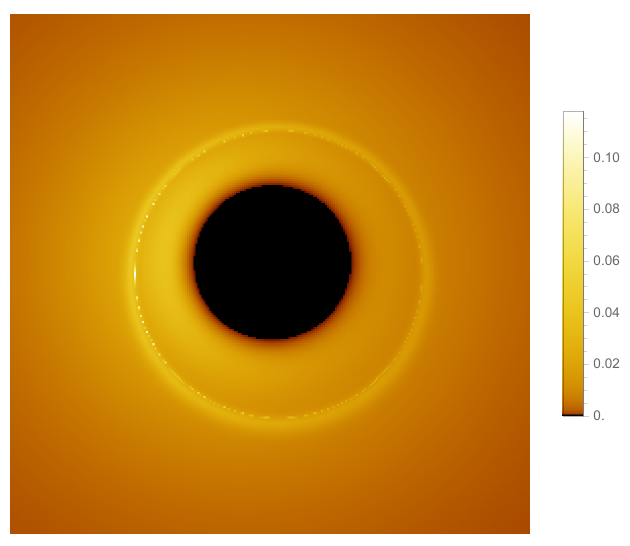}}
	\subfigure[$a=0.5,\theta_o=45^\circ$]{\includegraphics[scale=0.35]{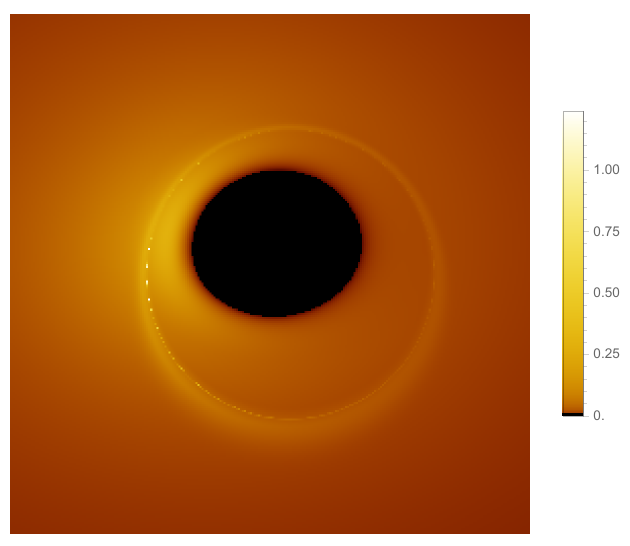}}
	\subfigure[$a=0.5,\theta_o=80^\circ$]{\includegraphics[scale=0.35]{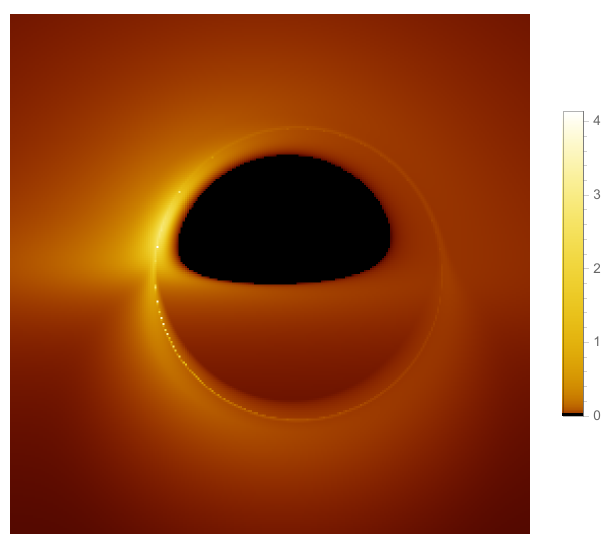}}
	
	\subfigure[$a=0.998,\theta_o=0^\circ$]{\includegraphics[scale=0.35]{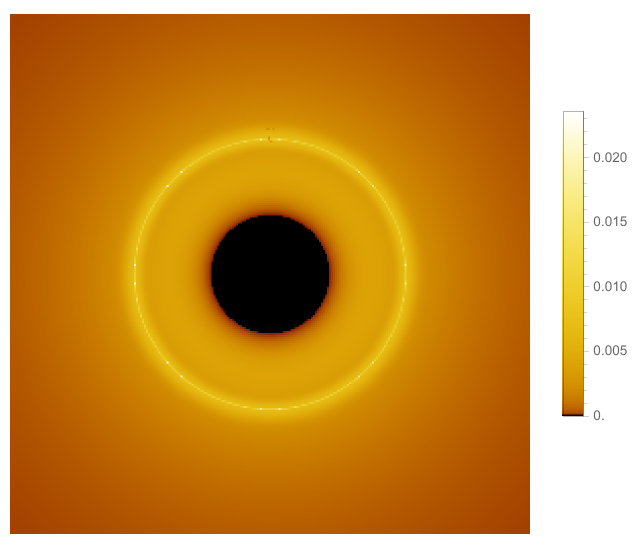}}
	\subfigure[$a=0.998,\theta_o=17^\circ$]{\includegraphics[scale=0.35]{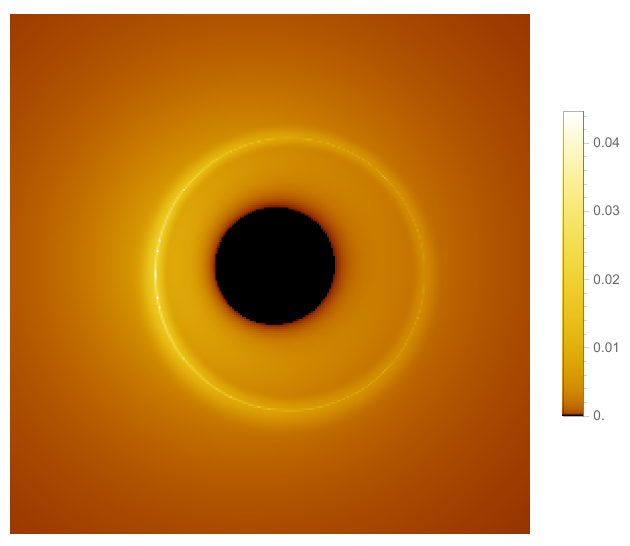}}
	\subfigure[$a=0.998,\theta_o=45^\circ$]{\includegraphics[scale=0.35]{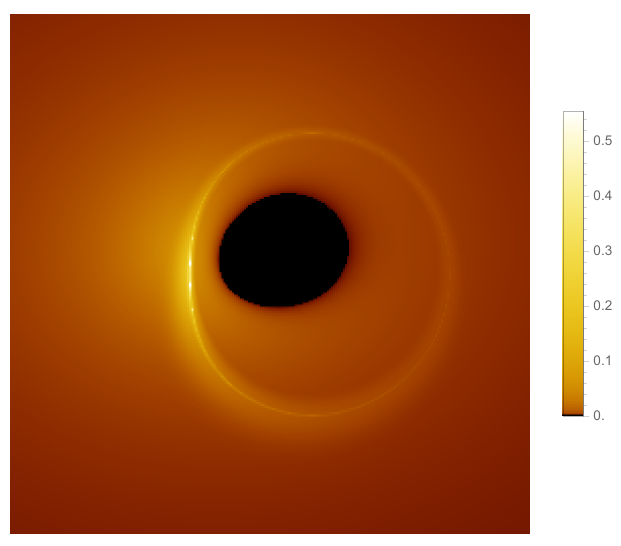}}
	\subfigure[$a=0.998,\theta_o=80^\circ$]{\includegraphics[scale=0.35]{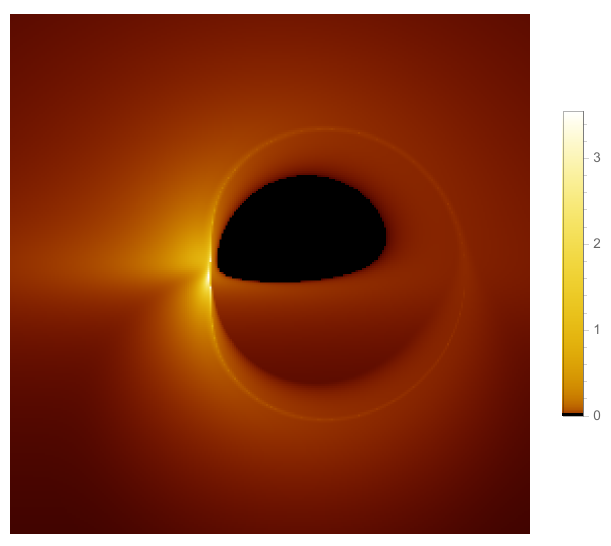}}
	
	\caption{Optical images of the Kerr-BR black hole under a thin accretion disk source with different values of the rotation parameter $a$ and observation inclination angle $\theta_o$. For all pictures, we set $B=0.002$ and $\alpha_{\mathrm{fov}}=3^\circ$}
	\label{fig3}
\end{figure}

Based on the provided thin accretion disk model, we can plot the optical images of a Kerr-BR black hole with varying relevant parameters and analyze the influence of these parameters on the optical images. For example, Fig.~\ref{fig3} illustrates the influence of the rotation parameter $a$ and observation inclination angle $\theta_o$ on the black hole shadow by fixing the magnetic field $B=0.002$ and the field angle $\alpha_{\mathrm{fov}}=3^\circ$. For other values of $B$ and $\alpha_{\mathrm{fov}}$, the results are similar, and so we do not discuss them here.

It can be observed that, regardless of the variations in $a$ and $\theta_o$, a black region always exists in the optical images, referred to as the inner shadow of the black hole~\cite{Zhang:2023bzv}, formed by photons that fail to reach the observer. If the photons emitted by the accretion disk fall directly into the event horizon, i.e., $n=0$, according to Eq.~(\ref{eq:io}), the corresponding light intensity is $\mathcal{I}{o}=0$. This reflects the black hole's strong gravity, which prevents light from escaping. As the observation inclination angle $\theta{o}$ varies, the shape of the inner shadow undergoes significant changes. When $\theta_{o}=0^\circ$, the inner shadow appears as a perfect circle, whereas when $\theta_{o}=80^\circ$, the inner shadow deforms into a ``D'' shape (see each row from left to right). Notably, a common feature exists in all optical images, i.e., there is always a bright closed curve at the outer boundary of the inner shadow, also known as the critical curve~\cite{Perlick:2021aok}. As the rotation parameter $a$ increases, the size of the inner shadow significantly decreases, but its shape remains nearly unchanged (see each column from top to bottom). For small observation inclination angles (e.g., $\theta_{o}=0^\circ$ and $17^\circ$, see the first two columns), as $a$ increases, the bright band corresponding to the critical curve becomes more concentrated, enhancing its distinguishability. In such cases, direct and lensed images become difficult to differentiate. For large observation inclination angles (e.g., $\theta_{o}=80^\circ$, see the last column), a ``crescent-shaped'' bright region appears on the left side of the critical curve, with significantly higher intensity than other regions. This phenomenon is attributed to the Doppler effect caused by the relative motion between the prograde accretion disk and the observer. The prograde accretion disk causes light on the left side to propagate toward the observer, producing a blueshift that increases photon energy, thereby brightening the corresponding region.

\begin{figure}[H]
	\centering 
	\subfigure[$a=0.001,\theta_o=0^\circ$]{\includegraphics[scale=0.4]{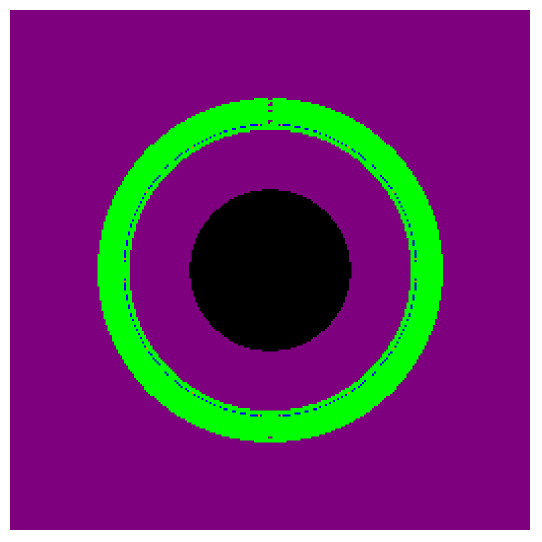}}
	\subfigure[$a=0.001,\theta_o=17^\circ$]{\includegraphics[scale=0.4]{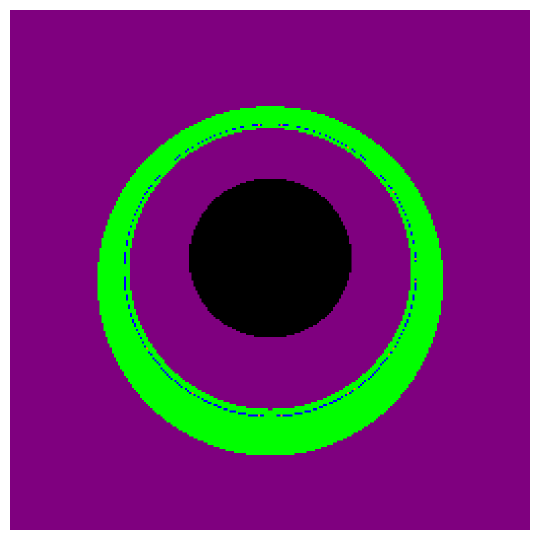}}
	\subfigure[$a=0.001,\theta_o=45^\circ$]{\includegraphics[scale=0.4]{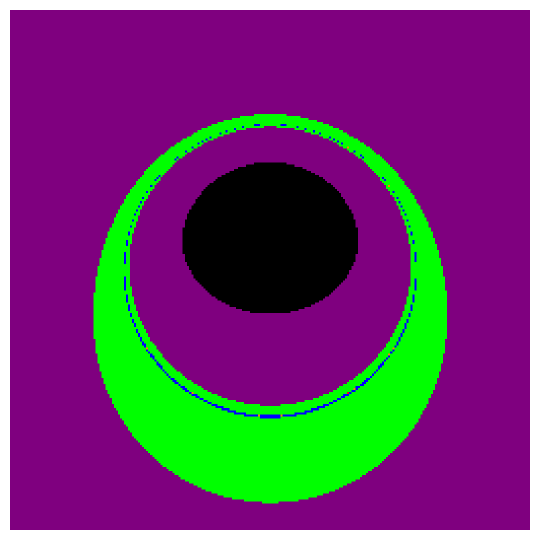}}
	\subfigure[$a=0.001,\theta_o=80^\circ$]{\includegraphics[scale=0.4]{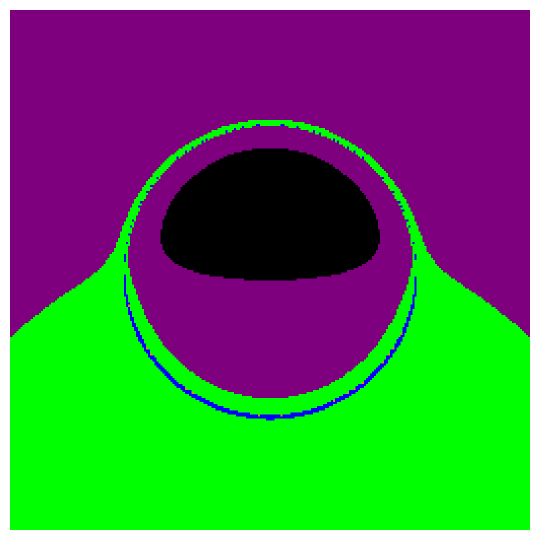}}
	
	\subfigure[$a=0.5,\theta_o=0^\circ$]{\includegraphics[scale=0.4]{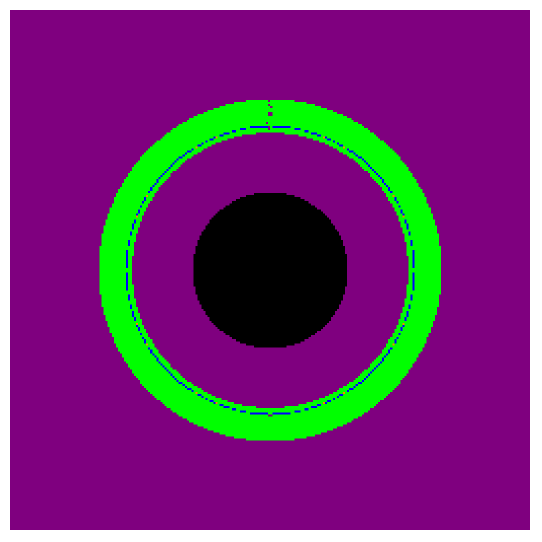}}
	\subfigure[$a=0.5,\theta_o=17^\circ$]{\includegraphics[scale=0.4]{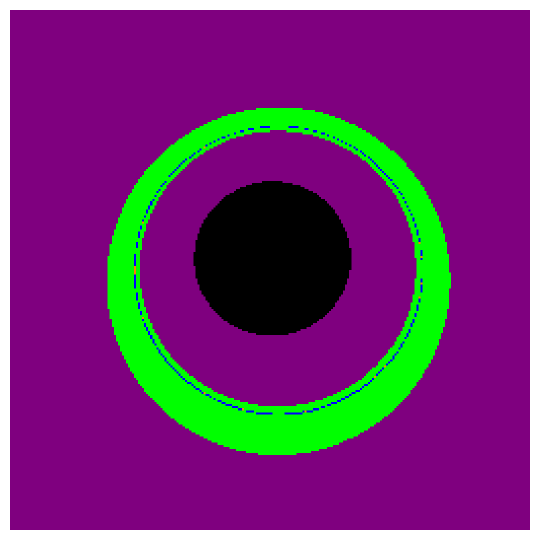}}
	\subfigure[$a=0.5,\theta_o=45^\circ$]{\includegraphics[scale=0.4]{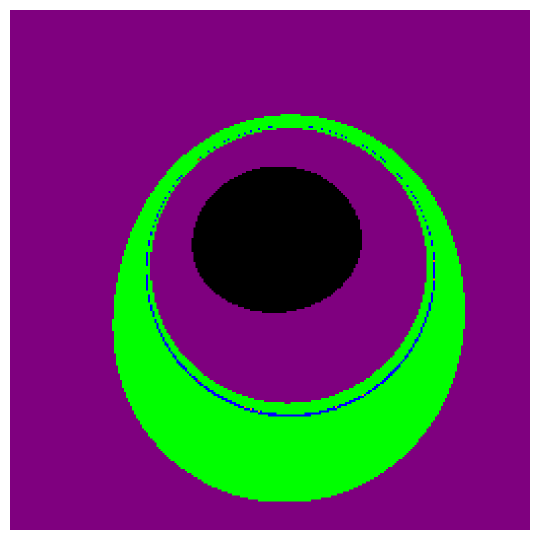}}
	\subfigure[$a=0.5,\theta_o=80^\circ$]{\includegraphics[scale=0.4]{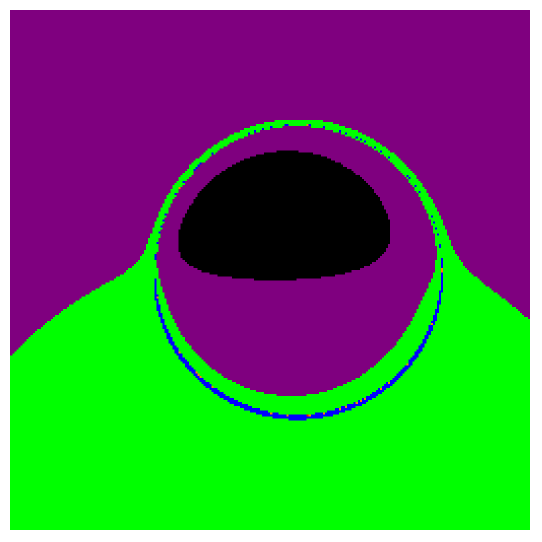}}
	
	\subfigure[$a=0.998,\theta_o=0^\circ$]{\includegraphics[scale=0.4]{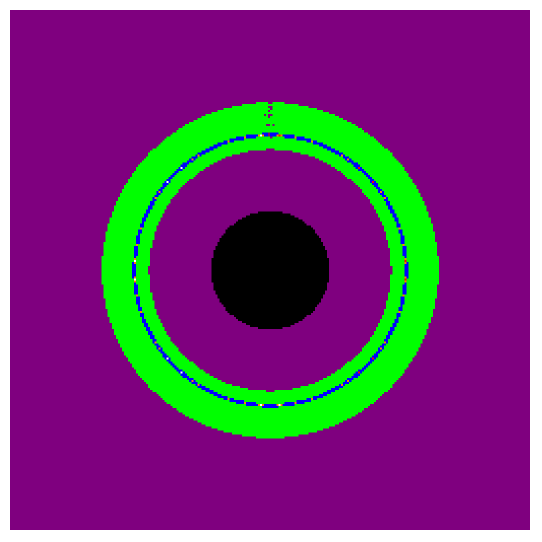}}
	\subfigure[$a=0.998,\theta_o=17^\circ$]{\includegraphics[scale=0.4]{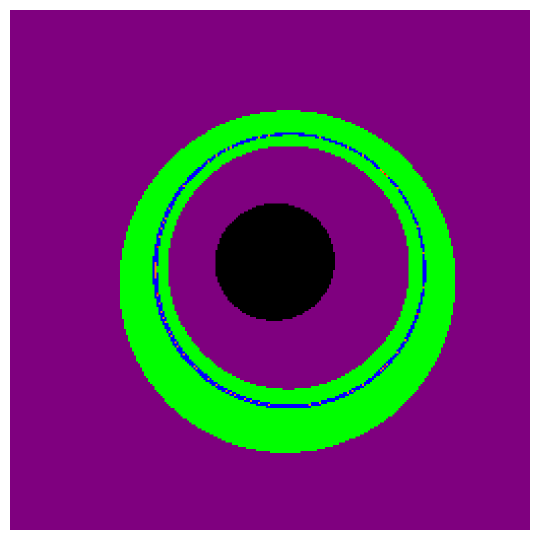}}
	\subfigure[$a=0.998,\theta_o=45^\circ$]{\includegraphics[scale=0.4]{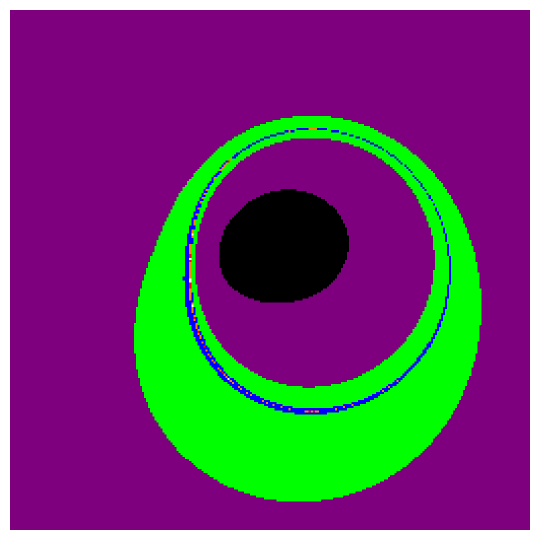}}
	\subfigure[$a=0.998,\theta_o=80^\circ$]{\includegraphics[scale=0.4]{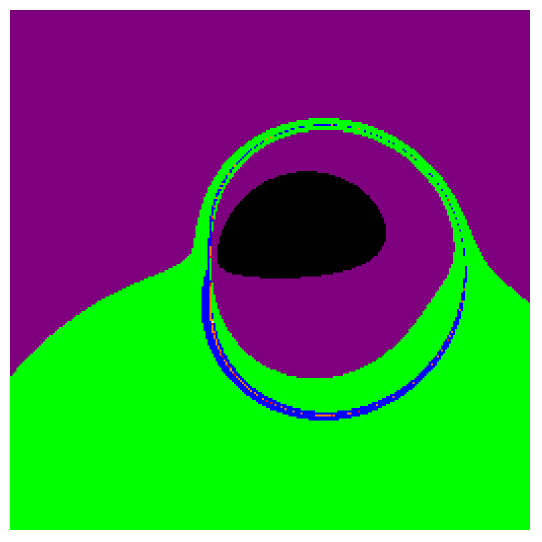}}
	
	\caption{Observed flux distribution of direct, lensed, and higher-order images of the Kerr-BR black hole under a thin accretion disk source.  The values of the parameters are exactly the same as those in Fig.~\ref{fig3}.}
	\label{fig4}
\end{figure}

To more intuitively distinguish between direct and lensed images, we present the observed flux distribution corresponding to Fig.~\ref{fig3} in Fig.~\ref{fig4}. In the images, purple, green, and blue represent the direct image, lensed image, and higher-order images of the black hole, respectively. These colors indicate the number of times photons cross the equatorial plane: purple for once, green for twice, and blue for three times or more. The black region at the center of each image remains the inner shadow of the black hole. It is noteworthy that all higher-order images are all located within the range of the lensed images. For the case of $\theta_{o}=0^\circ$ (see the first column), the inner shadow appears as a perfect circle, and the direct and lensed images form concentric rings. The changes in the rotation parameter $a$ can affect significantly the size of the inner shadow, which is consistent with the result in Fig.~\ref{fig3}. However, it is evident that the rotation parameter $a$ has no intuitive impact on the lensed image (see each column from top to bottom). As the observation inclination angle $\theta_{o}$ increase (see each row from left to right), the lensed image shifts downward on the observation screen. When $\theta_{o}$ increases to $80^\circ$ (see the fourth column), the inner shadow and lensed image undergo significant deformation.

\begin{figure}[H]
	\centering 
	\subfigure[$B=0.001,\theta_o=0^\circ$]{\includegraphics[scale=0.35]{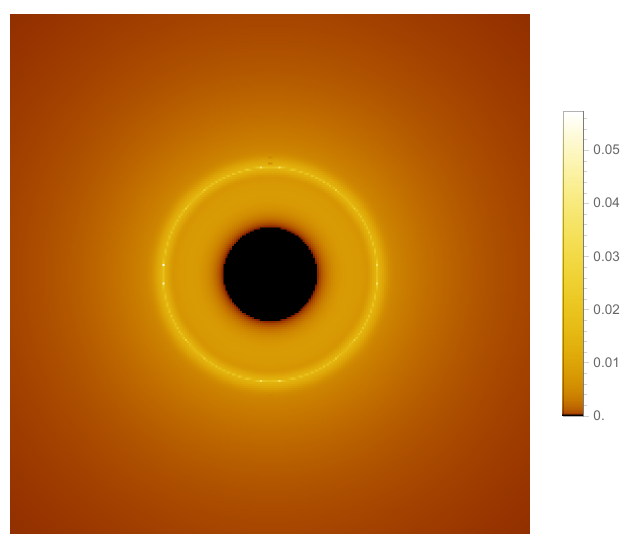}}
	\subfigure[$B=0.001,\theta_o=17^\circ$]{\includegraphics[scale=0.35]{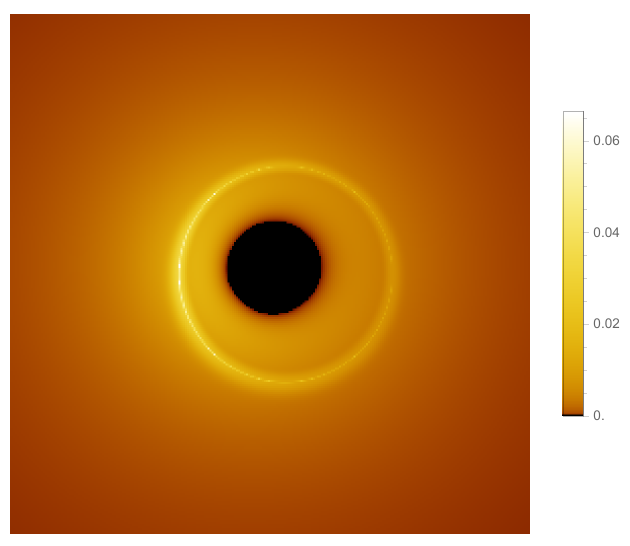}}
	\subfigure[$B=0.001,\theta_o=45^\circ$]{\includegraphics[scale=0.35]{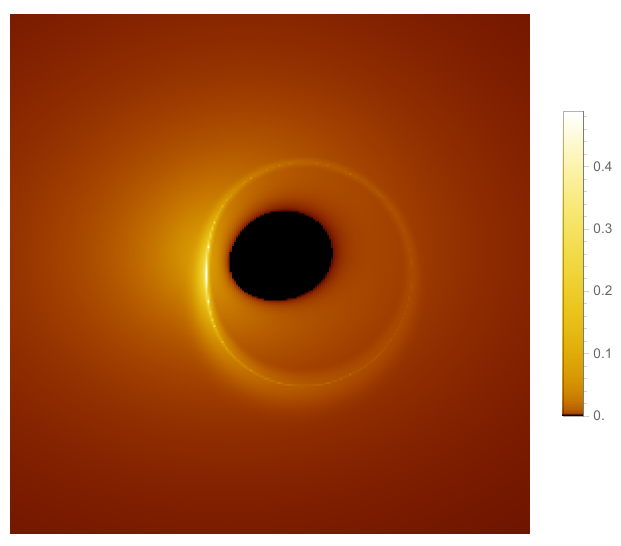}}
	\subfigure[$B=0.001,\theta_o=80^\circ$]{\includegraphics[scale=0.35]{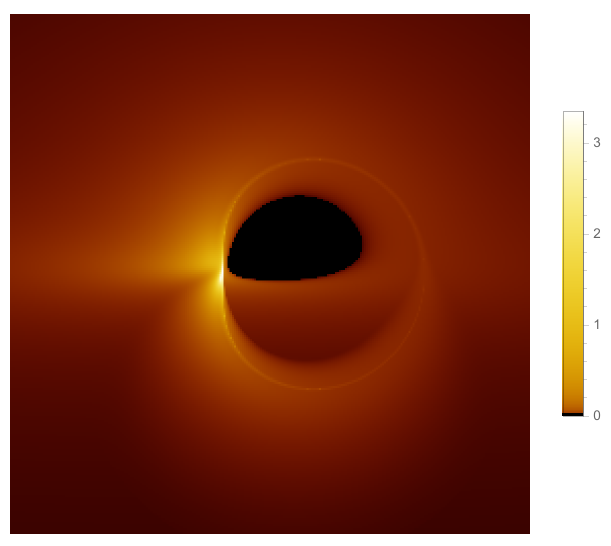}}
	
	\subfigure[$B=0.002,\theta_o=0^\circ$]{\includegraphics[scale=0.35]{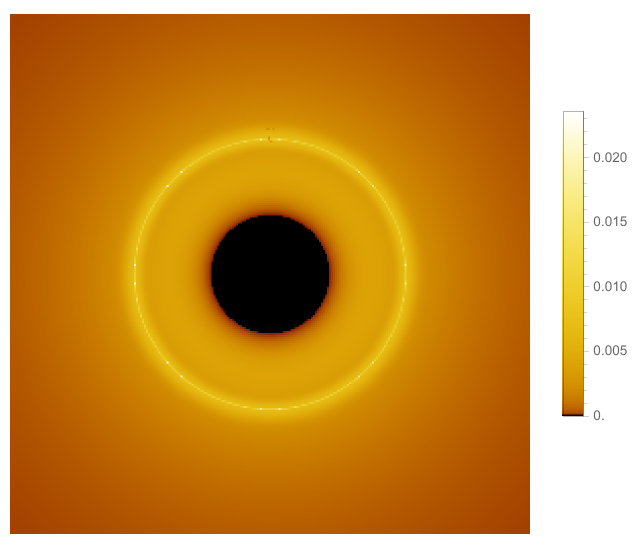}}
	\subfigure[$B=0.002,\theta_o=17^\circ$]{\includegraphics[scale=0.35]{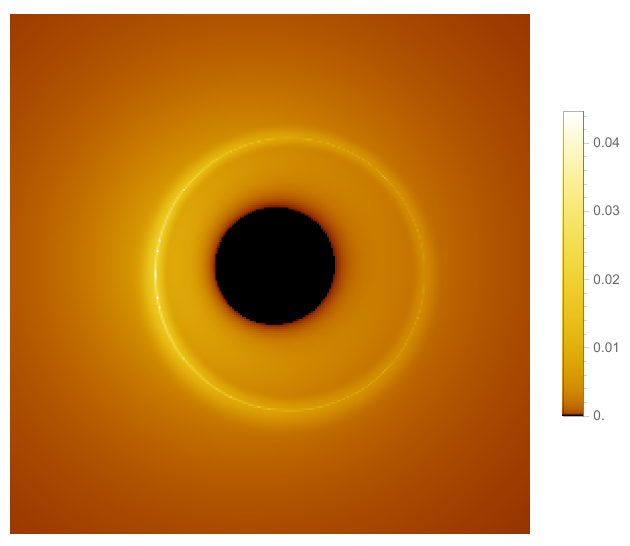}}
	\subfigure[$B=0.002,\theta_o=45^\circ$]{\includegraphics[scale=0.35]{Thin21_1.pdf}}
	\subfigure[$B=0.002,\theta_o=80^\circ$]{\includegraphics[scale=0.35]{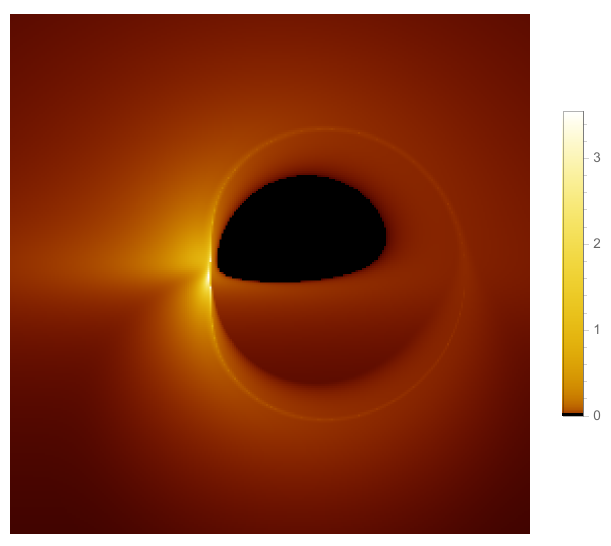}}
	
	\subfigure[$B=0.003,\theta_o=0^\circ$]{\includegraphics[scale=0.35]{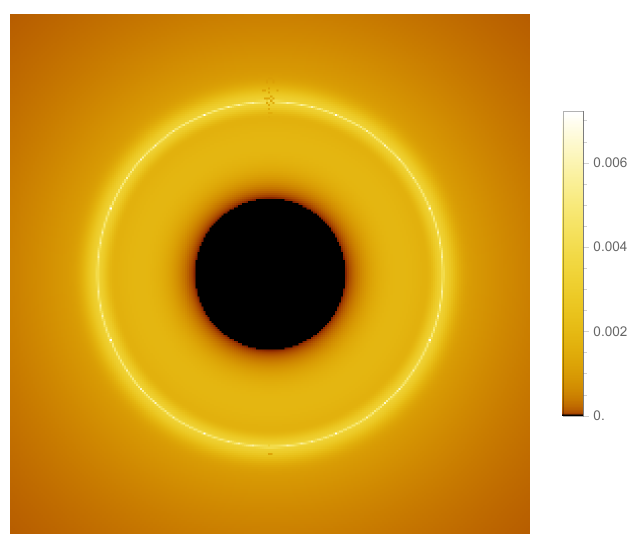}}
	\subfigure[$B=0.003,\theta_o=17^\circ$]{\includegraphics[scale=0.35]{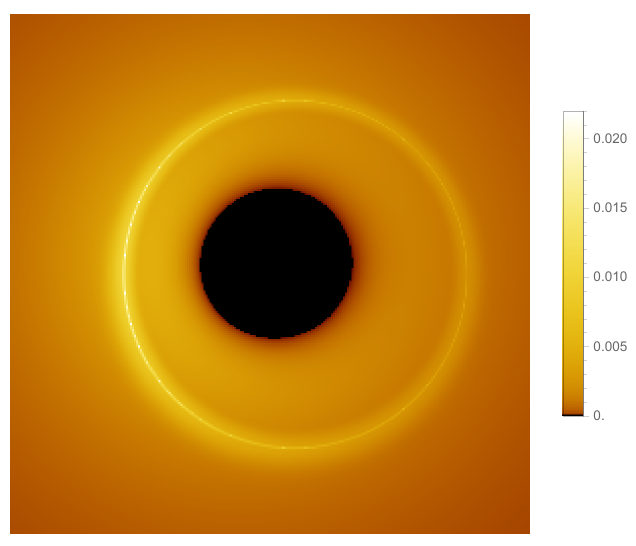}}
	\subfigure[$B=0.003,\theta_o=45^\circ$]{\includegraphics[scale=0.35]{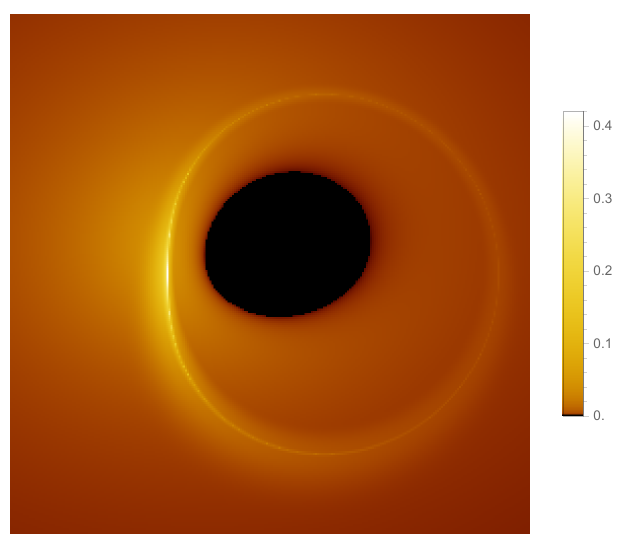}}
	\subfigure[$B=0.003,\theta_o=80^\circ$]{\includegraphics[scale=0.35]{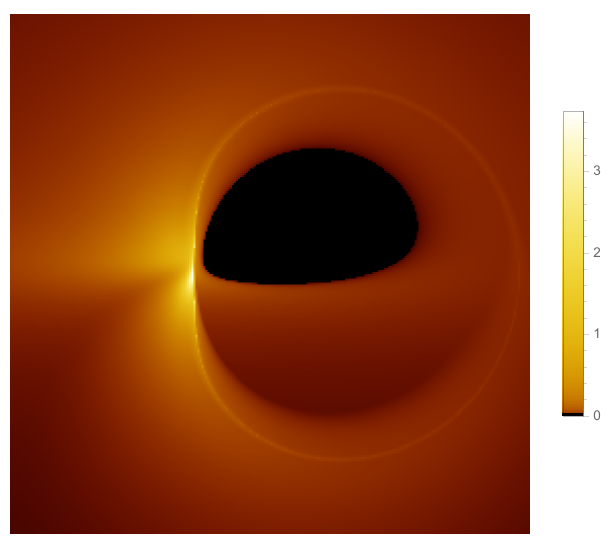}}
	
	\caption{Optical images of the Kerr-BR black hole under a thin accretion disk source with different values of the magnetic field $B$ and observation inclination angle $\theta_o$. For all pictures, we set $a=0.998$ and $\alpha_{\mathrm{fov}}=3^\circ$}
		\label{fig5}
\end{figure}

\begin{figure}[H]
	\centering 
	\subfigure[$B=0.001,\theta_o=0^\circ$]{\includegraphics[scale=0.4]{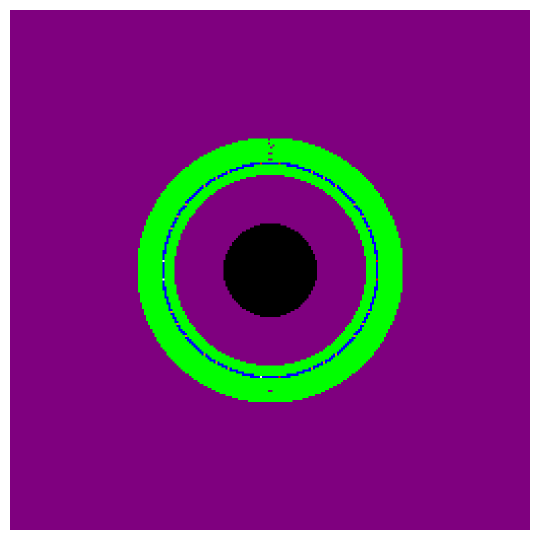}}
	\subfigure[$B=0.001,\theta_o=17^\circ$]{\includegraphics[scale=0.4]{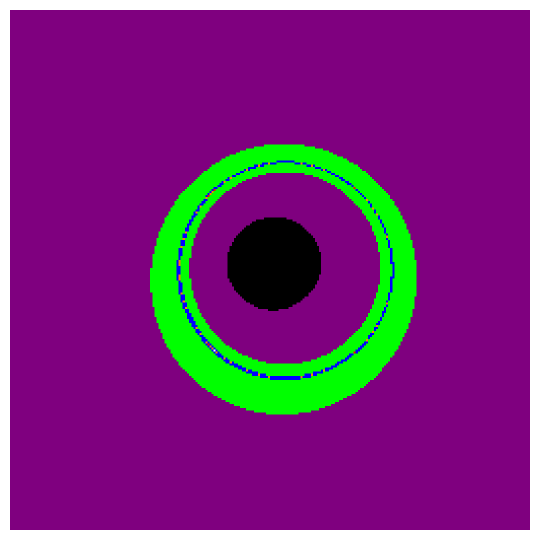}}
	\subfigure[$B=0.001,\theta_o=45^\circ$]{\includegraphics[scale=0.4]{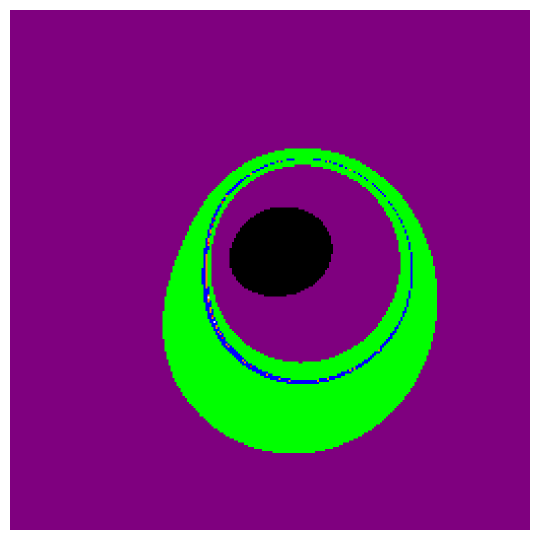}}
	\subfigure[$B=0.001,\theta_o=80^\circ$]{\includegraphics[scale=0.4]{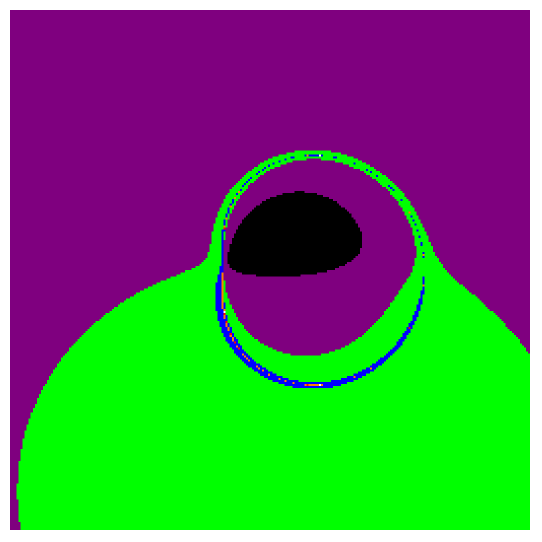}}
	
	\subfigure[$B=0.002,\theta_o=0^\circ$]{\includegraphics[scale=0.4]{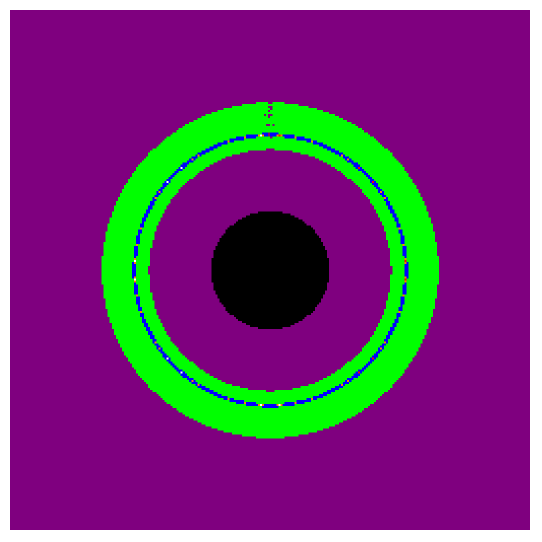}}
	\subfigure[$B=0.002,\theta_o=17^\circ$]{\includegraphics[scale=0.4]{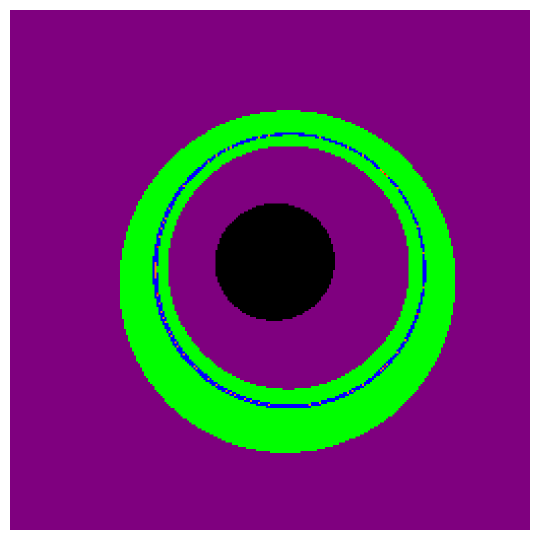}}
	\subfigure[$B=0.002,\theta_o=45^\circ$]{\includegraphics[scale=0.4]{Thin21_4.pdf}}
	\subfigure[$B=0.002,\theta_o=80^\circ$]{\includegraphics[scale=0.4]{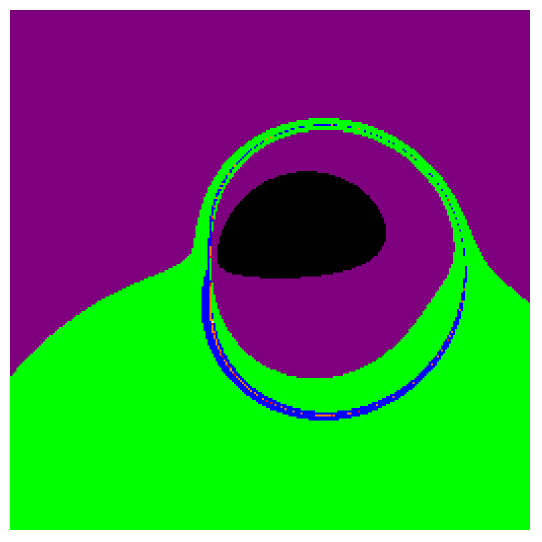}}
	
	\subfigure[$B=0.003,\theta_o=0^\circ$]{\includegraphics[scale=0.4]{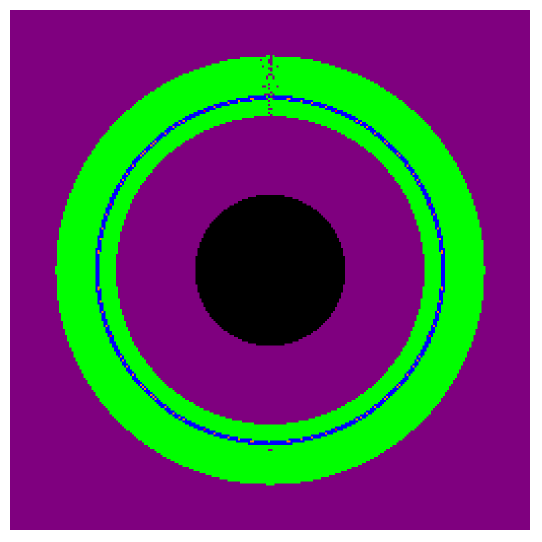}}
	\subfigure[$B=0.003,\theta_o=17^\circ$]{\includegraphics[scale=0.4]{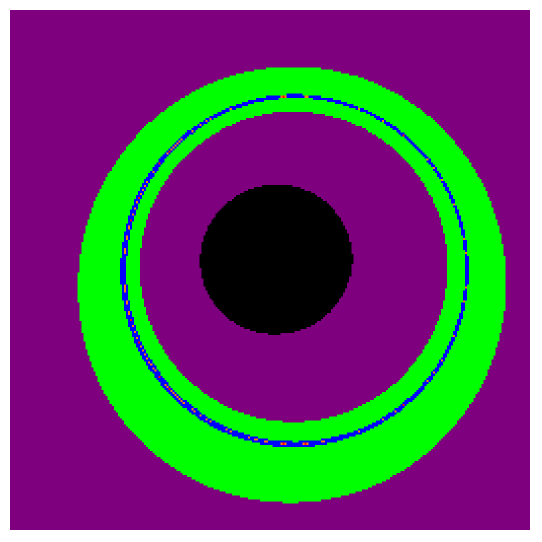}}
	\subfigure[$B=0.003,\theta_o=45^\circ$]{\includegraphics[scale=0.4]{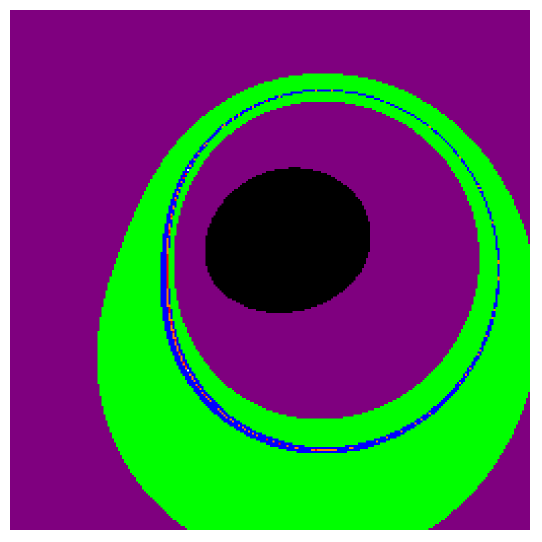}}
	\subfigure[$B=0.003,\theta_o=80^\circ$]{\includegraphics[scale=0.4]{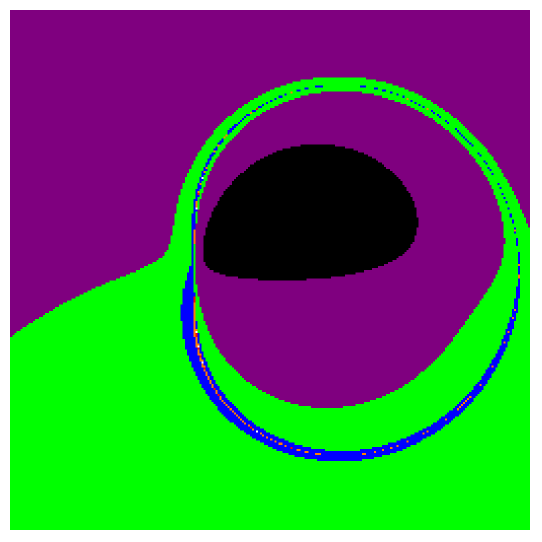}}
	\caption{Observed flux distribution of direct, lensed, and higher-order images of the Kerr-BR black hole under a thin accretion disk source.  The values of the parameters are exactly the same as those in Fig.~\ref{fig5}.}
	\label{fig6}
\end{figure}

Now, we examine the influence of the magnetic field $B$ on the optical images of the Kerr-BR black hole. For this purpose, we fix the rotation parameter $a=0.998$ and $\alpha_{\mathrm{fov}}=3^\circ$. As with other values of $a$ and $\alpha_{\mathrm{fov}}$, the results are similar, so we only discuss this specific case. In Fig.~\ref{fig5}, we present the optical images for different values of  the magnetic field $B$. The influence of the observation inclination angle $\theta_o$ is consistent with the conclusions drawn in Fig.~\ref{fig3} (see each row from left to right). By combining Figs.~\ref{fig5} and~\ref{fig6} (the observed flux distribution corresponding to Fig.~\ref{fig5}), we can identify two aspects that the influence of the magnetic field $B$ differs from that of the rotation parameter $a$. On one hand, the inner shadow exhibits a monotonic decrease with increasing rotation parameter $a$, while displaying a positive correlation with the magnetic field strength $B$. On the other hand, $B$ not only affects the size of the inner shadow (a property shared with the rotation parameter $a$) but also significantly alters the sizes of the lensed image and higher-order image (a property not exhibited by the rotation parameter $a$), as seen in each column from top to bottom in Figs.~\ref{fig5} and~\ref{fig6}.

\begin{figure}[H]
	\centering 
	\subfigure[$B=0.001$]{\includegraphics[scale=0.5]{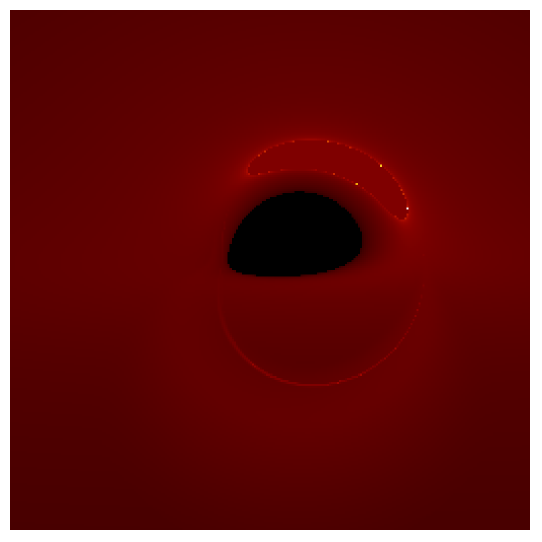}}
	\subfigure[$B=0.002$]{\includegraphics[scale=0.5]{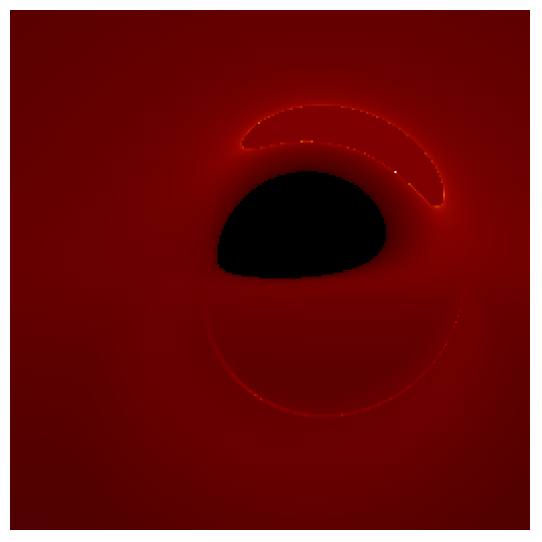}}
	\subfigure[$B=0.003$]{\includegraphics[scale=0.5]{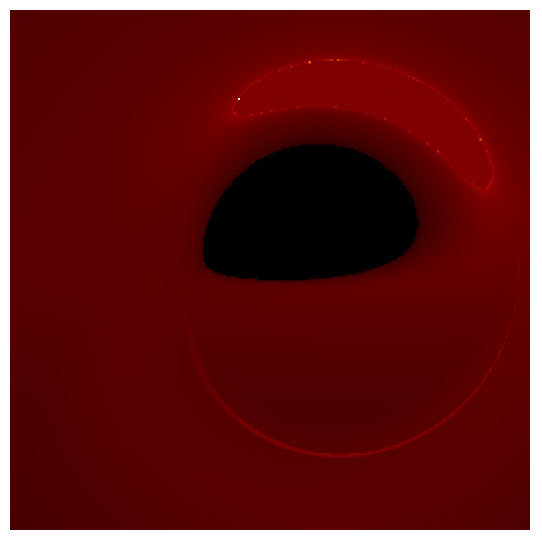}}	
	\caption{Optical images of the Kerr-BR black hole under a retrograde thin accretion disk source with different values of the magnetic field $B$. For all pictures, we set $a=0.998$, $\alpha_{\mathrm{fov}}=3^\circ$, and $\theta_o=80^\circ$}
	\label{fig7}
\end{figure}

\begin{figure}[H]
	\centering 
	\subfigure[$B=0.001$]{\includegraphics[scale=0.5]{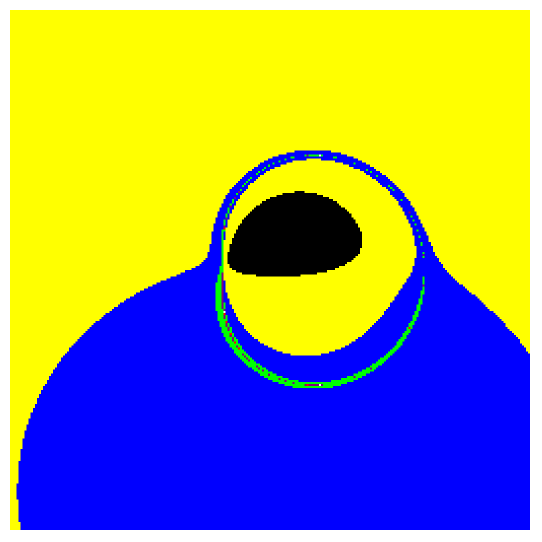}}
	\subfigure[$B=0.002$]{\includegraphics[scale=0.5]{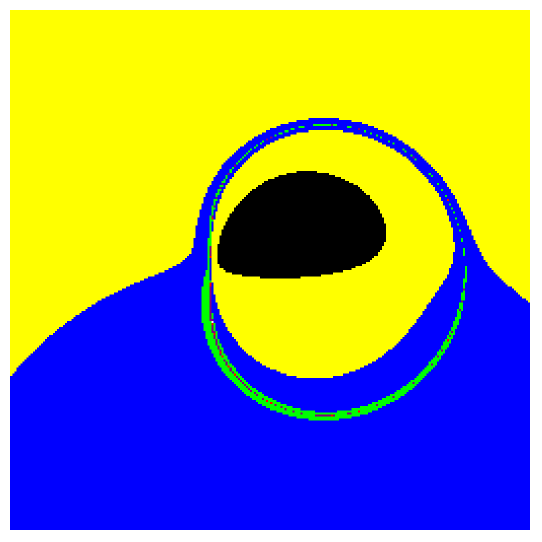}}
	\subfigure[$B=0.003$]{\includegraphics[scale=0.5]{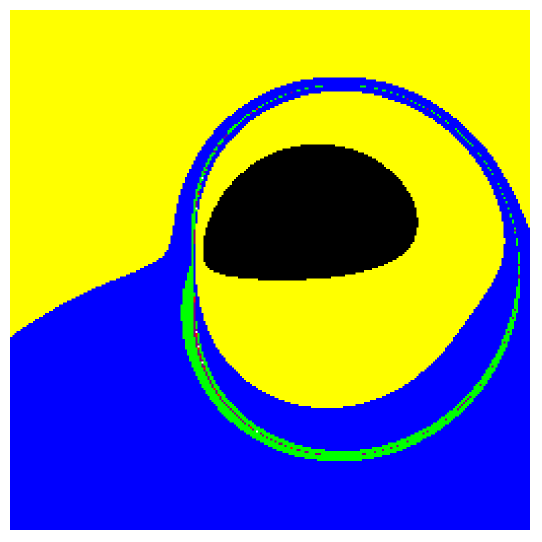}}	
	\caption{Observed flux distribution of direct, lensed, and higher-order images of the Kerr-BR black hole under a retrograde thin accretion disk source. The values of the parameters are exactly the same as those in Fig.~\ref{fig7}.}
	\label{fig8}
\end{figure}

At the end of this section, let us simply verify whether the influence of the magnetic field $B$ on the optical image of the Kerr-BR black hole is consistent in the case of a retrograde thin accretion disk. Unlike the prograde case, the gravitational redshift effect significantly reduces the observed brightness of the shadow image. The overall decrease in the light intensity makes it difficult to distinguish between the lensed image and higher-order image. Thus, the distinguishability of the critical curve is also reduced. In Fig.~\ref{fig7}, we display the optiacl images of the black hole under a retrograde thin accretion disk with different values of the magnetic field $B$. It can be observed that a ``crescent-shaped'' bright region appears on the right side of the observation screen. This phenomenon arises from the change in the direction of matter’s motion, which alters the direction of the Doppler-induced blueshift. Moreover, Fig.~\ref{fig8} shows the corresponding observed flux distribution, where yellow, blue, and green denote the direct image, lensed image, and higher-order image, respectively. Both figures confirm that as the magnetic field $B$ increases, the inner shadow, lensed image, and higher-order image all increase in size under a retrograde thin accretion disk source.

\section{Distribution of the Redshift Factor}

Under the thin accretion disk source, the light intensity on the observation screen is influenced by multiple factors, such as photon divergence, absorption, and redshift. In this section, based on the aforementioned accretion disk model, we study the distribution of the redshift factor on the observation screen and discuss the impact of relevant parameters.

\begin{figure}[H]
	\centering 
	\subfigure[$a=0.001,\theta_o=0^\circ$]{\includegraphics[scale=0.35]{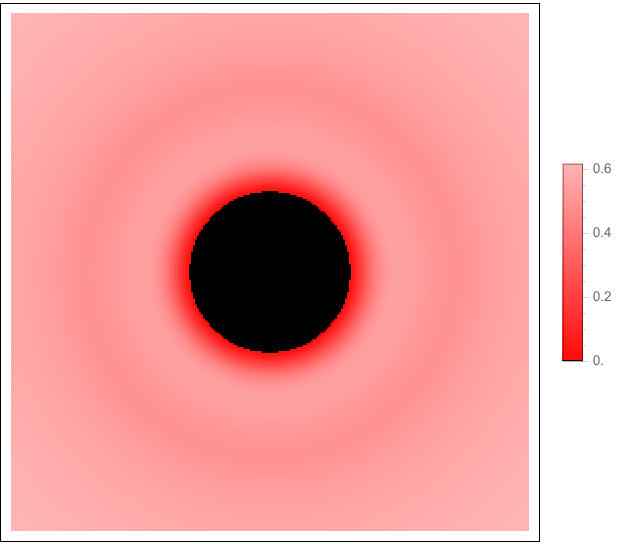}}
	\subfigure[$a=0.001,\theta_o=17^\circ$]{\includegraphics[scale=0.35]{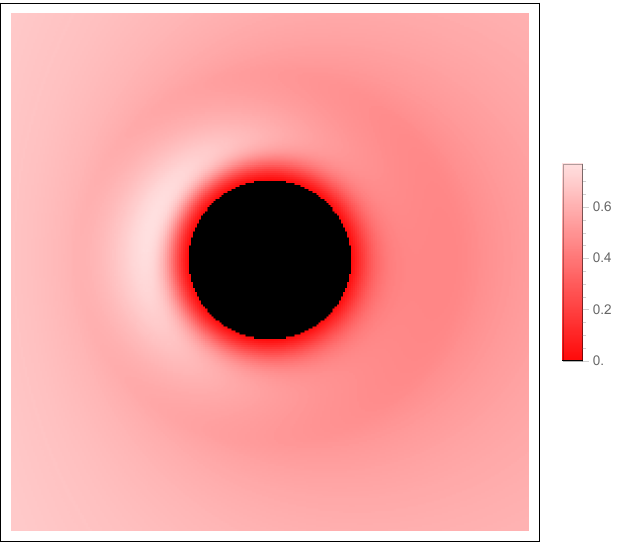}}
	\subfigure[$a=0.001,\theta_o=45^\circ$]{\includegraphics[scale=0.35]{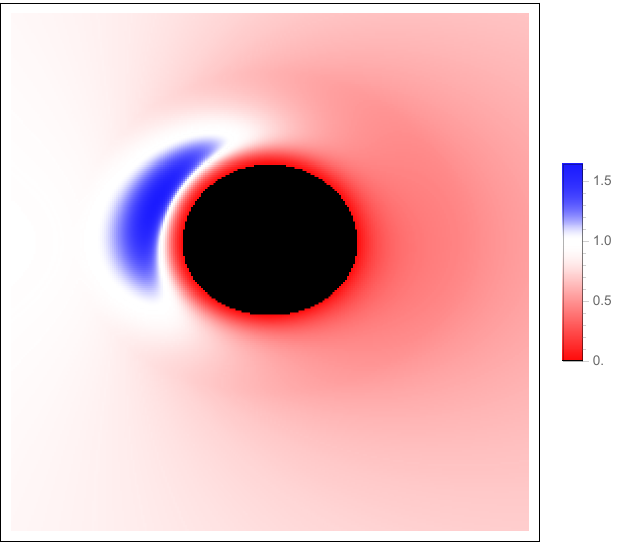}}
	\subfigure[$a=0.001,\theta_o=80^\circ$]{\includegraphics[scale=0.35]{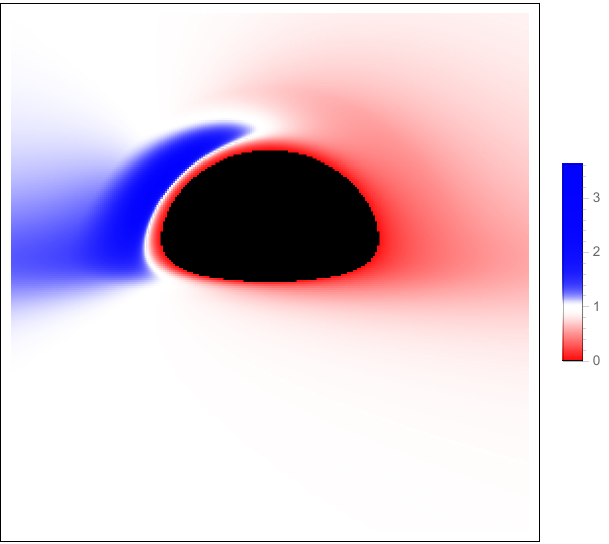}}
	
	\subfigure[$a=0.5,\theta_o=0^\circ$]{\includegraphics[scale=0.35]{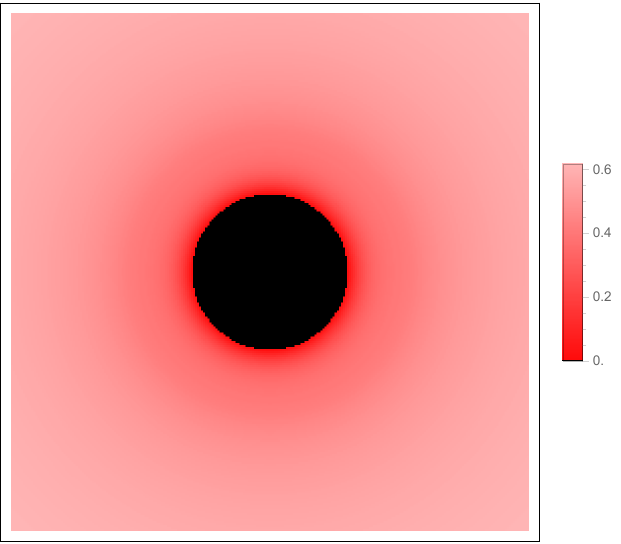}}
	\subfigure[$a=0.5,\theta_o=17^\circ$]{\includegraphics[scale=0.35]{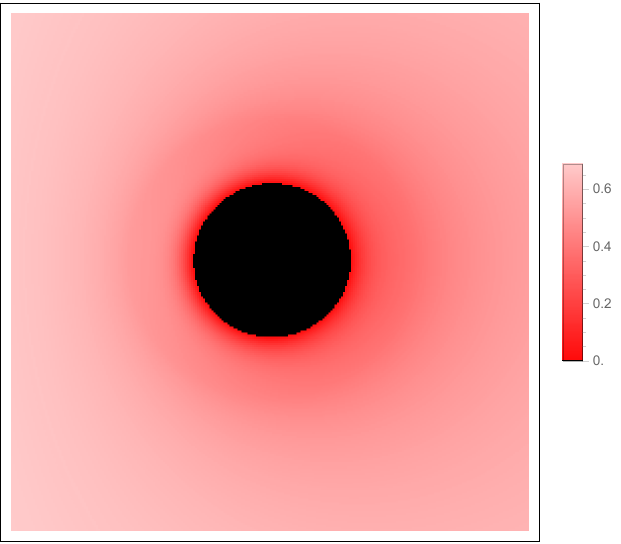}}
	\subfigure[$a=0.5,\theta_o=45^\circ$]{\includegraphics[scale=0.35]{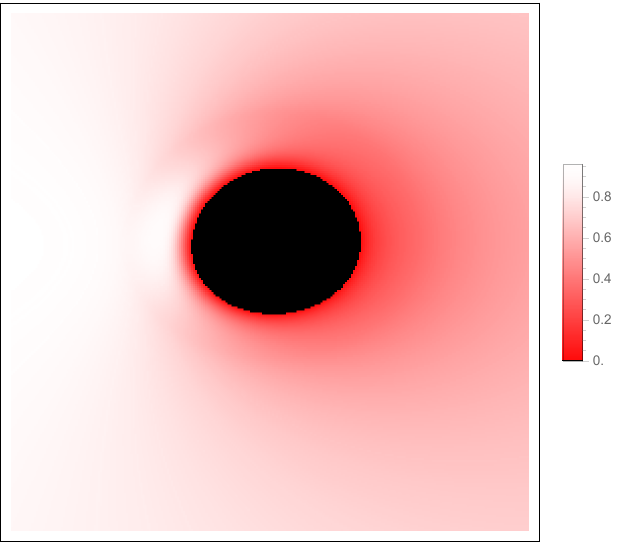}}
	\subfigure[$a=0.5,\theta_o=80^\circ$]{\includegraphics[scale=0.35]{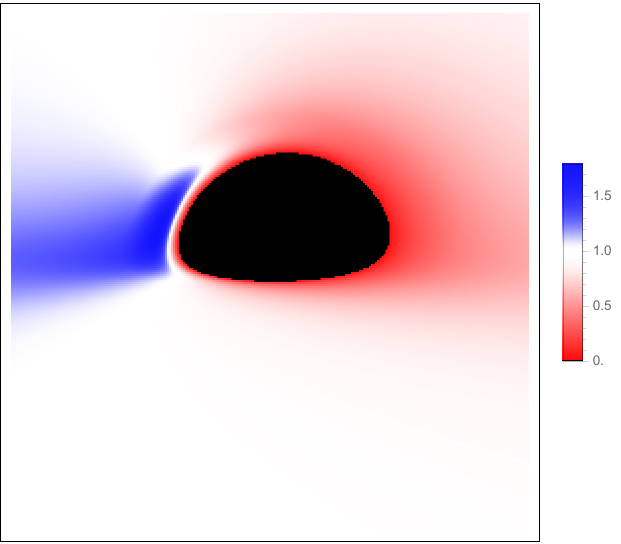}}
	
	\subfigure[$a=0.998,\theta_o=0^\circ$]{\includegraphics[scale=0.35]{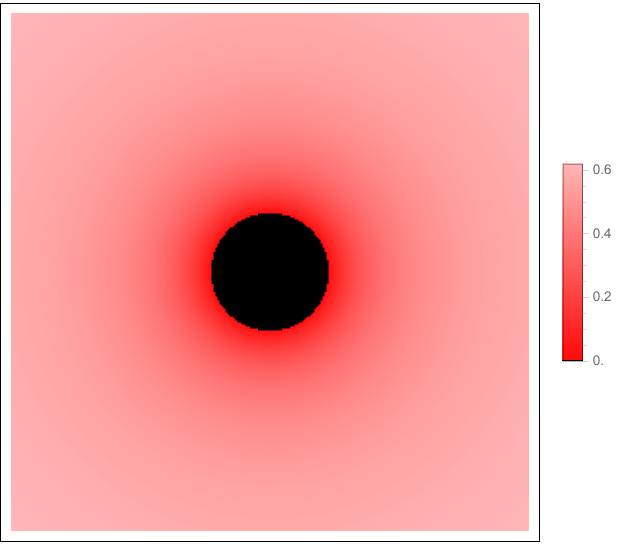}}
	\subfigure[$a=0.998,\theta_o=17^\circ$]{\includegraphics[scale=0.35]{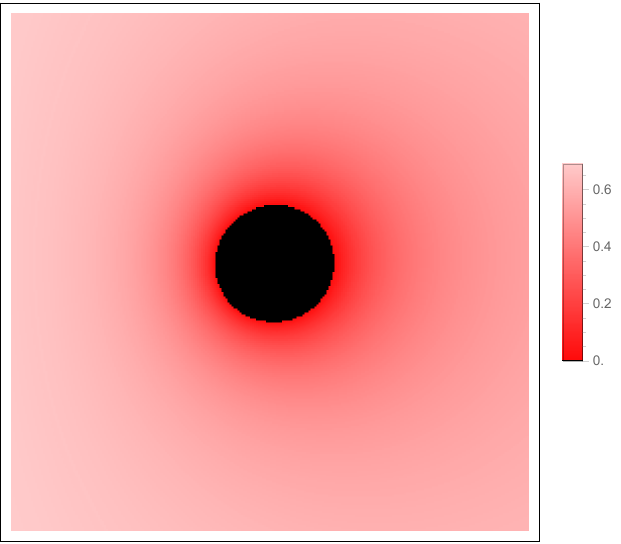}}
	\subfigure[$a=0.998,\theta_o=45^\circ$]{\includegraphics[scale=0.35]{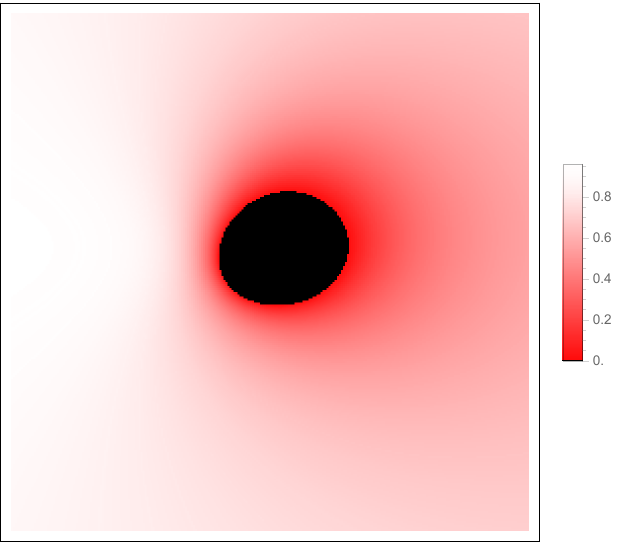}}
	\subfigure[$a=0.998,\theta_o=80^\circ$]{\includegraphics[scale=0.35]{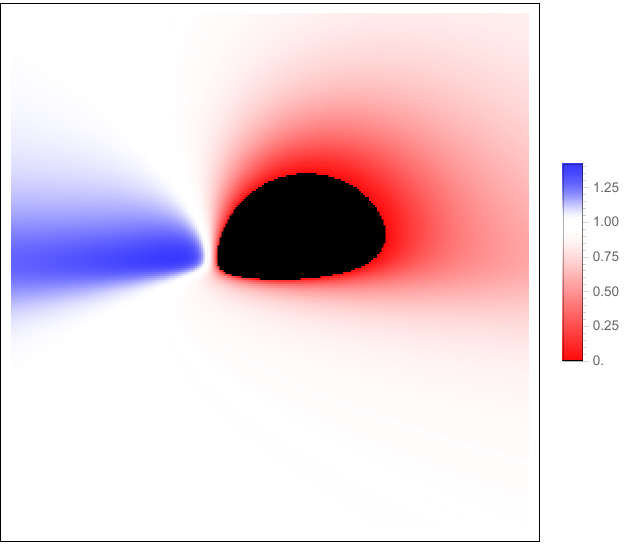}}
	
	\caption{Distribution of the redshift factor for the direct images with different values of the rotation parameter $a$ and observation inclination angle $\theta_o$. For all pictures, we set $B=0.002$ and $\alpha_{\mathrm{fov}}=3^\circ$}
	\label{fig9}
\end{figure}

\begin{figure}[H]
	\centering 
	\subfigure[$a=0.001,\theta_o=0^\circ$]{\includegraphics[scale=0.35]{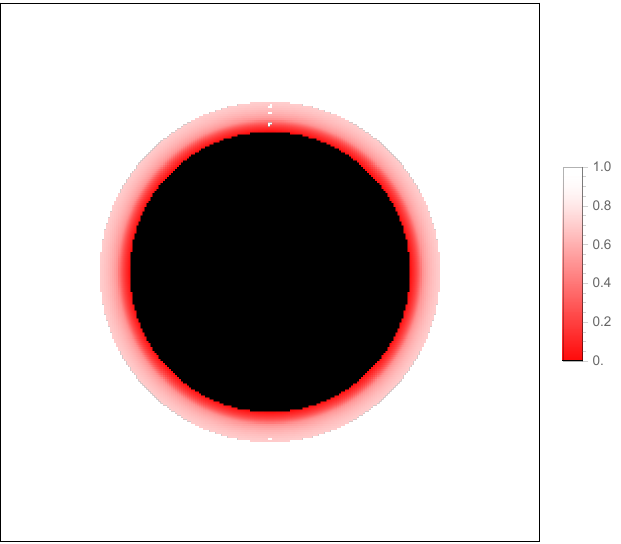}}
	\subfigure[$a=0.001,\theta_o=17^\circ$]{\includegraphics[scale=0.35]{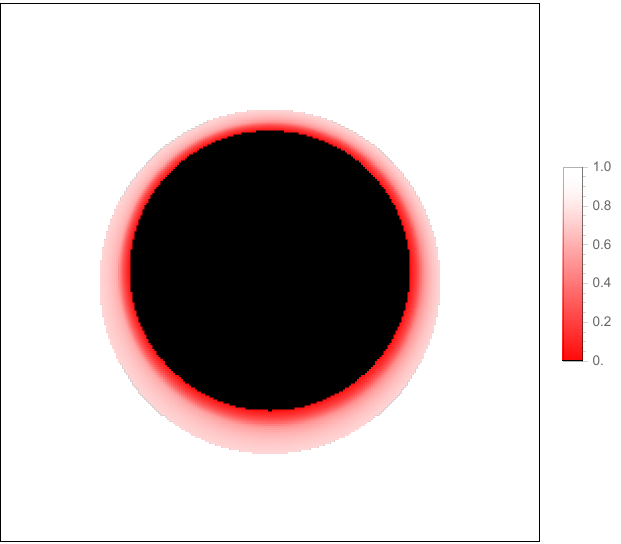}}
	\subfigure[$a=0.001,\theta_o=45^\circ$]{\includegraphics[scale=0.35]{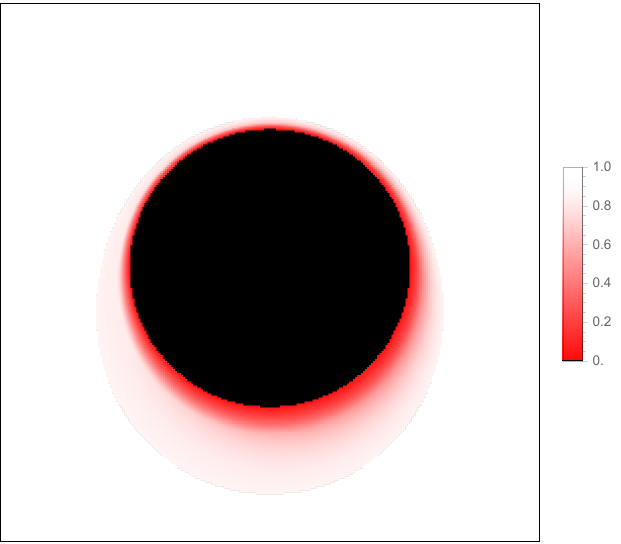}}
	\subfigure[$a=0.001,\theta_o=80^\circ$]{\includegraphics[scale=0.35]{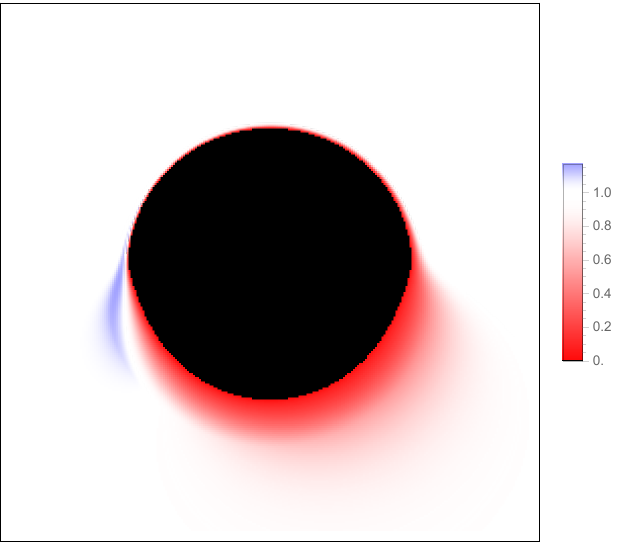}}
	
	\subfigure[$a=0.5,\theta_o=0^\circ$]{\includegraphics[scale=0.35]{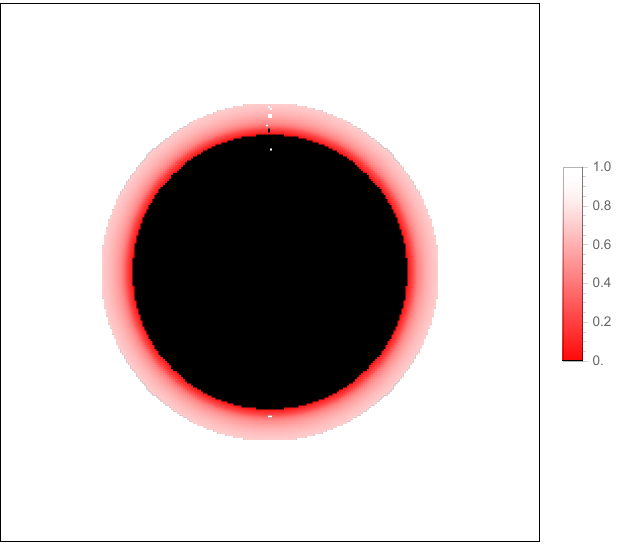}}
	\subfigure[$a=0.5,\theta_o=17^\circ$]{\includegraphics[scale=0.35]{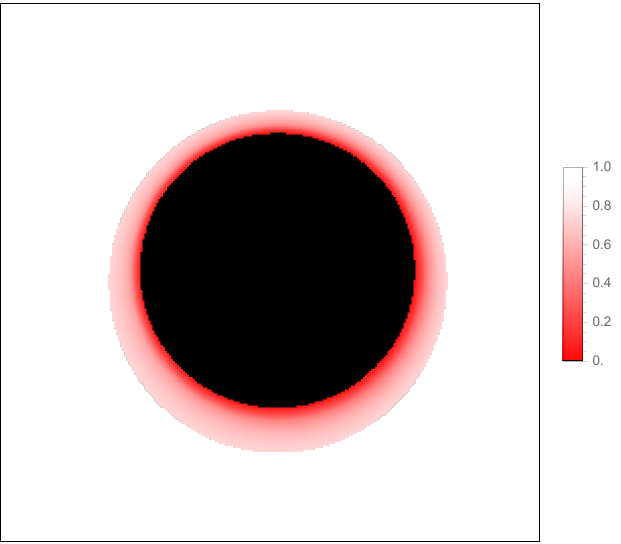}}
	\subfigure[$a=0.5,\theta_o=45^\circ$]{\includegraphics[scale=0.35]{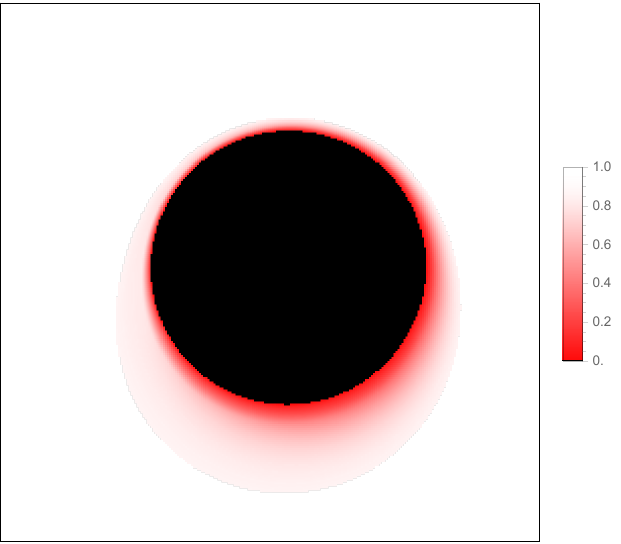}}
	\subfigure[$a=0.5,\theta_o=80^\circ$]{\includegraphics[scale=0.35]{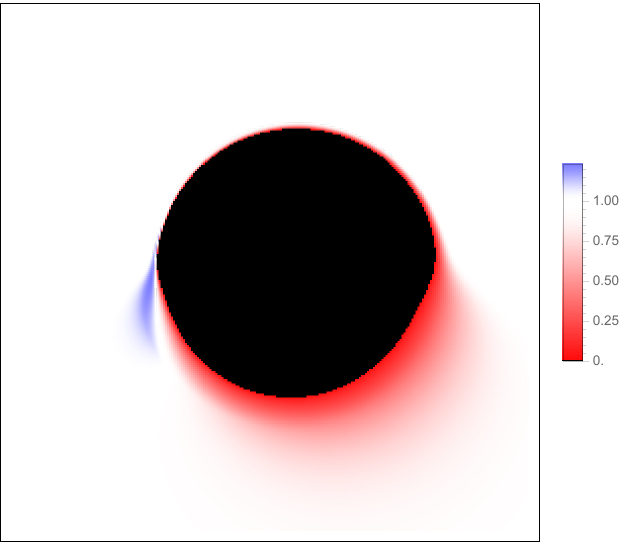}}
	
	\subfigure[$a=0.998,\theta_o=0^\circ$]{\includegraphics[scale=0.35]{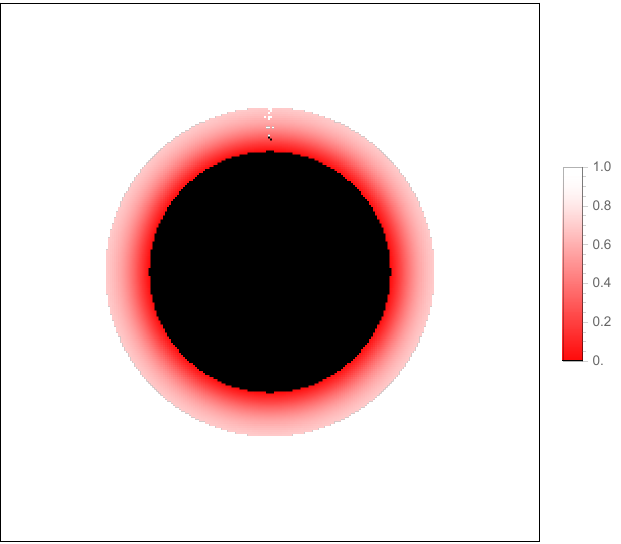}}
	\subfigure[$a=0.998,\theta_o=17^\circ$]{\includegraphics[scale=0.35]{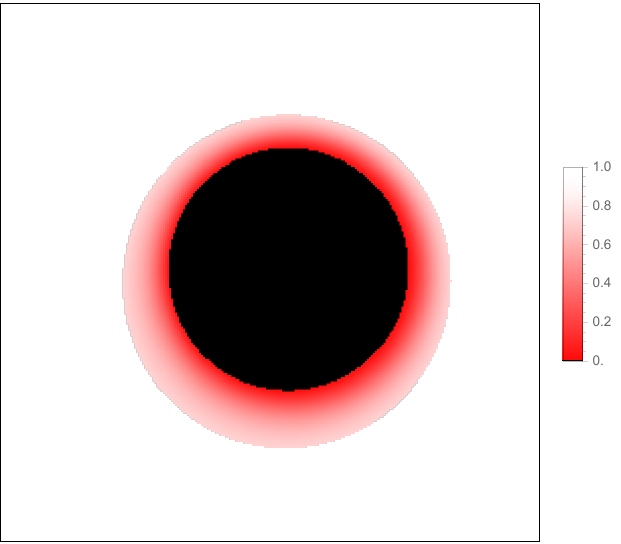}}
	\subfigure[$a=0.998,\theta_o=45^\circ$]{\includegraphics[scale=0.35]{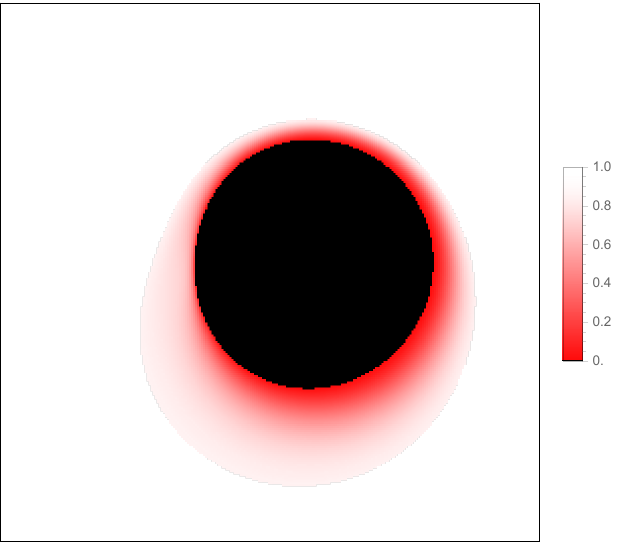}}
	\subfigure[$a=0.998,\theta_o=80^\circ$]{\includegraphics[scale=0.35]{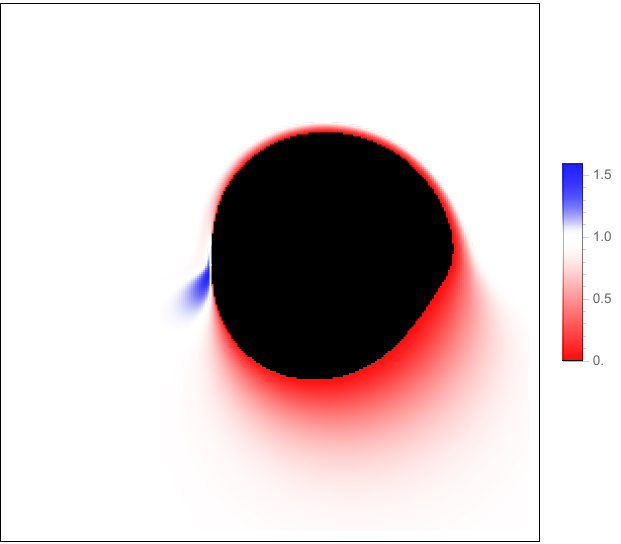}}
	
	\caption{Distribution of the redshift factor for the lensed images with different values of the rotation parameter $a$ and observation inclination angle $\theta_o$. For all pictures, we set $B=0.002$ and $\alpha_{\mathrm{fov}}=3^\circ$}
	\label{fig10}
\end{figure}

In Figs.~\ref{fig9} and~\ref{fig10}, we present the redshift (factor) distribution for the direct image and lensed image under a prograde thin accretion disk, varying with the rotation parameter $a$ and observation inclination angle $\theta_o$. In the figures, the magnetic field and the field angle are fixed at $B=0.002$ and $\alpha_{\mathrm{fov}}=3^\circ$, respectively. The direct image corresponds to the redshift factor $\chi_1$, and the lensed image corresponds to $\chi_2$. The red and blue colors represent redshift and blueshift, respectively. For the direct image, the black region at the center of the image is the inner shadow (see Fig.~\ref{fig9}), while for the lensed image, the black region includes the inner shadow and part of the direct image (see Fig.~\ref{fig10}).

For the direct image, it can be observed that when the observation inclination angle $\theta_o$ is small (e.g., $\theta_o=0^\circ$ and $\theta_o=17^\circ$), only redshift is present. The redshift becomes stronger closer to the inner shadow, where gravitational redshift dominates. The variation of the rotation parameter $a$ (see each column from top to bottom) has a less significant influence on the overall redshift distribution compared to the influence of $\theta_o$ (each row  from left to right). Generally, the larger the observation inclination angle and the smaller the raotation parameter, the more pronounced the blueshift. Notably, the blueshift appearing on the left side of the image and the redshift on the right are determined by the motion direction of the thin accretion disk. For a retrograde thin accretion disk, these two regions swap (see Fig.~\ref{fig13}). For the lensed image, the variation of $a$ (see each column from top to bottom) has a relatively weak effect on the overall redshift distribution, and the observation inclination angle $\theta_o$ only produces blueshift at larger values (e.g., $\theta_o=80^\circ$). Generally, the rotation parameter and observation inclination angle have a weaker influence on the redshift distribution in the lensed image compared to their impact on the direct image.

\begin{figure}[H]
	\centering 
	\subfigure[$B=0.001,\theta_o=0^\circ$]{\includegraphics[scale=0.35]{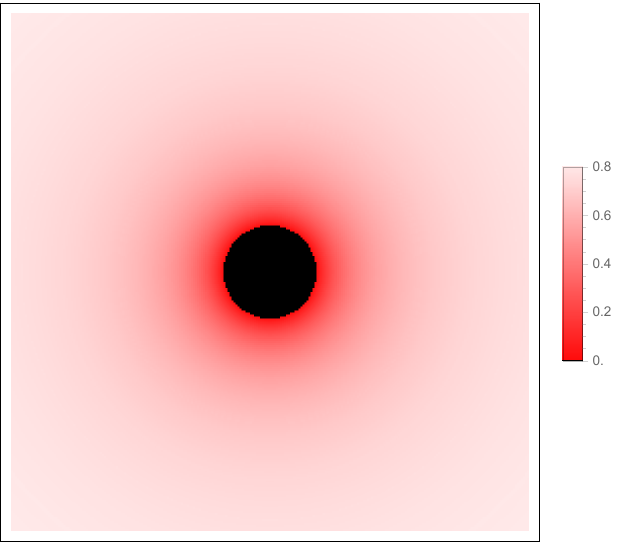}}
	\subfigure[$B=0.001,\theta_o=17^\circ$]{\includegraphics[scale=0.35]{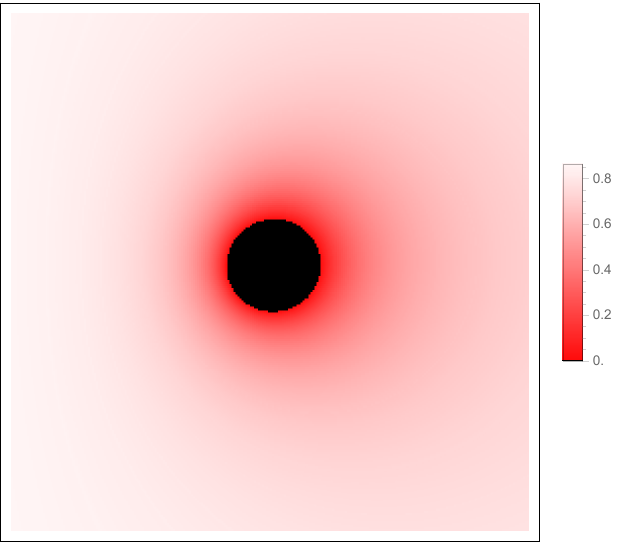}}
	\subfigure[$B=0.001,\theta_o=45^\circ$]{\includegraphics[scale=0.35]{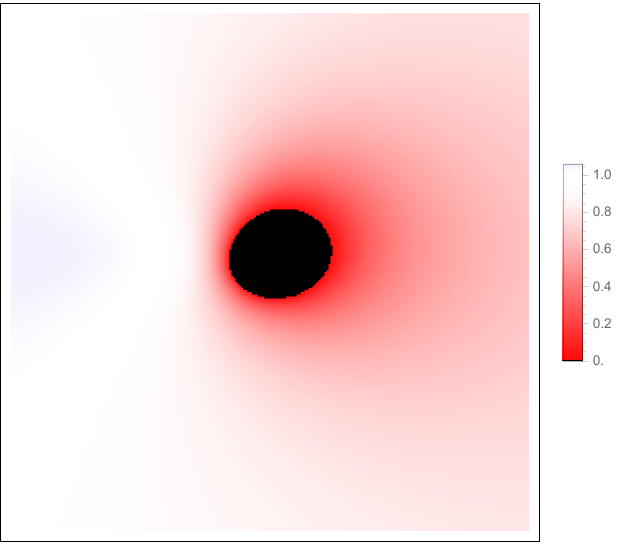}}
	\subfigure[$B=0.001,\theta_o=80^\circ$]{\includegraphics[scale=0.35]{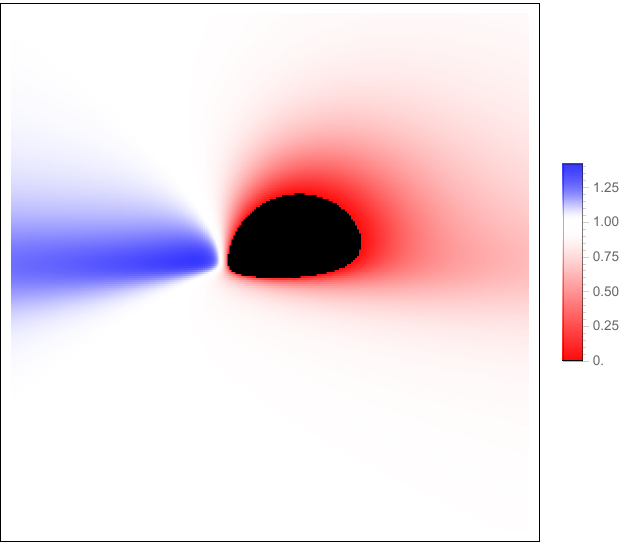}}
	
	\subfigure[$B=0.002,\theta_o=0^\circ$]{\includegraphics[scale=0.35]{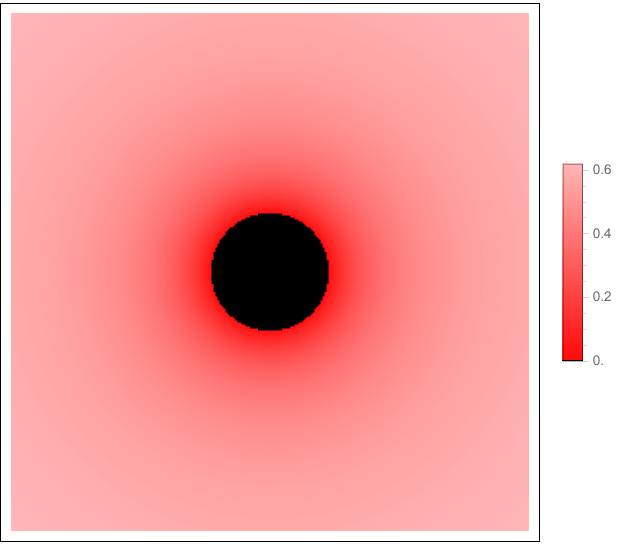}}
	\subfigure[$B=0.002,\theta_o=17^\circ$]{\includegraphics[scale=0.35]{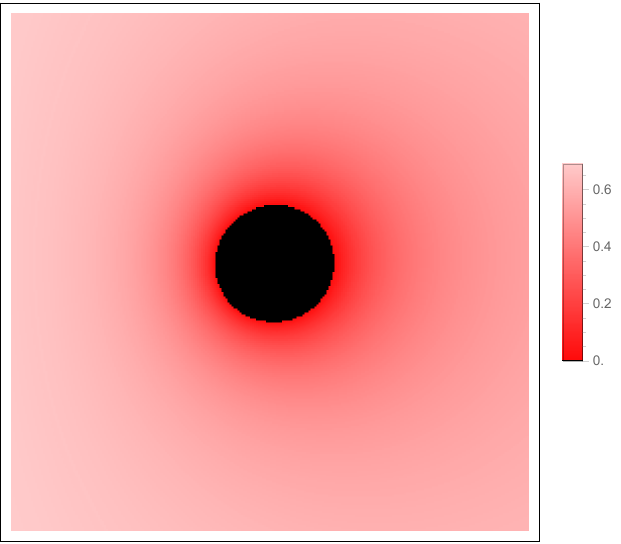}}
	\subfigure[$B=0.002,\theta_o=45^\circ$]{\includegraphics[scale=0.35]{Thin21_2.pdf}}
	\subfigure[$B=0.002,\theta_o=80^\circ$]{\includegraphics[scale=0.35]{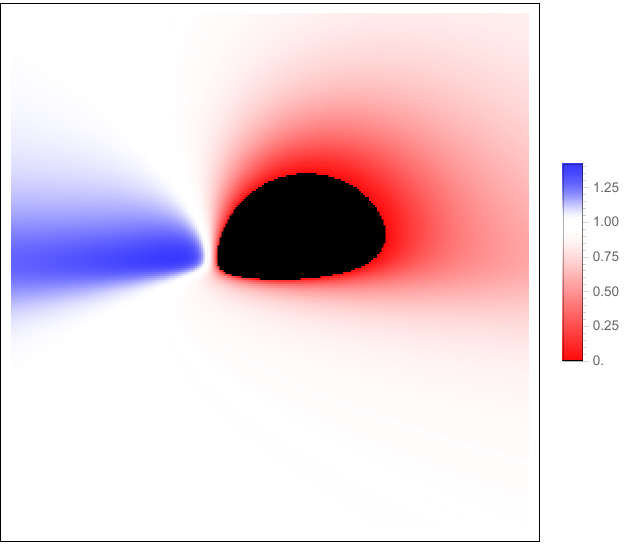}}
	
	\subfigure[$B=0.003,\theta_o=0^\circ$]{\includegraphics[scale=0.35]{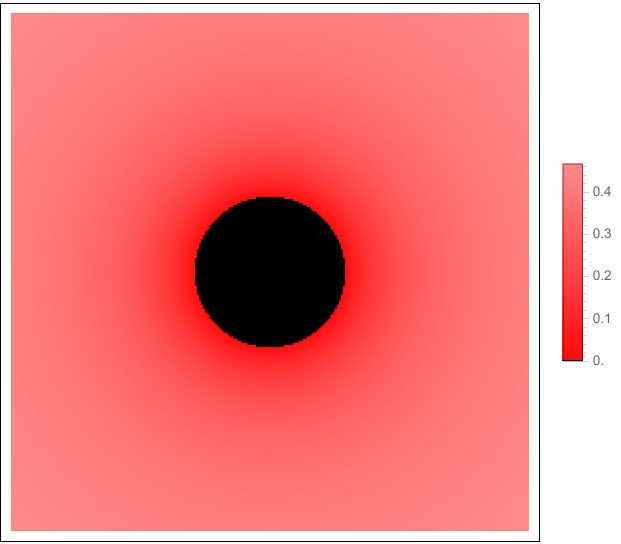}}
	\subfigure[$B=0.003,\theta_o=17^\circ$]{\includegraphics[scale=0.35]{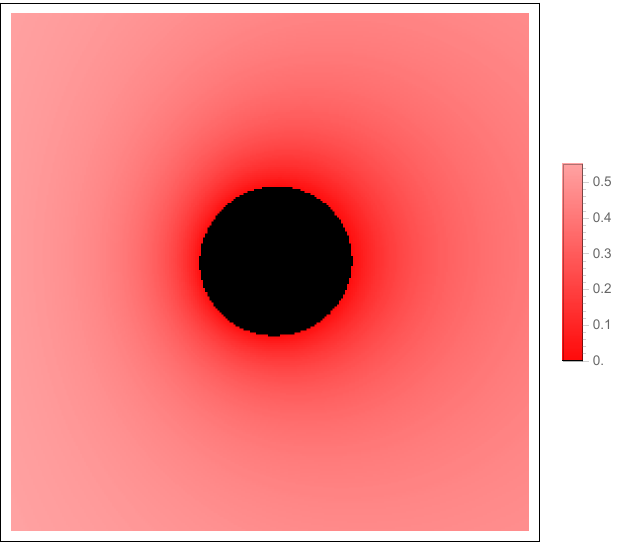}}
	\subfigure[$B=0.003,\theta_o=45^\circ$]{\includegraphics[scale=0.35]{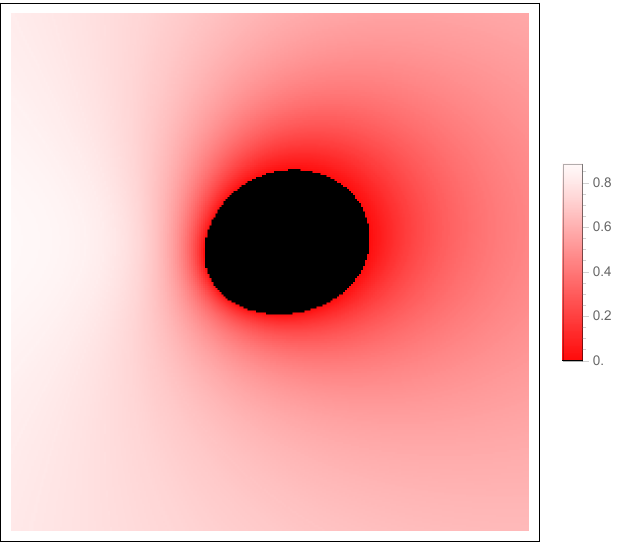}}
	\subfigure[$B=0.003,\theta_o=80^\circ$]{\includegraphics[scale=0.35]{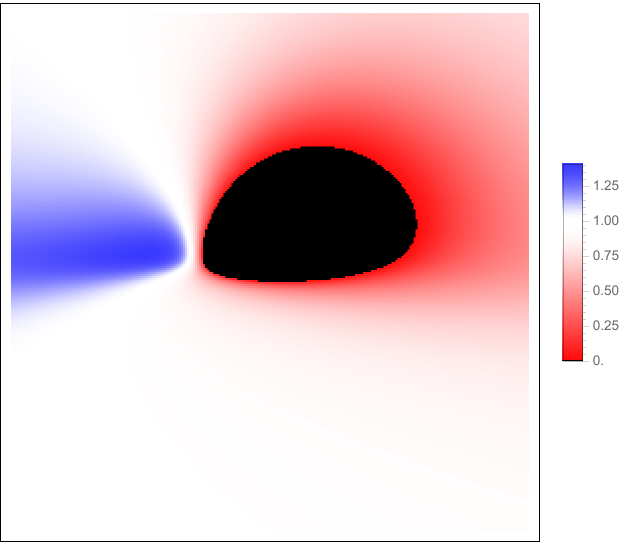}}
	
	\caption{Distribution of the redshift factor for the direct images with different values of the magnetic field $B$ and observation inclination angle $\theta_o$. For all pictures, we set $a=0.998$ and $\alpha_{\mathrm{fov}}=3^\circ$}
		\label{fig11}
\end{figure}

\begin{figure}[H]
	\centering 
	\subfigure[$B=0.001,\theta_o=0^\circ$]{\includegraphics[scale=0.35]{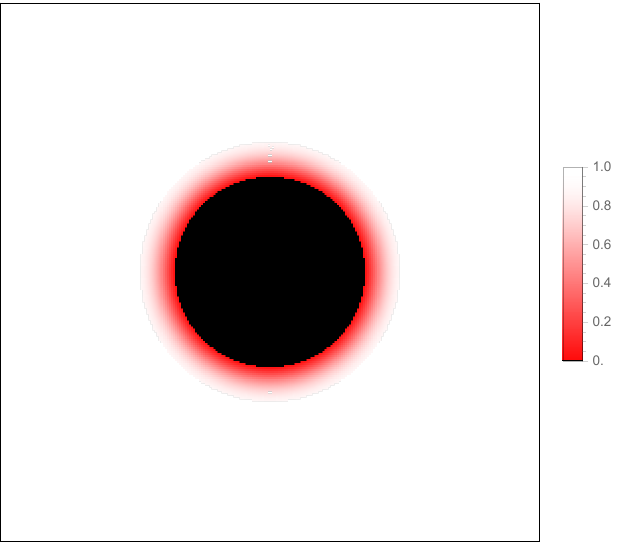}}
	\subfigure[$B=0.001,\theta_o=17^\circ$]{\includegraphics[scale=0.35]{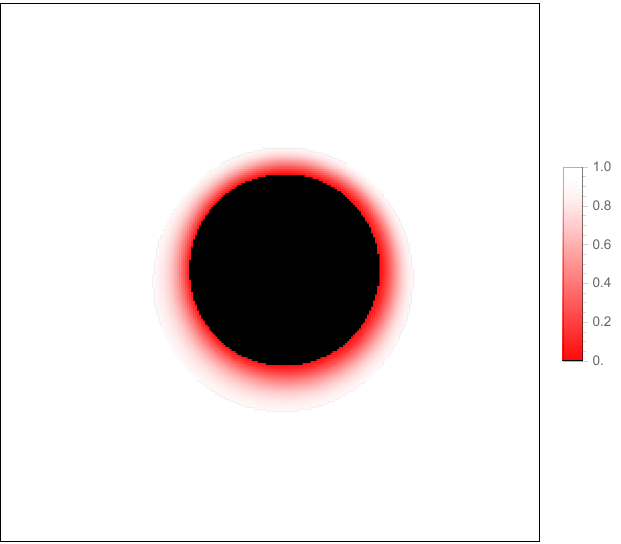}}
	\subfigure[$B=0.001,\theta_o=45^\circ$]{\includegraphics[scale=0.35]{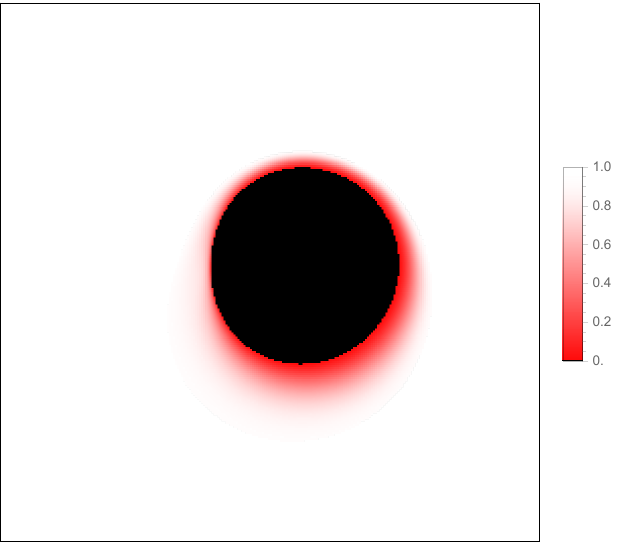}}
	\subfigure[$B=0.001,\theta_o=80^\circ$]{\includegraphics[scale=0.35]{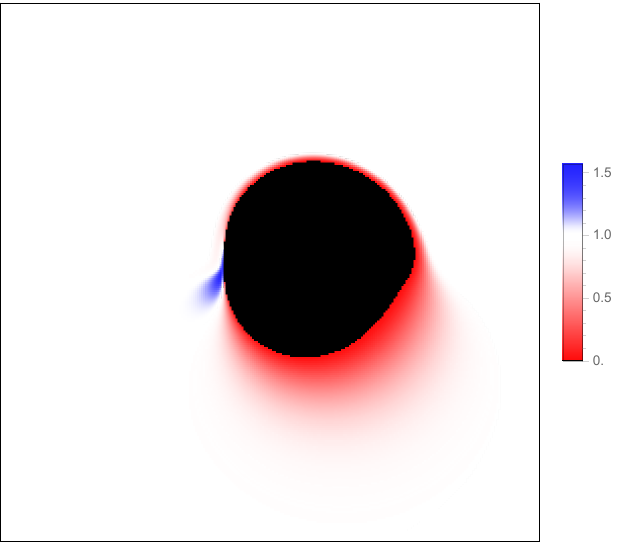}}
	
	\subfigure[$B=0.002,\theta_o=0^\circ$]{\includegraphics[scale=0.35]{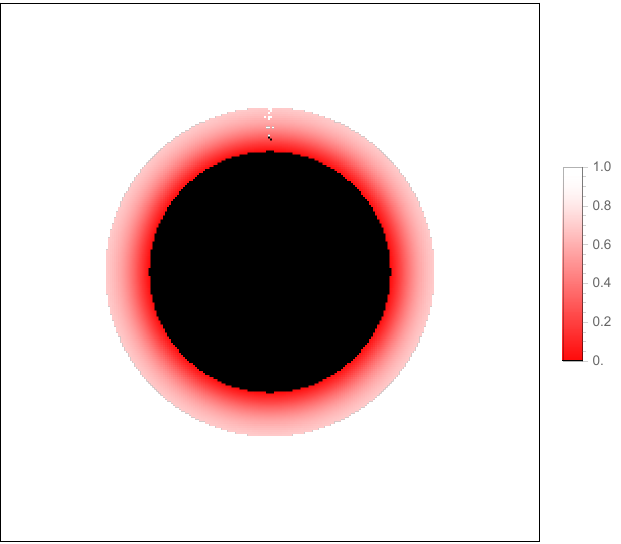}}
	\subfigure[$B=0.002,\theta_o=17^\circ$]{\includegraphics[scale=0.35]{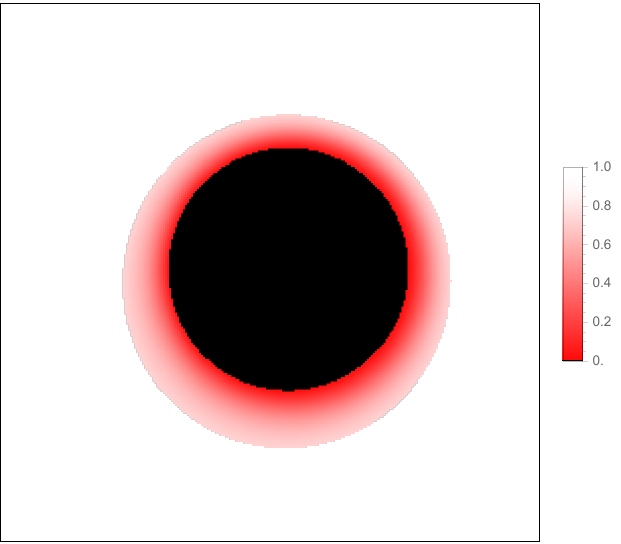}}
	\subfigure[$B=0.002,\theta_o=45^\circ$]{\includegraphics[scale=0.35]{Thin21_3.pdf}}
	\subfigure[$B=0.002,\theta_o=80^\circ$]{\includegraphics[scale=0.35]{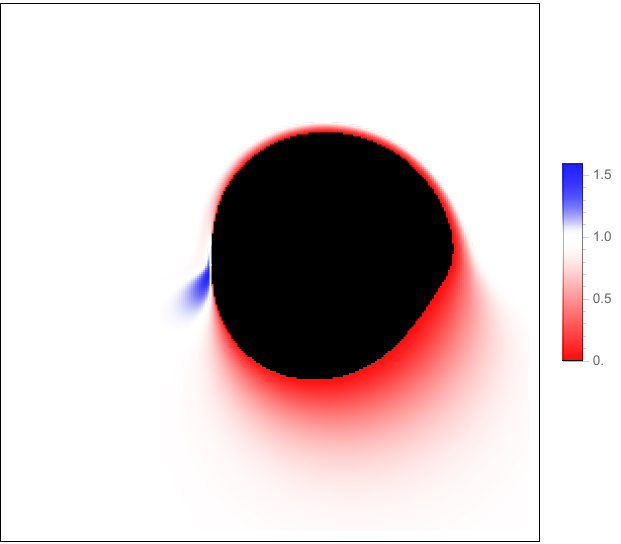}}
	
	\subfigure[$B=0.003,\theta_o=0^\circ$]{\includegraphics[scale=0.35]{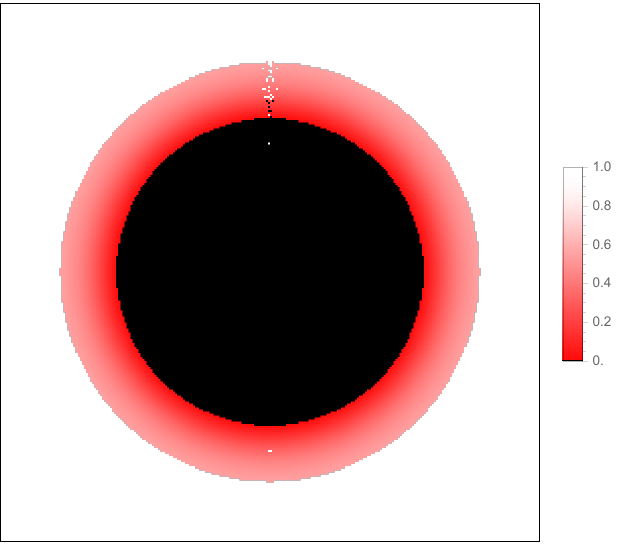}}
	\subfigure[$B=0.003,\theta_o=17^\circ$]{\includegraphics[scale=0.35]{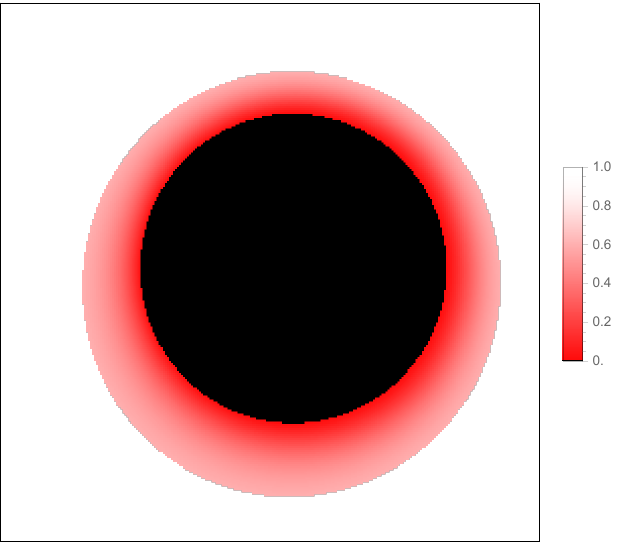}}
	\subfigure[$B=0.003,\theta_o=45^\circ$]{\includegraphics[scale=0.35]{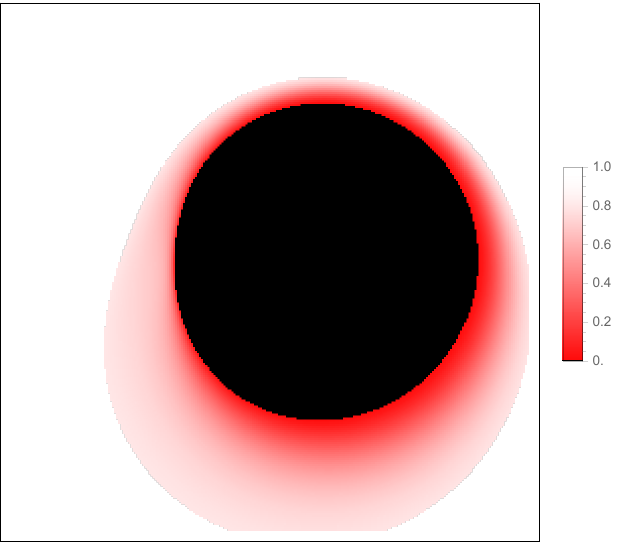}}
	\subfigure[$B=0.003,\theta_o=80^\circ$]{\includegraphics[scale=0.35]{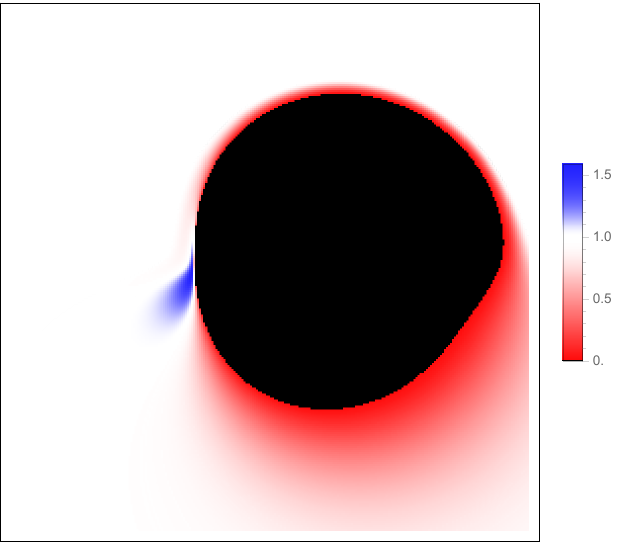}}
	
	\caption{Distribution of the redshift factor for the lensed images with different values of the magnetic field $B$ and observation inclination angle $\theta_o$. For all pictures, we set $a=0.998$ and $\alpha_{\mathrm{fov}}=3^\circ$}
		\label{fig12}
\end{figure}

\begin{figure}[H]
	\centering 
	\subfigure[$B=0.001$]{\includegraphics[scale=0.5]{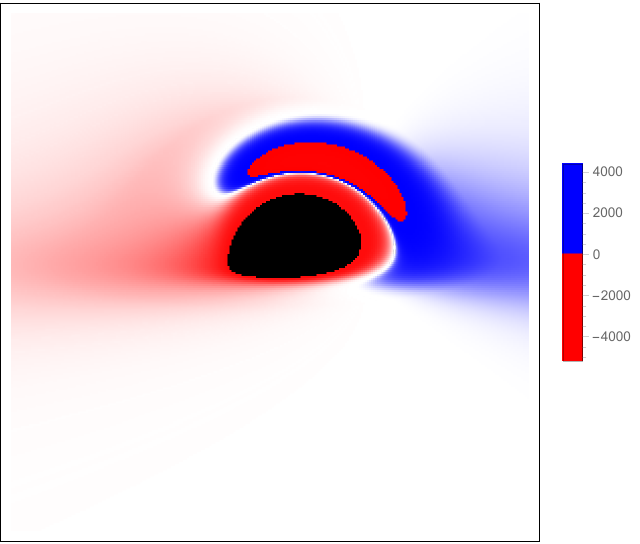}}
	\subfigure[$B=0.002$]{\includegraphics[scale=0.5]{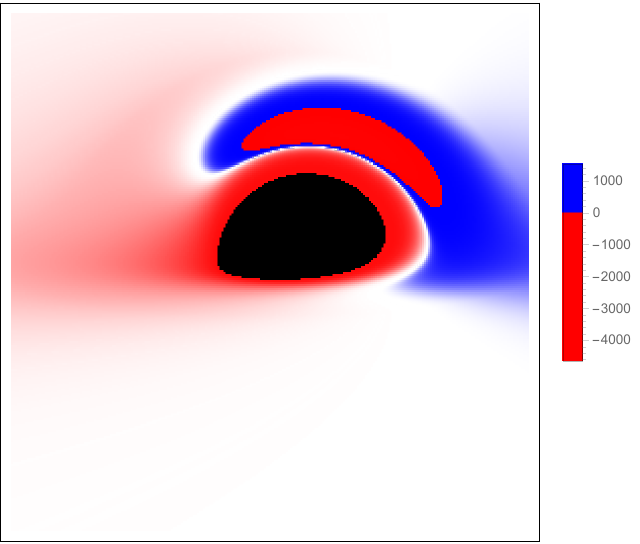}}
	\subfigure[$B=0.003$]{\includegraphics[scale=0.5]{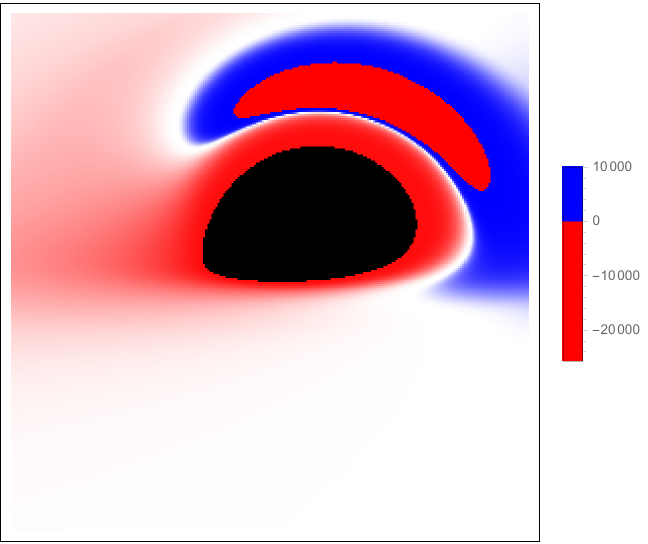}}
	
	\caption{Distribution of the redshift factor for the direct images with different values of the magnetic field $B$ (under a retrograde thin accretion disk source). For all pictures, we set $a=0.998$, $\alpha_{\mathrm{fov}}=3^\circ$, and $\theta_o=80^\circ$}
		\label{fig13}
\end{figure}

In Figs.~\ref{fig11} and~\ref{fig12}, we present the redshift distribution for the direct image and lensed image, varying with the magnetic field $B$ and observation inclination angle $\theta_o$. In the figures, the rotation parameter and field angle are fixed at $a=0.998$ and $\alpha_{\mathrm{fov}}=3^\circ$, respectively. The influence of the observation inclination angle on the redshift distribution is consistent with previous discussions, so we will not elaborate further. Here, we focus on the impact of the magnetic field $B$ on the redshift distribution and how it differs from the effect of the rotation parameter $a$. For the direct image (see Fig.~\ref{fig11}), the influence of $B$ on the redshift distribution (see each column from top to bottom) is even weaker than that of $a$. Even at large observation inclination angles (see the fourth column, $\theta_o=80^\circ$), changes in $B$ do not lead to significant variations in the redshift distribution. The same phenomenon is also observed in the lensed image (see Fig.~\ref{fig12}). Although the rotation parameter $a$ has a relatively weak effect on the redshift distribution, the influence of the magnetic field $B$ is almost negligible. Therefore, in terms of the redshift distribution, the Kerr-BR black hole behaves similarly to a purely rotating black hole, with the redshift distribution primarily determined by the rotation parameter $a$.

\begin{figure}[H]
	\centering 
	\subfigure[$B=0.001$]{\includegraphics[scale=0.5]{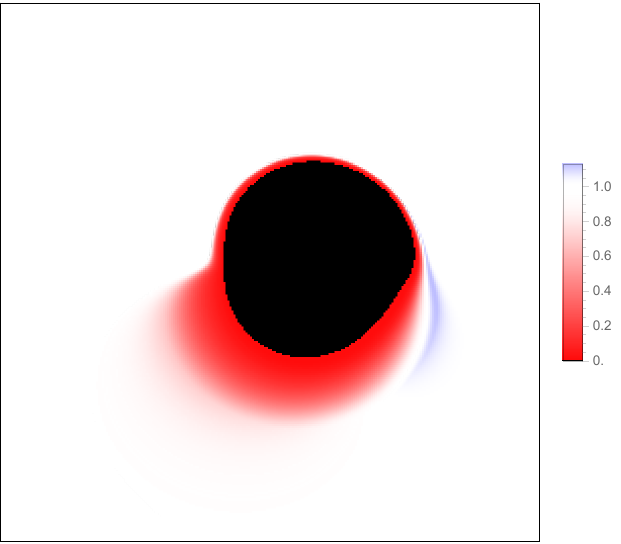}}
	\subfigure[$B=0.002$]{\includegraphics[scale=0.5]{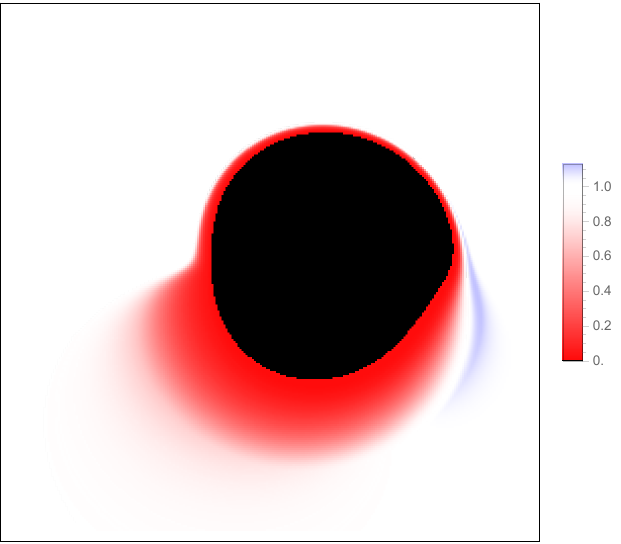}}
	\subfigure[$B=0.003$]{\includegraphics[scale=0.5]{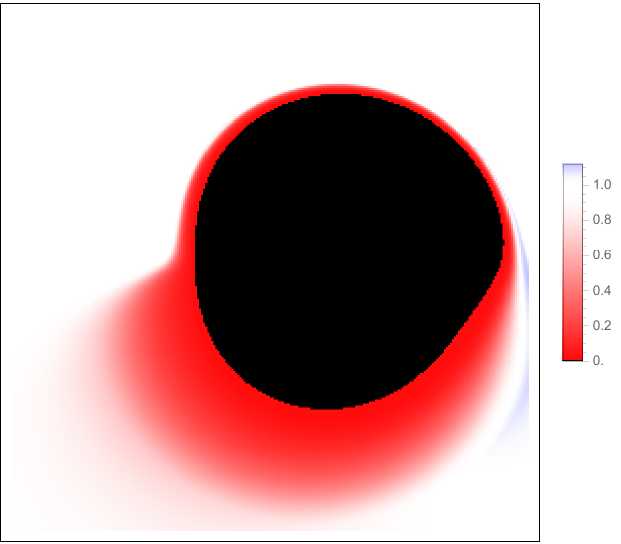}}
	
	\caption{Distribution of the redshift factor for the lensed images with different values of the magnetic field $B$ (under a retrograde thin accretion disk source). For all pictures, we set $a=0.998$, $\alpha_{\mathrm{fov}}=3^\circ$, and $\theta_o=80^\circ$}
	\label{fig14}
\end{figure}

Finally, we present the redshift distribution for the direct image and lensed image under a retrograde thin accretion disk, varying with the magnetic field $B$ (see Figs.~\ref{fig13} and~\ref{fig14}). It is evident that, in terms of the overall contour of the images, the influence of the magnetic field $B$ on the redshift distribution is indeed negligible.

%%%%%%%%%%%%%%%%%%%%%%%%%%%%%%%%%%%%%%%%%%%%%%%%%%%%%%%%%%%%%%%%%%%%%%%%
%%%%%%%%%%%%%%%%%%%%%%%%%%%%%%%%%%%%%%%%%%%%%%%%%%%%%%%%%%%%%%%%%%%%%%%%
%%%%%%%%%%%%%%%%%%%%%%%%%%%%%%%%%%%%%%%%%%%%%%%%%%%%%%%%%%%%%%%%%%%%%%%%

\section{Constraints on the magnetic field $B$}
Finally, we try to constrain the magnetic field $B$ of the Kerr-BR black hole based on the date from M87* and Sgr A*. In astronomical observations, the physical quantity characterizing the size of the black hole shadow is the angular diameter $D$. When the black hole is distant from the observer, the angular diameter can be quantitatively expressed as~\cite{Amarilla:2011fx,Li:2024ctu}
\begin{equation}\label{333}
	D=2 \times 9.87098\tilde{R}_{d}\left(\frac{\mathcal{M}}{M_{\odot}}\right)\left(\frac{1kpc}{D_{o}}\right)\mathrm{\mu as},
\end{equation} 
where $D_o$ is the distance between the black hole and the observer, and $\mathcal{M}$ denotes the black hole mass. We assume the background spacetime of M87* and Sgr A* to be BR spacetime, which means that both M87* and Sgr A* are the Kerr-BR black holes. Using the formula~(\ref{333}), we can calculate their theoretical angular diameters under different parameter configurations. These are then compared with actual astronomical observations. 

For M87*, the distance from Earth is about $D_o = 16.8\ \mathrm{kpc}$, and the estimated black hole mass is $\mathcal{M} = (6.5 \pm 0.7) \times 10^6 M_\odot$. The shadow diameter obtained from astronomical observations is $D_{\mathrm{M87^*}} = (37.8 \pm 2.7)\ \mathrm{\mu as}$~\cite{Capozziello:2023tbo}. For Sgr A*, the distance from Earth is about $D_o = 8\ \mathrm{kpc}$, and the estimated black hole mass is $\mathcal{M} = \left(4.0_{-0.6}^{+1.1}\right) \times 10^6 M_\odot$. The shadow diameter obtained from astronomical observations is $D_{\mathrm{Sgr A^*}} = \left(48.7 \pm 7\right)\ \mathrm{\mu as}$\cite{KumarWalia:2022aop}.

\begin{figure}[H]
	\centering 
	\subfigure[Angular diameter of M87* shadow]{\includegraphics[scale=0.88]{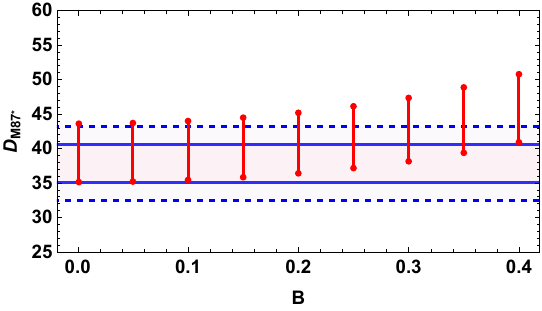}}
	\subfigure[Angular diameter of Sgr A* shadow]{\includegraphics[scale=0.88]{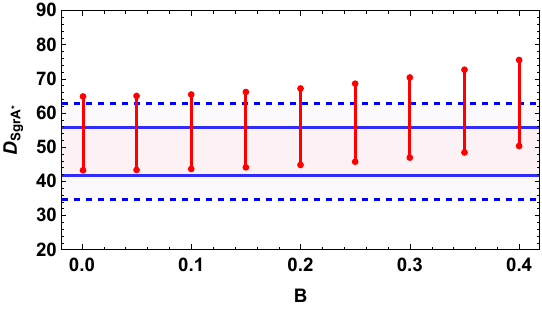}}
	\caption{Estimated ranges of the angular diameter $D$. The solid and dashed blue lines represent the $1\sigma$ and $2\sigma$ confidence intervals of $D$, respectively, while the red segments indicate the estimated ranges. The fixed parameter is $a=0.5$.}
	\label{fig15}
\end{figure}

In Fig.~\ref{fig15}, we present the theoretical variation of the angular diameter $D$ for M87* and Sgr A* with respect to the magnetic field $B$ in the Kerr-BR spacetime background. The rotation parameter is fixed at $a=0.5$. The left panel corresponds to M87*, while the right panel corresponds to Sgr A*. The solid blue line represents the $1\sigma$ confidence interval for $D$, the dashed blue line indicates the $2\sigma$ confidence interval, and the red segment denotes the theoretically estimated range of $D$ in the BR spacetime. As can be seen from Fig.~\ref{fig15}, for Sgr A*, the estimated interval remains entirely within the $1\sigma$ confidence bounds. However, for M87*, when $B=0.4$, the estimated interval exceeds the $1\sigma$ confidence range. This indicates that M87* imposes stricter constraints on the magnetic field $B$.

Overall, the constraints on $B$ provided by both M87* and Sgr A* are very limited. Even considering the $1\sigma$ confidence interval of M87*, the upper limit of $B$ reaches 0.4. Moreover, since the value of the rotation parameter $a$ also affects the angular diameter $D$, the constraints derived here are incomplete. Only with more comprehensive and precise observational data of black holes in the future will it be possible to further refine the constraints on the magnetic field $B$.

\section{Conclusion}

In this study, we conducted a comprehensive investigation into the optical characteristics of the Kerr-BR black hole, with a particular focus on its shadow contour and behavior under diverse illumination conditions. Through a combination of theoretical analysis and numerical simulations, we mainly studied the effects of the rotation parameter $a$ and the magnetic field $B$ on the optical properties of the Kerr-BR black hole.

Our results indicate that the shadow contour of the Kerr-BR black hole undergoes significant changes as the variations of the rotation parameter $a$ and  the magnetic field $B$. Specifically, an increase in $a$ causes the shadow contour to transition from a circular configuration to a ``D'' shape, reflecting a pronounced frame-dragging effect. Conversely, an increase $B$ results in a measurable expansion of both the shadow contour and the Einstein ring radius. This indicates that these two parameters have significantly different effects on the shadow contour of the black hole.

Additionally, we analyzed the optical images of the Kerr-BR black hole under celestial light source illumination and thin accretion disk illumination. The findings reveal that the shape and size of the optical image are related to variations in the rotation parameter $a$, the magnetic field $B$, and the observation inclination angle $\theta_o$. However, the variations in optical images exhibit distinct responses to different parameters. Generally, all optical images show significant sensitivity to the observation inclination angle $\theta_o$. Under celestial light source illumination, the magnetic field $B$ exerts notable influence on both the inner features and Einstein rings of the images, while the rotation parameter $a$ primarily affects the inner shadow. Consequently, under celestial light source illumination, differences in the impact of these two parameters on the optical image of the Kerr-BR black hole can be clearly distinguished. For thin accretion disk illumination, the optical images remain highly responsive to the observation inclination angle $\theta_o$. The rotation parameter $a$ continues to predominantly influence the inner shadow, whereas the magnetic field $B$ additionally affects the size of the lensed image and higher-order image. These results suggest that, in the case of the thin accretion disk source, it may indeed be feasible to determine the magnitude of the magnetic field $B$ in BR spacetime using actual observational images of black holes. We further analyze the relationship between the redshift factor distribution and three key parameters: the rotation parameter $a$, the magnetic field $B$, and the observation inclination angle $\theta_o$. Additionally, we present some results under a retrograde thin accretion disk source as supporting evidence for our study on the prograde thin accretion disk source. 

In the final section, we attempted to constrain the magnetic field $B$ using the data from M87* and Sgr A*, but the results proved unsatisfactory. In summary, this work establishes a robust theoretical framework and identifies key diagnostics for probing the magnetized environments (BR-like magnetic field) of rotating black holes through their optical characteristics. 
In the future, as the optical image of the black hole becomes more refined, we may be able to determine whether BR magnetic fields exist around black holes and even estimate their strength. This would significantly aid in distinguishing between rotating BR black holes and non-rotating, extremally charged black holes~\cite{Maldacena:1998uz,Clement:2000ms,Clement:2001gia,deCesare:2024csp}.

\vspace{10pt}
\noindent {\bf Acknowledgments}

\noindent
This work is supported by the National Natural Science Foundation of China (Grant No. 12375043), the Natural Science Foundation of Chongqing (CSTB2023NSCQ-MSX0594), and the China Postdoctoral Science Foundation (Grant No. 2024M753825).


\begin{thebibliography}{99}
	
%\cite{EventHorizonTelescope:2019dse}
\bibitem{EventHorizonTelescope:2019dse}
K.~Akiyama \textit{et al.} [Event Horizon Telescope],
%``First M87 Event Horizon Telescope Results. I. The Shadow of the Supermassive Black Hole,''
Astrophys. J. Lett. \textbf{875}, L1 (2019)
%doi:10.3847/2041-8213/ab0ec7
[arXiv:1906.11238 [astro-ph.GA]].
%4076 citations counted in INSPIRE as of 02 Aug 2025

%\cite{EventHorizonTelescope:2022wkp}
\bibitem{EventHorizonTelescope:2022wkp}
K.~Akiyama \textit{et al.} [Event Horizon Telescope],
%``First Sagittarius A* Event Horizon Telescope Results. I. The Shadow of the Supermassive Black Hole in the Center of the Milky Way,''
Astrophys. J. Lett. \textbf{930}, no.2, L12 (2022)
%doi:10.3847/2041-8213/ac6674
[arXiv:2311.08680 [astro-ph.HE]].
%1650 citations counted in INSPIRE as of 02 Aug 2025	

%\cite{Gralla:2019xty}
\bibitem{Gralla:2019xty}
S.~E.~Gralla, D.~E.~Holz and R.~M.~Wald,
%``Black Hole Shadows, Photon Rings, and Lensing Rings,''
Phys. Rev. D \textbf{100}, no.2, 024018 (2019)
%doi:10.1103/PhysRevD.100.024018
[arXiv:1906.00873 [astro-ph.HE]].
%519 citations counted in INSPIRE as of 02 Aug 2025


%\cite{Falcke:1999pj}
\bibitem{Falcke:1999pj}
H.~Falcke, F.~Melia and E.~Agol,
%``Viewing the shadow of the black hole at the galactic center,''
Astrophys. J. Lett. \textbf{528}, L13 (2000)
%doi:10.1086/312423
[arXiv:astro-ph/9912263 [astro-ph]].
%905 citations counted in INSPIRE as of 02 Aug 2025


%\cite{EventHorizonTelescope:2020qrl}
\bibitem{EventHorizonTelescope:2020qrl}
D.~Psaltis \textit{et al.} [Event Horizon Telescope],
%``Gravitational Test Beyond the First Post-Newtonian Order with the Shadow of the M87 Black Hole,''
Phys. Rev. Lett. \textbf{125}, no.14, 141104 (2020)
%%doi:10.1103/PhysRevLett.125.141104
[arXiv:2010.01055 [gr-qc]].
%312 citations counted in INSPIRE as of 02 Aug 2025

\bibitem{Zeng:2022pvb}
X.~X.~Zeng, K.~J.~He, G.~P.~Li, E.~W.~Liang and S.~Guo,
%``QED and accretion flow models effect on optical appearance of Euler{\textendash}Heisenberg black holes,''
Eur. Phys. J. C \textbf{82}, no.8, 764 (2022)
%%doi:10.1140/epjc/s10052-022-10733-y
[arXiv:2209.05938 [gr-qc]].

\bibitem{Zeng:2021mok}
X.~X.~Zeng, K.~J.~He and G.~P.~Li,
%``Effects of dark matter on shadows and rings of Brane-World black holes illuminated by various accretions,''
Sci. China Phys. Mech. Astron. \textbf{65}, no.9, 290411  (2022)
%%doi:10.1007/s11433-022-1896-0
[arXiv:2111.05090 [gr-qc]].

\bibitem{Zeng:2021dlj}
X.~X.~Zeng, G.~P.~Li and K.~J.~He,
%``The shadows and observational appearance of a noncommutative black hole surrounded by various profiles of accretions,''
Nucl. Phys. B \textbf{974}, 115639 (2022)
%%doi:10.1016/j.nuclphysb.2021.115639
[arXiv:2106.14478 [hep-th]].


\bibitem{Zeng:2025nmu}
X.~X.~Zeng, H.~Ye, K.~J.~He and H.~Yu,
%``Optical images of massive boson stars with nonlinear electrodynamics,''
[arXiv:2507.11583 [gr-qc]].

%\cite{Kerr:1963ud}
\bibitem{Kerr:1963ud}
R.~P.~Kerr,
%``Gravitational field of a spinning mass as an example of algebraically special metrics,''
Phys. Rev. Lett. \textbf{11}, 237-238  (1963).
%%doi:10.1103/PhysRevLett.11.237
%2700 citations counted in INSPIRE as of 02 Aug 2025

%\cite{Chandrasekhar:1985kt}
\bibitem{Chandrasekhar:1985kt}
S.~Chandrasekhar, \textit{The mathematical theory of black holes} (Clarendon Press, 1985).
%369 citations counted in INSPIRE as of 30 Jul 2025

%\cite{Synge:1966okc}
\bibitem{Synge:1966okc}
J.~L.~Synge,
%``The Escape of Photons from Gravitationally Intense Stars,''
Mon. Not. Roy. Astron. Soc. \textbf{131}, no.3, 463-466  (1966).
%%doi:10.1093/mnras/131.3.463
%564 citations counted in INSPIRE as of 02 Aug 2025

%\cite{Hioki:2009na}
\bibitem{Hioki:2009na}
K.~Hioki and K.~i.~Maeda,
%``Measurement of the Kerr Spin Parameter by Observation of a Compact Object's Shadow,''
Phys. Rev. D \textbf{80}, 024042 (2009)
%%doi:10.1103/PhysRevD.80.024042
[arXiv:0904.3575 [astro-ph.HE]].
%472 citations counted in INSPIRE as of 02 Aug 2025

%\cite{Atamurotov:2013sca}
\bibitem{Atamurotov:2013sca}
F.~Atamurotov, A.~Abdujabbarov and B.~Ahmedov,
%``Shadow of rotating non-Kerr black hole,''
Phys. Rev. D \textbf{88}, no.6, 064004 (2013).
%%doi:10.1103/PhysRevD.88.064004
%273 citations counted in INSPIRE as of 02 Aug 2025

%\cite{Gralla:2017ufe}
\bibitem{Gralla:2017ufe}
S.~E.~Gralla, A.~Lupsasca and A.~Strominger,
%``Observational Signature of High Spin at the Event Horizon Telescope,''
Mon. Not. Roy. Astron. Soc. \textbf{475}, no.3, 3829-3853 (2018)
%%doi:10.1093/mnras/sty039
[arXiv:1710.11112 [astro-ph.HE]].
%89 citations counted in INSPIRE as of 02 Aug 2025

%\cite{Wang:2017hjl}
\bibitem{Wang:2017hjl}
M.~Wang, S.~Chen and J.~Jing,
%``Shadow casted by a Konoplya-Zhidenko rotating non-Kerr black hole,''
JCAP \textbf{10}, 051 (2017)
%%doi:10.1088/1475-7516/2017/10/051
[arXiv:1707.09451 [gr-qc]].
%125 citations counted in INSPIRE as of 02 Aug 2025

%\cite{Chen:2020qyp}
\bibitem{Chen:2020qyp}
S.~Chen, M.~Wang and J.~Jing,
%``Polarization effects in Kerr black hole shadow due to the coupling between photon and bumblebee field,''
JHEP \textbf{07}, 054 (2020)
%%doi:10.1007/JHEP07(2020)054
[arXiv:2004.08857 [gr-qc]].
%48 citations counted in INSPIRE as of 02 Aug 2025

%\cite{Tsukamoto:2017fxq}
\bibitem{Tsukamoto:2017fxq}
N.~Tsukamoto,
%``Black hole shadow in an asymptotically-flat, stationary, and axisymmetric spacetime: The Kerr-Newman and rotating regular black holes,''
Phys. Rev. D \textbf{97}, no.6, 064021  (2018)
%%doi:10.1103/PhysRevD.97.064021
[arXiv:1708.07427 [gr-qc]].
%228 citations counted in INSPIRE as of 02 Aug 2025

%\cite{Wei:2019pjf}
\bibitem{Wei:2019pjf}
S.~W.~Wei, Y.~C.~Zou, Y.~X.~Liu and R.~B.~Mann,
%``Curvature radius and Kerr black hole shadow,''
JCAP \textbf{08}, 030 (2019)
%%doi:10.1088/1475-7516/2019/08/030
[arXiv:1904.07710 [gr-qc]].
%146 citations counted in INSPIRE as of 02 Aug 2025

%\cite{Chang:2021ngy}
\bibitem{Chang:2021ngy}
Z.~Chang and Q.~H.~Zhu,
%``The observer-dependent shadow of the Kerr black hole,''
JCAP \textbf{09}, 003 (2021)
%%doi:10.1088/1475-7516/2021/09/003
[arXiv:2104.14221 [gr-qc]].
%16 citations counted in INSPIRE as of 02 Aug 2025

%\cite{Afrin:2021imp}
\bibitem{Afrin:2021imp}
M.~Afrin, R.~Kumar and S.~G.~Ghosh,
%``Parameter estimation of hairy Kerr black holes from its shadow and constraints from M87*,''
Mon. Not. Roy. Astron. Soc. \textbf{504}, no.4, 5927-5940 (2021)
%%doi:10.1093/mnras/stab1260
[arXiv:2103.11417 [gr-qc]].
%172 citations counted in INSPIRE as of 02 Aug 2025

%\cite{Kuang:2022xjp}
\bibitem{Kuang:2022xjp}
X.~M.~Kuang and A.~{\"O}vg{\"u}n,
%``Strong gravitational lensing and shadow constraint from M87* of slowly rotating Kerr-like black hole,''
Annals Phys. \textbf{447}, 169147  (2022)
%%doi:10.1016/j.aop.2022.169147
[arXiv:2205.11003 [gr-qc]].
%106 citations counted in INSPIRE as of 02 Aug 2025

%\cite{Zheng:2024ftk}
\bibitem{Zheng:2024ftk}
H.~B.~Zheng, M.~Q.~Wu, G.~P.~Li and Q.~Q.~Jiang,
%``Shadows and accretion disk images of charged rotating black hole in modified gravity theory,''
Eur. Phys. J. C \textbf{85}, no.1, 46  (2025)
%%doi:10.1140/epjc/s10052-025-13791-0
[arXiv:2411.10315 [gr-qc]].
%12 citations counted in INSPIRE as of 02 Aug 2025

%\cite{Luminet:1979nyg}
\bibitem{Luminet:1979nyg}
J.~P.~Luminet,
%``Image of a spherical black hole with thin accretion disk,''
Astron. Astrophys. \textbf{75}, 228-235 (1979).
%846 citations counted in INSPIRE as of 02 Aug 2025

%\cite{Cui:1997zs}
\bibitem{Cui:1997zs}
W.~Cui, S.~N.~Zhang and W.~Chen,
%``Evidence for frame-dragging around spinning black holes in x-ray binaries,''
Astrophys. J. Lett. \textbf{492}, L53 (1998)
%%doi:10.1086/311092
[arXiv:astro-ph/9710352 [astro-ph]].
%96 citations counted in INSPIRE as of 02 Aug 2025

%\cite{Bisnovatyi-Kogan:2017kii}
\bibitem{Bisnovatyi-Kogan:2017kii}
G.~S.~Bisnovatyi-Kogan and O.~Y.~Tsupko,
%``Gravitational Lensing in Presence of Plasma: Strong Lens Systems, Black Hole Lensing and Shadow,''
Universe \textbf{3}, no.3, 57 (2017)
%%doi:10.3390/universe3030057
[arXiv:1905.06615 [gr-qc]].
%125 citations counted in INSPIRE as of 02 Aug 2025

%\cite{Kraniotis:2019ked}
\bibitem{Kraniotis:2019ked}
G.~V.~Kraniotis,
%``Gravitational redshift/blueshift of light emitted by geodesic test particles, frame-dragging and pericentre-shift effects, in the Kerr{\textendash}Newman{\textendash}de Sitter and Kerr{\textendash}Newman black hole geometries,''
Eur. Phys. J. C \textbf{81}, no.2, 147 (2021)
%%doi:10.1140/epjc/s10052-021-08911-5
[arXiv:1912.10320 [gr-qc]].
%31 citations counted in INSPIRE as of 02 Aug 2025

%\cite{Wong:2020ziu}
\bibitem{Wong:2020ziu}
G.~N.~Wong,
%``Black Hole Glimmer Signatures of Mass, Spin, and Inclination,''
Astrophys. J. \textbf{909}, no.2, 217 (2021)
%%doi:10.3847/1538-4357/abdd2d
[arXiv:2009.06641 [astro-ph.HE]].
%41 citations counted in INSPIRE as of 02 Aug 2025

%\cite{Kuang:2022ojj}
\bibitem{Kuang:2022ojj}
X.~M.~Kuang, Z.~Y.~Tang, B.~Wang and A.~Wang,
%``Constraining a modified gravity theory in strong gravitational lensing and black hole shadow observations,''
Phys. Rev. D \textbf{106}, no.6, 064012 (2022)
%%doi:10.1103/PhysRevD.106.064012
[arXiv:2206.05878 [gr-qc]].
%88 citations counted in INSPIRE as of 02 Aug 2025

%\cite{Blandford:1977ds}
\bibitem{Blandford:1977ds}
R.~D.~Blandford and R.~L.~Znajek,
%``Electromagnetic extractions of energy from Kerr black holes,''
Mon. Not. Roy. Astron. Soc. \textbf{179}, 433-456 (1977).
%%doi:10.1093/mnras/179.3.433
%3837 citations counted in INSPIRE as of 02 Aug 2025

%\cite{Begelman:1984mw}
\bibitem{Begelman:1984mw}
M.~C.~Begelman, R.~D.~Blandford and M.~J.~Rees,
%``Theory of extragalactic radio sources,''
Rev. Mod. Phys. \textbf{56}, 255-351 (1984).
%%doi:10.1103/RevModPhys.56.255
%621 citations counted in INSPIRE as of 02 Aug 2025

%\cite{Gammie:2003rj}
\bibitem{Gammie:2003rj}
C.~F.~Gammie, J.~C.~McKinney and G.~Toth,
%``HARM: A Numerical scheme for general relativistic magnetohydrodynamics,''
Astrophys. J. \textbf{589}, 444-457 (2003)
%%doi:10.1086/374594
[arXiv:astro-ph/0301509 [astro-ph]].
%667 citations counted in INSPIRE as of 02 Aug 2025



%\cite{Junior:2021dyw}
\bibitem{Junior:2021dyw}
H.~C.~D.~L.~Junior, P.~V.~P.~Cunha, C.~A.~R.~Herdeiro and L.~C.~B.~Crispino,
%``Shadows and lensing of black holes immersed in strong magnetic fields,''
Phys. Rev. D \textbf{104}, no.4, 044018 (2021)
%%doi:10.1103/PhysRevD.104.044018
[arXiv:2104.09577 [gr-qc]].
%71 citations counted in INSPIRE as of 02 Aug 2025


%\cite{Wang:2021ara}
\bibitem{Wang:2021ara}
M.~Wang, S.~Chen and J.~Jing,
%``Kerr black hole shadows in Melvin magnetic field with stable photon orbits,''
Phys. Rev. D \textbf{104}, no.8, 084021 (2021)
%%doi:10.1103/PhysRevD.104.084021
[arXiv:2104.12304 [gr-qc]].
%46 citations counted in INSPIRE as of 02 Aug 2025

%\cite{Zhong:2021mty}
\bibitem{Zhong:2021mty}
Z.~Zhong, Z.~Hu, H.~Yan, M.~Guo and B.~Chen,
%``QED effects on Kerr black hole shadows immersed in uniform magnetic fields,''
Phys. Rev. D \textbf{104}, no.10, 104028 (2021)
%%doi:10.1103/PhysRevD.104.104028
[arXiv:2108.06140 [gr-qc]].
%54 citations counted in INSPIRE as of 02 Aug 2025


%\cite{Junior:2021svb}
\bibitem{Junior:2021svb}
H.~C.~D.~L.~Junior, J.~Z.~Yang, L.~C.~B.~Crispino, P.~V.~P.~Cunha and C.~A.~R.~Herdeiro,
%``Einstein-Maxwell-dilaton neutral black holes in strong magnetic fields: Topological charge, shadows, and lensing,''
Phys. Rev. D \textbf{105}, no.6, 064070 (2022)
%%doi:10.1103/PhysRevD.105.064070
[arXiv:2112.10802 [gr-qc]].
%55 citations counted in INSPIRE as of 02 Aug 2025

%\cite{Taylor:2025ixw}
\bibitem{Taylor:2025ixw}
K.~J.~Taylor and A.~Ritz,
%``Null orbits and shadows in the Ernst-Wild geometry: insights for black holes immersed in a magnetic field,''
[arXiv:2505.11629 [gr-qc]].
%0 citations counted in INSPIRE as of 02 Aug 2025



%\cite{Yumoto:2012kz}
\bibitem{Yumoto:2012kz}
A.~Yumoto, D.~Nitta, T.~Chiba and N.~Sugiyama,
%``Shadows of Multi-Black Holes: Analytic Exploration,''
Phys. Rev. D \textbf{86}, 103001 (2012)
%%doi:10.1103/PhysRevD.86.103001
[arXiv:1208.0635 [gr-qc]].
%145 citations counted in INSPIRE as of 02 Aug 2025



	%\cite{Hu:2020usx}
\bibitem{Hu:2020usx}
Z.~Hu, Z.~Zhong, P.~C.~Li, M.~Guo and B.~Chen,
%``QED effect on a black hole shadow,''
Phys. Rev. D \textbf{103}, no.4, 044057 (2021)
%%doi:10.1103/PhysRevD.103.044057
[arXiv:2012.07022 [gr-qc]].
%103 citations counted in INSPIRE as of 30 Jul 2025


%\cite{Bertotti:1959pf}
\bibitem{Bertotti:1959pf}
B.~Bertotti,
%``Uniform electromagnetic field in the theory of general relativity,''
Phys. Rev. \textbf{116}, 1331 (1959).
%%doi:10.1103/PhysRev.116.1331
%372 citations counted in INSPIRE as of 02 Aug 2025

%\cite{Robinson:1959ev}
\bibitem{Robinson:1959ev}
I.~Robinson,
%``A Solution of the Maxwell-Einstein Equations,''
Bull. Acad. Pol. Sci. Ser. Sci. Math. Astron. Phys. \textbf{7}, 351-352 (1959).
%184 citations counted in INSPIRE as of 02 Aug 2025

%\cite{Kunduri:2013gce}
\bibitem{Kunduri:2013gce}
H.~K.~Kunduri and J.~Lucietti,
%``Classification of near-horizon geometries of extremal black holes,''
Living Rev. Rel. \textbf{16}, 8 (2013)
%%doi:10.12942/lrr-2013-8
[arXiv:1306.2517 [hep-th]].
%225 citations counted in INSPIRE as of 02 Aug 2025


%\cite{Podolsky:2025tle}
\bibitem{Podolsky:2025tle}
J.~Podolsky and H.~Ovcharenko,
%``Kerr black hole in a uniform magnetic field: An exact solution,''
[arXiv:2507.05199 [gr-qc]].
%1 citations counted in INSPIRE as of 30 Jul 2025

%\cite{Griffiths:2005qp}
\bibitem{Griffiths:2005qp}
J.~B.~Griffiths and J.~Podolsky,
%``A New look at the Plebanski-Demianski family of solutions,''
Int. J. Mod. Phys. D \textbf{15}, 335-370 (2006)
%%doi:10.1142/S0218271806007742
[arXiv:gr-qc/0511091 [gr-qc]].
%244 citations counted in INSPIRE as of 02 Aug 2025



%\cite{Podolsky:2006px}
\bibitem{Podolsky:2006px}
J.~Podolsky and J.~B.~Griffiths,
%``Accelerating Kerr-Newman black holes in (anti-)de Sitter space-time,''
Phys. Rev. D \textbf{73}, 044018 (2006)
%%doi:10.1103/PhysRevD.73.044018
[arXiv:gr-qc/0601130 [gr-qc]].
%69 citations counted in INSPIRE as of 02 Aug 2025

%\cite{Podolsky:2022xxd}
\bibitem{Podolsky:2022xxd}
J.~Podolsky and A.~Vratny,
%``New form of all black holes of type D with a cosmological constant,''
Phys. Rev. D \textbf{107}, no.8, 084034 (2023)
[erratum: Phys. Rev. D \textbf{108}, no.12, 129902 (2023)]
%%doi:10.1103/PhysRevD.107.084034
[arXiv:2212.08865 [gr-qc]].
%24 citations counted in INSPIRE as of 02 Aug 2025

%\cite{Wu:2024tuh}
\bibitem{Wu:2024tuh}
S.~Q.~Wu and D.~Wu,
%``Is the type-D NUT C metric really missing from the most general Pleba{\'n}ski-Demia{\'n}ski solution?,''
Phys. Rev. D \textbf{110}, no.10, 104072 (2024)
%%doi:10.1103/PhysRevD.110.104072
[arXiv:2409.06733 [gr-qc]].
%5 citations counted in INSPIRE as of 02 Aug 2025

%\cite{Ovcharenko:2025fxg}
\bibitem{Ovcharenko:2025fxg}
H.~Ovcharenko, J.~Podolsky and M.~Astorino,
%``Revisiting black holes of algebraic type D with a cosmological constant,''
Phys. Rev. D \textbf{111}, no.8, 084016 (2025)
%%doi:10.1103/PhysRevD.111.084016
[arXiv:2501.07537 [gr-qc]].
%2 citations counted in INSPIRE as of 02 Aug 2025

%\cite{Wald:1974np}
\bibitem{Wald:1974np}
R.~M.~Wald,
%``Black hole in a uniform magnetic field,''
Phys. Rev. D \textbf{10}, 1680-1685 (1974).
%%doi:10.1103/PhysRevD.10.1680
%691 citations counted in INSPIRE as of 02 Aug 2025

%\cite{Garfinkle:2011mp}
\bibitem{Garfinkle:2011mp}
D.~Garfinkle and E.~N.~Glass,
%``Bertotti-Robinson and Melvin Spacetimes,''
Class. Quant. Grav. \textbf{28}, 215012 (2011)
%doi:10.1088/0264-9381/28/21/215012
[arXiv:1109.1535 [gr-qc]].
%15 citations counted in INSPIRE as of 04 Aug 2025

%\cite{Melvin:1963qx}
\bibitem{Melvin:1963qx}
M.~A.~Melvin,
%``Pure magnetic and electric geons,''
Phys. Lett. \textbf{8}, 65-70 (1964).
%doi:10.1016/0031-9163(64)90801-7
%418 citations counted in INSPIRE as of 04 Aug 2025

%\cite{Melvin:1965zza}
\bibitem{Melvin:1965zza}
M.~A.~Melvin,
%``Dynamics of Cylindrical Electromagnetic Universes,''
Phys. Rev. \textbf{139}, B225-B243 (1965).
%doi:10.1103/PhysRev.139.B225
%109 citations counted in INSPIRE as of 04 Aug 2025

%\cite{Ernst:1976bsr}
\bibitem{Ernst:1976bsr}
F.~J.~Ernst and W.~J.~Wild,
%``Kerr black holes in a magnetic universe,''
J. Math. Phys. \textbf{17}, no.2, 182 (1976).
%doi:10.1063/1.522875
%114 citations counted in INSPIRE as of 04 Aug 2025



%\cite{Maldacena:1998uz}
\bibitem{Maldacena:1998uz}
J.~M.~Maldacena, J.~Michelson and A.~Strominger,
%``Anti-de Sitter fragmentation,''
JHEP \textbf{02}, 011 (1999)
%doi:10.1088/1126-6708/1999/02/011
[arXiv:hep-th/9812073 [hep-th]].
%547 citations counted in INSPIRE as of 04 Aug 2025

%\cite{Clement:2000ms}
\bibitem{Clement:2000ms}
G.~Clement and D.~Gal'tsov,
%``The Near horizon geometry of dilaton axion black holes,''
%doi:10.1142/9789812777386{\_}0207
[arXiv:gr-qc/0101040 [gr-qc]].
%2 citations counted in INSPIRE as of 04 Aug 2025

%\cite{Clement:2001gia}
\bibitem{Clement:2001gia}
G.~Clement and D.~Gal'tsov,
%``Bertotti-Robinson type solutions to dilaton - axion gravity,''
Phys. Rev. D \textbf{63}, 124011 (2001)
%doi:10.1103/PhysRevD.63.124011
[arXiv:gr-qc/0102025 [gr-qc]].
%18 citations counted in INSPIRE as of 04 Aug 2025

%\cite{deCesare:2024csp}
\bibitem{deCesare:2024csp}
M.~de Cesare, R.~Oliveri and A.~P.~Porfyriadis,
%``Connecting gravitational perturbations: From Bertotti-Robinson to extreme Reissner-Nordstr{\"o}m black holes,''
Phys. Rev. D \textbf{111}, no.4, 044028 (2025)
%doi:10.1103/PhysRevD.111.044028
[arXiv:2410.23446 [gr-qc]].
%2 citations counted in INSPIRE as of 04 Aug 2025


%\cite{Strominger:1998yg}
\bibitem{Strominger:1998yg}
A.~Strominger,
%``AdS(2) quantum gravity and string theory,''
JHEP \textbf{01}, 007 (1999)
%doi:10.1088/1126-6708/1999/01/007
[arXiv:hep-th/9809027 [hep-th]].
%336 citations counted in INSPIRE as of 04 Aug 2025

%\cite{Spradlin:1999bn}
\bibitem{Spradlin:1999bn}
M.~Spradlin and A.~Strominger,
%``Vacuum states for AdS(2) black holes,''
JHEP \textbf{11}, 021 (1999)
%doi:10.1088/1126-6708/1999/11/021
[arXiv:hep-th/9904143 [hep-th]].
%172 citations counted in INSPIRE as of 04 Aug 2025




%\cite{Cadoni:1994uf}
\bibitem{Cadoni:1994uf}
M.~Cadoni and S.~Mignemi,
%``Nonsingular four-dimensional black holes and the Jackiw-Teitelboim theory,''
Phys. Rev. D \textbf{51}, 4319-4329 (1995)
%doi:10.1103/PhysRevD.51.4319
[arXiv:hep-th/9410041 [hep-th]].
%89 citations counted in INSPIRE as of 04 Aug 2025

%\cite{Navarro-Salas:1999zer}
\bibitem{Navarro-Salas:1999zer}
J.~Navarro-Salas and P.~Navarro,
%``AdS(2) / CFT(1) correspondence and near extremal black hole entropy,''
Nucl. Phys. B \textbf{579}, 250-266 (2000)
%doi:10.1016/S0550-3213(00)00165-6
[arXiv:hep-th/9910076 [hep-th]].
%105 citations counted in INSPIRE as of 04 Aug 2025

%\cite{Caldarelli:2000xk}
\bibitem{Caldarelli:2000xk}
M.~Caldarelli, G.~Catelani and L.~Vanzo,
%``Action, Hamiltonian and CFT for 2-D black holes,''
JHEP \textbf{10}, 005 (2000)
%doi:10.1088/1126-6708/2000/10/005
[arXiv:hep-th/0008058 [hep-th]].
%22 citations counted in INSPIRE as of 04 Aug 2025

%\cite{Ottewill:2012mq}
\bibitem{Ottewill:2012mq}
A.~C.~Ottewill and P.~Taylor,
%``Quantum field theory on the Bertotti-Robinson space-time,''
Phys. Rev. D \textbf{86}, 104067 (2012)
%doi:10.1103/PhysRevD.86.104067
[arXiv:1209.6080 [gr-qc]].
%13 citations counted in INSPIRE as of 04 Aug 2025

%\cite{Zeng:2025olq}
\bibitem{Zeng:2025olq}
X.~X.~Zeng and K.~Wang,
%``Energy Extraction From the Kerr-Bertotti-Robinson Black Hole via Magnetic Reconnection under Circular Plasma and Plunging Plasma,''
[arXiv:2507.21777 [gr-qc]].
%1 citations counted in INSPIRE as of 02 Aug 2025

%\cite{Wang:2025vsx}
\bibitem{Wang:2025vsx}
X.~Wang, Y.~Hou, X.~Wan, M.~Guo and B.~Chen,
%``Geodesics and Shadows in the Kerr-Bertotti-Robinson Black Hole Spacetime,''
[arXiv:2507.22494 [gr-qc]].
%0 citations counted in INSPIRE as of 02 Aug 2025


\bibitem{Cunha:2018acu}
P.~V.~P.~Cunha and C.~A.~R.~Herdeiro,
%``Shadows and strong gravitational lensing: a brief review,''
Gen. Rel. Grav. \textbf{50}, no.4, 42 (2018)
%%doi:10.1007/s10714-018-2361-9
[arXiv:1801.00860 [gr-qc]].


%\cite{Lee:2022rtg}
\bibitem{Lee:2022rtg}
T.~C.~Lee, Z.~Hu, M.~Guo and B.~Chen,
%``Circular orbits and polarized images of charged particles orbiting a Kerr black hole with a weak magnetic field,''
Phys. Rev. D \textbf{108}, no.2, 024008 (2023)
%%doi:10.1103/PhysRevD.108.024008
[arXiv:2211.04143 [gr-qc]].
%20 citations counted in INSPIRE as of 31 Jul 2025

%\cite{He:2024qka}
\bibitem{He:2024qka}
K.~J.~He, J.~T.~Yao, X.~Zhang and X.~Li,
%``Shadows and photon motions in the axially symmetric Finslerian extension of a Schwarzschild black hole,''
Phys. Rev. D \textbf{109}, no.6, 064049 (2024).
%%doi:10.1103/PhysRevD.109.064049
%15 citations counted in INSPIRE as of 31 Jul 2025



%\cite{Guo:2024mij}
\bibitem{Guo:2024mij}
S.~Guo, Y.~X.~Huang, E.~W.~Liang, Y.~Liang, Q.~Q.~Jiang and K.~Lin,
%``Image of the Kerr{\textendash}Newman Black Hole Surrounded by a Thin Accretion Disk,''
Astrophys. J. \textbf{975}, no.2, 237 (2024)
%%doi:10.3847/1538-4357/ad7d85
[arXiv:2411.07914 [astro-ph.HE]].
%15 citations counted in INSPIRE as of 31 Jul 2025


%\cite{Hou:2022eev}
\bibitem{Hou:2022eev}
Y.~Hou, Z.~Zhang, H.~Yan, M.~Guo and B.~Chen,
%``Image of a Kerr-Melvin black hole with a thin accretion disk,''
Phys. Rev. D \textbf{106}, no.6, 064058 (2022)
%%doi:10.1103/PhysRevD.106.064058
[arXiv:2206.13744 [gr-qc]].
%90 citations counted in INSPIRE as of 31 Jul 2025

%\cite{Zhang:2023bzv}
\bibitem{Zhang:2023bzv}
Z.~Zhang, Y.~Hou and M.~Guo,
%``Observational signatures of rotating black holes in the semiclassical gravity with trace anomaly*,''
Chin. Phys. C \textbf{48}, no.8, 085106 (2024)
%%doi:10.1088/1674-1137/ad432b
[arXiv:2305.14924 [gr-qc]].
%11 citations counted in INSPIRE as of 30 Jul 2025


%\cite{Yang:2024nin}
\bibitem{Yang:2024nin}
C.~Y.~Yang, M.~I.~Aslam, X.~X.~Zeng and R.~Saleem,
%``Shadow images of Ghosh-Kumar rotating black hole illuminated by spherical light sources and thin accretion disks,''
JHEAp \textbf{46}, 345 (2025)
%%doi:10.1016/j.jheap.2025.01.017
[arXiv:2411.11807 [astro-ph.HE]].
%14 citations counted in INSPIRE as of 01 Aug 2025


%\cite{Perlick:2021aok}
\bibitem{Perlick:2021aok}
V.~Perlick and O.~Y.~Tsupko,
%``Calculating black hole shadows: Review of analytical studies,''
Phys. Rept. \textbf{947}, 1-39 (2022)
%%doi:10.1016/j.physrep.2021.10.004
[arXiv:2105.07101 [gr-qc]].
%452 citations counted in INSPIRE as of 01 Aug 2025



%\cite{Amarilla:2011fx}
\bibitem{Amarilla:2011fx}
L.~Amarilla and E.~F.~Eiroa,
%``Shadow of a rotating braneworld black hole,''
Phys. Rev. D \textbf{85}, 064019 (2012)
%%doi:10.1103/PhysRevD.85.064019
[arXiv:1112.6349 [gr-qc]].
%304 citations counted in INSPIRE as of 02 Aug 2025

%\cite{Li:2024ctu}
\bibitem{Li:2024ctu}
G.~P.~Li, H.~B.~Zheng, K.~J.~He and Q.~Q.~Jiang,
%``The shadow and observational images of the non-singular rotating black holes in loop quantum gravity,''
Eur. Phys. J. C \textbf{85}, no.3, 249 (2025)
%%doi:10.1140/epjc/s10052-025-13997-2
[arXiv:2410.17295 [gr-qc]].
%19 citations counted in INSPIRE as of 02 Aug 2025

%\cite{Capozziello:2023tbo}
\bibitem{Capozziello:2023tbo}
S.~Capozziello, S.~Zare, L.~M.~Nieto and H.~Hassanabadi,
%``Modified Kerr black holes surrounded by dark matter spike,''
[arXiv:2311.12896 [gr-qc]].
%32 citations counted in INSPIRE as of 02 Aug 2025

%\cite{KumarWalia:2022aop}
\bibitem{KumarWalia:2022aop}
R.~Kumar Walia, S.~G.~Ghosh and S.~D.~Maharaj,
%``Testing Rotating Regular Metrics with EHT Results of Sgr A*,''
Astrophys. J. \textbf{939}, no.2, 77 (2022)
%%doi:10.3847/1538-4357/ac9623
[arXiv:2207.00078 [gr-qc]].
%80 citations counted in INSPIRE as of 02 Aug 2025




\end{thebibliography}
\end{document}